 \documentclass[11pt]{article}

\usepackage{amsmath,amsfonts,amssymb}

 \usepackage[numbers,sort&compress]{natbib}

\usepackage{hyperref}
\hypersetup{
   colorlinks   =  true,
    linkcolor    = blue,
    citecolor    = red,
     urlcolor	=blue,     
	urlbordercolor = {1 1 0}
}

\newcommand\be{\begin{equation}}
\newcommand\ee{\end{equation}}
\newcommand\bea{\begin{eqnarray}}
\newcommand\eea{\end{eqnarray}}

\usepackage{lmodern}
\usepackage[english]{babel}
\usepackage[utf8]{inputenc}
\usepackage[T1]{fontenc}

\usepackage{doi}
\usepackage{pgf,tikz,pgfplots}
\usepackage{tikz-cd}
\usetikzlibrary{babel}
\usepackage{svg}
\pgfplotsset{compat=1.15}
\usetikzlibrary{matrix,arrows}
\usepackage{graphicx}
\usepackage{adjustbox}
\usepackage{subcaption,booktabs}
\graphicspath{{graficos/}}

\usepackage{amsmath}
\usepackage{amsthm}
\usepackage{physics}
\usepackage{amssymb}
\usepackage{stmaryrd}
\usepackage{bm}
\usepackage{framed}

\newcommand{\I}{\mathrm{I}}

\newcommand{\set}[1]{\left\{ #1 \right\}} 

\newcommand{\R}{\mathbb{R}} 
\newcommand{\C}{\mathbb{C}} 

\newcommand{\cH}{\mathcal{H}} 

\newcommand{\cS}{\mathcal{S}}

\newcommand{\E}{\mathbf{E}} 
\newcommand{\M}{\mathbf{M}} 

\newcommand{\g}{\mathfrak{g}}
\newcommand{\h}{\mathfrak{h}}

\newcommand{\cN}{\mathcal{N}}
\newcommand{\gl}{\mathfrak{gl}}
\renewcommand{\sl}{\mathfrak{s}\mathfrak{l}}
\newcommand{\so}{\mathfrak{s}\mathfrak{o}}

\newcommand{\cL}{\mathcal{L}}

\renewcommand{\sp}{\mathfrak{s}\mathfrak{p}}

\newcommand{\ad}{\textup{ad}}

\newcommand{\cv}{\mathfrak{X}} 
\newcommand{\X}{\mathbf{X}} 
\newcommand{\Y}{\mathbf{Y}} 
\newcommand{\x}{\mathbf{x}} 
\newcommand{\cD}{\mathcal{D}}
\renewcommand{\div}{\textup{div}\:}

\newcommand{\Id}{\mathrm{Id}}

\renewcommand{\S}{\mathbb{S}}
\newcommand{\pr}{\mathrm{pr}}

\newcommand{\red}{\mathrm{red}}

\newcommand{\e}{\mathrm{e}}
\newcommand{\sgn}{\mathrm{sgn}}

\newcommand{\cR}{\mathcal{R}}
\newcommand{\Ann}{\mathrm{Ann}}
\theoremstyle{definition}
\newtheorem{defi}{Definition}[section] 
\newtheorem{example}[defi]{Example} 

\theoremstyle{plain}
\newtheorem{thm}[defi]{Theorem} 
\newtheorem{prop}[defi]{Proposition} 

\theoremstyle{remark}
\newtheorem*{remark}{Remark}

\numberwithin{equation}{section} 

\begin{document}
\
  \vskip0.5cm

  \begin{center}     
 \noindent
 {\Large \bf  
A representation-theoretical approach to higher-dimensional \\ [5pt] Lie--Hamilton systems:  The symplectic Lie algebra $\sp(4, \R)$} 
 
   \end{center}

\medskip

\begin{center}

{\sc  Rutwig Campoamor-Stursberg$^{1,2}$, Oscar Carballal$^{2}$\\[2pt] and Francisco J.~Herranz$^3$}

\end{center}

\medskip

 \noindent
$^1$ Instituto de Matem\'atica Interdisciplinar, Universidad Complutense de Madrid, E-28040 Madrid,  Spain

\noindent 
$^2$ Departamento de \'Algebra, Geometr\'{\i}a y Topolog\'{\i}a,  Facultad de Ciencias 
Matem\'aticas, Universidad Complutense de Madrid, Plaza de Ciencias 3, E-28040 Madrid, Spain

\noindent
{$^3$ Departamento de F\'isica, Universidad de Burgos, 
E-09001 Burgos, Spain}

 \medskip
 
\noindent  E-mail: {\small
 \href{mailto:rutwig@ucm.es}{\texttt{rutwig@ucm.es}}, \href{mailto:oscarbal@ucm.es}{\texttt{oscarbal@ucm.es}}, \href{mailto:fjherranz@ubu.es}{\texttt{fjherranz@ubu.es}} 
}
 
\medskip

\begin{abstract}
\noindent

A new procedure for the construction of higher-dimensional Lie--Hamilton systems is proposed. This method is  based on techniques belonging to the representation theory of Lie algebras and their realization by vector fields. The notion of intrinsic Lie--Hamilton system is defined, and a sufficiency criterion for this property given. Novel four-dimensional Lie--Hamilton  systems arising from the fundamental representation of the symplectic Lie algebra $\sp(4, \R)$ are obtained and proved to be intrinsic. Two distinguished subalgebras, the two-photon Lie algebra $\h_{6}$ and the Lorentz Lie algebra $\so(1,3)$, are also considered in detail. As applications, coupled time-dependent systems which generalize the Bateman oscillator and the one-dimensional Caldirola--Kanai models are constructed, as well as systems depending on a time-dependent electromagnetic field and generalized coupled oscillators. A superposition rule for these systems, exhibiting interesting symmetry properties, is obtained using the coalgebra method. 
\end{abstract}
\medskip
\medskip


 
\smallskip
\noindent
\textbf{PACS 2010 codes}:  {02.30.Hq, 02.20.Sv, 45.20.Jj, 45.10.Na}\\

\medskip

 \newpage

\tableofcontents


\section{Introduction}

While \textit{Lie systems}, i.e., first-order systems of ordinary differential equations (ODEs in short)  possessing a \textit{superposition rule} have been extensively studied \cite{LS,Ves,SH,BOU,CGM07}, the notion of \textit{Lie--Hamilton system} (LH system in short), that is, of Lie systems compatible with a certain symplectic structure, is relatively recent (see \cite{Lucas} and references therein). It turns out that these enriched Lie systems, which appear naturally in many physical applications, mainly in the context of Mechanics, have relevant geometric properties that allow an algorithmic construction of time-independent constants of the motion of the associated LH system \cite{Ballesteros2013}, hence eventually simplifying the deduction of superposition principles, which constitutes one a salient property of Lie systems, when compared to the classical analytical methods \cite{Carinena2009,CGM00}. In this sense, it is worthy to be mentioned that techniques developed in the context of superintegrability, such as the \textit{coalgebra formalism} \cite{Rag}, can be used to simplify the computation of the constants of the motion and algorithmize the problem \cite{Ballesteros2013}. So far, LH systems have been completely classified on the plane \cite{Ballesteros2015,Blasco2015}, while a systematic analysis of LH systems in higher dimensions, in the absence of complete classifications of Lie algebras of vector fields for dimensions greater than three, currently constitutes an open problem. One of the relevant questions that arises in this context is whether a given LH system admits a coordinate frame in which it decouples, i.e., it can be written as the union of lower-dimensional LH systems. Instead of using the classical approach by vector fields, in \cite{Campoamor2018,Campoamor2018a} an ansatz that uses the \textit{representation theory of Lie algebras} \cite{Wan1975} was proposed, providing an alternative method to construct nonlinear realizations and their associated Lie systems.       

\medskip  In this work, these ideas are refined and extended to the case of LH systems, in order to provide a systematization towards the construction of LH systems in higher dimensions. Using a reformulation in terms of (generalized) involutive distributions \cite{Cartan1945,FLAN,Sussman1973,Stefan1974,Palais1957}, a criterion that allows to decide whether a LH system is (locally) diffeomorphic to another one in lower dimension is proposed. In addition, the notion of coupling of low-dimensional LH systems is introduced, and some properties to determine if a system corresponds to an uncoupled sum are analyzed. The main focus of this work is on LH systems arising from the four-dimensional \textit{fundamental representation} of the \textit{symplectic Lie algebra} $\mathfrak{sp}(4,\mathbb{R})$, that provides genuine new types of LH systems in four dimensions that are \textit{intrinsic}. Two relevant subalgebras, the \textit{Lorentz algebra} $\mathfrak{so}(1,3)$ and the \textit{two-photon Lie algebra} $\h_{6}$ (isomorphic to the \textit{extended Schr\"odinger algebra} $\widehat{\mathcal{S}}(1)$) are also considered, and it is shown how differential invariants arising from the corresponding embedding of Lie algebras can be profitably used to construct superposition principles where the application of the coalgebra formalism becomes too cumbersome.

\medskip  

This paper is structured as follows: after recalling the main properties of LH systems and the coalgebra method, in Section~\ref{section:representations}, we analyze Lie systems determined by a faithful representation of a Lie algebra. Of particular interest are those LH systems of lower dimension that are constructed by the so-called \textit{reduction by invariants}, that amounts to use differential invariants of a linear realization by vector fields of a Lie algebra \cite{Campoamor2018}. This property can be reformulated in terms of differential forms, formally corresponding to the annihilator of the (generalized) distribution determined by the vector fields spanning the Vessiot--Guldberg Lie algebra associated to the system \cite{Cartan1945,FLAN}. This allows us to introduce the notion of local diffeomorphism between Lie systems defined in spaces of different dimension, as well as to analyze whether a Lie system is intrinsic, i.e., if it is not locally diffeomorphic to a lower-dimensional Lie system or isomorphic to an uncoupled sum of two Lie systems. 
As an illustration, it is shown that the $\sl(2,\R)$-Lie systems on the real plane arise from the Lie system associated to the adjoint representation of $\mathfrak{sl}(2,\mathbb{R})$. In Section~\ref{section:twophoton} we consider the two-photon Lie algebra $\h_{6} \simeq \widehat{\cS}(1)$ and show that the Lie system associated to a faithful representation in $\mathbb{R}^4$ actually possesses the structure of a LH system. Its reduction by invariants leads to a three-dimensional Lie system with the salient property that its projection on one of its coordinates recovers the LH system on the plane associated to the P$_5$-class \cite{Lucas,Ballesteros2015}. Using the coalgebra formalism and a differential invariant of the prolongation, a superposition principle for these systems is deduced. In Section~\ref{section:Lorentz}, we study the LH system that corresponds to the irreducible four-dimensional representation of the Lorentz algebra $\mathfrak{so}(1,3)$. It is shown that this system is intrinsic, providing genuine new examples of higher-dimensional LH systems. Although a superposition rule can be obtained using the coalgebra formalism, we skip its explicit computation, as it is observed that this LH system is a special case of LH systems associated to the symplectic Lie algebra $\mathfrak{sp}(4,\mathbb{R})$. As applications of the $\mathfrak{so}(1,3)$-LH system, we construct several coupled systems, such as a $t$-dependent Bateman and a coupled Caldirola--Kanai Hamiltonians, that generalize the Bateman oscillator  and the one-dimensional Caldirola--Kanai models, respectively. In Section~\ref{section:symplectic} we analyze the LH system that arises from the four-dimensional fundamental representation of $\mathfrak{sp}(4,\mathbb{R})$, showing its intrinsic character. A superposition principle exhibiting some interesting symmetry properties is derived by means of the coalgebra method. It is further shown that this superposition rule can be applied to the $\mathfrak{so}(1,3)$-LH systems previously studied, as well as to any other LH system obtained from a proper subalgebra of $\mathfrak{sp}(4,\mathbb{R})$. As applications, various generalized coupled oscillators and a time-dependent Hamiltonian that describes the motion of particles constrained by a time-dependent electromagnetic field are constructed. Finally, in Section~\ref{section:concluding}, some conclusions and potential extensions of the work are drawn. In particular, the relation between LH systems reduced to odd dimensions by means of the reduction by invariants and the existence of contact structures is addressed to. For completeness in the exposition, in Appendix~\ref{section:app} we briefly recall the main algebraic notions and structures used in the coalgebra formalism.

\medskip
Unless otherwise stated, the Einstein summation convention is used in this work.

\section{Lie--Hamilton systems and the coalgebra formalism} \label{subsection:coalgebra}

Any non-autonomous first-order ODE system on an $n$-dimensional smooth manifold  $M$  
\begin{equation}
 \frac{{\rm d} x^j}{{\rm d} t  }=\psi^j(t,{\bf x}), \qquad \mathbf{x} \in M, \qquad 1\leq j\leq n,  
 \label{system}
\end{equation}
can formally be rewritten by means of a $t$-dependent vector field ${\bf X}:\mathbb{R}\times M \rightarrow TM$ locally given by 
\begin{equation}\label{Vect}
{\bf X}(t,{\bf x}):= \psi^j(t,{\bf x})\frac{\partial}{\partial x^j},
\end{equation}
allowing us to identify the system \eqref{system} with the $t$-dependent vector field ${\bf X}$. We say that \eqref{system} is a \textit{Lie system} if the general solution can be written as 
\begin{equation}\label{spp}
{\bf x}(t)= \Psi\left({\bf x}_1(t),\dots ,{\bf x}_s(t);k_1,\dots ,k_n\right),
\end{equation}
where ${\bf x}_\ell(t)$ is a particular solution ($1\leq \ell\leq s$) and $(k_1,\dots ,k_n) \in M$ is a point related to the initial conditions. Equation \eqref{spp} is usually called a (nonlinear) \textit{superposition rule}, and generalizes the well-known notion of superposition valid for linear (non-autonomous) ODE systems to more wide classes  \cite{LS,CGM07,Hur,Ovs}. In his pioneering work, S. Lie showed that the existence of a superposition principle can be reformulated in purely algebraic terms, by means of Lie algebras of vector fields: 

\smallskip
\noindent
{\bf Lie--Scheffers Theorem:} A system \eqref{system} is a Lie system if and only if the $t$-dependent vector field $\bf X$ in \eqref{Vect} admits the decomposition 
\begin{equation*}\label{LieS}
{\bf X}(t,{\bf x})=\sum_{\alpha = 1}^{r} b_{\alpha}(t){\bf X}_{\alpha}({\bf x}),
\end{equation*}
for some functions $b_1(t),\ldots,b_{r}(t) \in C^{\infty}(\R)$ and vector fields ${\bf X}_1,\ldots,{\bf X}_{r}$ on $M$ spanning an $r$-dimensional real Lie algebra $V^{X}$, usually called a {\it Vessiot--Guldberg Lie algebra} (VG Lie algebra in short) of ${\bf X}$. In addition, the constraint 
\begin{equation*}
n\;s \geq \dim V^{X}
\end{equation*}
 must be satisfied (Lie's condition). 

Thus, a Lie system $\X$ can be regarded as a curve taking values in a Lie algebra of vector fields of finite dimension, a VG Lie algebra $V^{X}$. Even if the Lie--Scheffers Theorem provides a criterion that guarantees the existence of a superposition principle, it does not give any procedure on how to find it. In this context, the compatibility of Lie systems with additional geometrical structures, such as symplectic ones, enables to apply the so-called coalgebra formalism for an algorithmic construction of time-independent constants of the motion \cite{Ballesteros2013}.  

  A Lie system ${\bf X}$ on a symplectic manifold $(M, \omega)$ is  a LH system if it admits a VG algebra $V^{X}$ of Hamiltonian vector fields with respect to a symplectic form $\omega$. This implies that the invariance condition
\begin{equation*}\label{der}
\mathcal{L}_{{\bf X}_ {\alpha}}\omega=0,\qquad 1\leq  \alpha\leq r
\end{equation*}
is satisfied for the generators ${\bf X}_1,\ldots,{\bf X}_{r}$ of $V^{X}$ (see e.g. \cite{Ballesteros2015}). Each vector field ${\bf X}_{\alpha}$  is associated to a Hamiltonian function $h_{\alpha}$ determined by the prescription 
\begin{equation*}
\iota_{{\bf X}_{\alpha}}\omega={\rm d}h_ {\alpha}, \qquad 1 \leq  \alpha \leq r,
\label{contract}
\end{equation*}
where $\iota_{{\bf X}_ {\alpha}}\omega$ denotes the inner product of ${\bf X}_ {\alpha}$ with the symplectic form $\omega$. As $\omega$ is non-degenerate, any function $f \in C^{\infty}(M)$ induces a unique Hamiltonian vector field ${\bf X}_f$. This allows us to define a Poisson bracket on $C^\infty(M)$  by
 \begin{equation*}\label{LB}
 \{\cdot,\cdot\}_\omega\ :\ C^{\infty}(M) \times C^{\infty}(M)\ni (f_1,f_2)\mapsto \X_{f_2} f_1\in C^{\infty}(M),
 \end{equation*}
hence turning $(C^\infty(M),\{\cdot,\cdot\}_\omega)$ into a Lie algebra. Now, the space ${\rm Ham}(\omega)$ of  Hamiltonian vector fields on $M$ (with respect to $\omega$) also inherits the structure of a Lie algebra. Actually, provided that $M$ is connected, the following exact sequence of Lie algebras holds (see \cite{Vaisman1994}):
\begin{equation*}\label{seq}
0\hookrightarrow \mathbb{R}\hookrightarrow (C^\infty(M),\{\cdot,\cdot\}_\omega)\stackrel{\varphi}{\longrightarrow} ({\rm Ham}(\omega),[\cdot,\cdot])\stackrel{\pi}{\longrightarrow} 0,
\end{equation*} 
where $\pi$ denotes the projection and $\varphi$ maps each $f\in C^\infty(M)$ onto the Hamiltonian vector field ${\bf X}_{-f}$.
It follows that the Hamiltonian functions  $h_1,\ldots,h_{r}$ generate a finite-dimensional (functional) Lie algebra contained in $\varphi^{-1}(V^{X})$. We call it a \textit{Lie--Hamilton algebra} (LH  algebra in short) $\cH_\omega$  of  ${\bf X}$.  

\medskip  
We briefly recall how coalgebra method can be used for the construction of $t$-independent constants of the motion of LH systems (see \cite{Blasco2015,Lucas,Ballesteros2013} for further details). 

Let $\cH_{\omega}$ be the LH algebra of a LH system $\X$ on $(M, \omega)$ with generators $\set{h_{1}, \ldots, h_{r}}$ and commutation rules 
\[ \set{h_{\alpha}, h_{\beta}}_{\omega} = C_{\alpha \beta}^{\gamma} h_{\gamma}, \qquad 1 \leq \alpha, \beta, \gamma \leq r. \]
Now, considering  $\mathcal{H}_{\omega}$ as a Lie--Poisson algebra with generators $\{ v_1,\dots,v_r \}$, these satisfy the brackets (see Appendix~\ref{section:app}) 
\begin{equation}\label{ca}	
\{v_{\alpha},v_{\beta}\} =  C_{\alpha \beta}^{\gamma}v_{\gamma},\qquad 1 \leq \alpha, \beta, \gamma\leq r.
\end{equation}
Let  $S\left({\cal H}_\omega\right)$ be the symmetric algebra  of ${\cal H}_\omega$, seen as a Poisson algebra, with the Poisson bracket given by (\ref{ca}) (the technical details can be found in the Appendix~\ref{section:app} and \cite{Lucas}). Recall that $S\left({\cal H}_\omega\right)$ can be endowed with  structure of a coalgebra by means of the primitive coproduct map $\Delta$ given  by
\begin{equation}
 {\Delta} :S\left({\cal H}_\omega \right)\rightarrow
S\left({\cal H}_\omega\right) \otimes S\left({\cal H}_\omega\right)    ,\qquad      {\Delta}(v_{\alpha}):=v_{\alpha}\otimes 1+1\otimes v_{\alpha},  \qquad   1 \leq \alpha \leq r,
\label{cb}
\end{equation}
defining a Poisson algebra morphism of (\ref{ca}). The trivial counit and the antipode can also be defined, endowing $S(\cH_{\omega})$ with a Hopf algebra structure \cite{majid}. Now the coassociativity condition 
\begin{equation*}
(\Delta \otimes \mathrm{Id}) \circ \Delta = ( \mathrm{Id} \otimes \Delta) \circ \Delta
\end{equation*}
is used to extend the 2-coproduct $\Delta\equiv \Delta^{(2)}$ to a coproduct of third-order
\begin{equation*}
\Delta^{(3)} : S(\cH_{\omega}) \to S(\cH_{\omega}) \otimes S(\cH_{\omega}) \otimes S(\cH_{\omega}) \equiv S^{(3)} (\cH_{\omega})
\end{equation*}
given by  
\begin{equation*}\label{cc}
\begin{split}
&\Delta^{(3)}:=(\Delta \otimes {\rm Id}) \circ \Delta=({\rm Id} \otimes \Delta) \circ \Delta ,\\
&{\Delta}^{(3)}(v_{\alpha})=v_{\alpha}\otimes 1\otimes 1 +1\otimes v_{\alpha}\otimes 1+1\otimes 1\otimes v_{\alpha}, \qquad 1 \leq \alpha \leq r.
\end{split}
\end{equation*}
By recursion, the procedure can be extended to a  $k$th-order coproduct  defined by
\begin{equation*}\label{cd}
\begin{split}
&\Delta ^{(k)}: \ S({\cal H}_\omega) \rightarrow S(\cH_{\omega}) \otimes \overset{k}{\cdots} \otimes S(\cH_{\omega}) \equiv S^{(k)}(\cH_{\omega}) ,\\
&{\Delta}^{(k)}:= \bigr( {\Id}   \otimes  \overset{(k-2)}{\cdots} \otimes {\rm Id} \otimes {\Delta^{(2)}}  \bigr)\circ \Delta^{(k-1)}.
\end{split}
\end{equation*}

Now, elements in $S({\cal H}_\omega)$ can be interpreted as polynomial functions on the dual ${\cal H}_\omega^*$ (see Appendix~\ref{section:app}), implying that the coproduct (\ref{cb}) in $S(\mathcal{H}_\omega)$ can be expanded to  
\begin{equation*}
 {\Delta} :C^\infty\left({\cal H}_\omega^*\right)\rightarrow
C^\infty\left({\cal H}_\omega^*\right) \otimes C^\infty\left({\cal H}_\omega^*\right),
\label{ce}
\end{equation*}
and iterating the procedure for the higher-order coproduct. It follows that $C^\infty(\mathcal{H}^*_\omega)$ is endowed with the structure of a Poisson coalgebra. 

\smallskip
If the   Hamiltonian functions  $\{ h_1,\dots,h_r \}$ span the LH algebra ${\cal H}_\omega$, we define the Lie algebra morphism  
\begin{equation} \label{corre}
\phi:{\cal H}_\omega\rightarrow C^\infty(N),\qquad \phi(v_\alpha):= h_{\alpha},\qquad 1 \leq \alpha \leq r,
\end{equation}
where $N \subset M$ is a domain ensuring that all the functions $h_{\alpha}$ are well defined. This allows further to construct the following morphisms of Poisson algebras:
\begin{equation}\label{cg}
\begin{array}[c]{ll}
D: C^\infty\left( {\cal H}_\omega^* \right) \rightarrow C^\infty(N), \\[0.1cm]
 D^{(k)} : C^{\infty}(\cH_{\omega}^{*}) \otimes \overset{k}{\cdots} \otimes C^{\infty}(\cH_{\omega}^{*})   \rightarrow C^{\infty}(N) \otimes \overset{k}{\cdots} \otimes C^{\infty}(N) \subset C^{\infty}(N^{k}) ,\\[0.1cm]
D( v_{\alpha}):= h_{\alpha}({\bf x}_1), \qquad D^{(k)} \left( {\Delta}^{(k)}(v_{\alpha}) \right):= h_{\alpha}({\bf x}_1)+\cdots + h_{\alpha}({\bf x}_k):= h_{\alpha}^{(k)},\qquad 1 \leq \alpha \leq r,
\end{array}
\end{equation}
where  ${\bf x}_j=\left( (x_1)_j,\dots ,(x_n)_j\right)$ denotes the coordinates in the $j^\textup{th}$-copy submanifold $N$ within the product $N \times \overset{k}{\cdots} \times N \equiv N^{k}$. 
In the generic case of a $k${{th}}-order tensor product of elements $u_{i}(v_{1}, \ldots, v_{r}) \in C^{\infty}(\cH_{\omega}^{*})$ $(1 \leq i \leq k)$, the morphism $D^{(k)}$ acts as the product of functions (see \cite{Ballesteros2013,Ballesterosrev2021}) 
\[ D^{(k)}(u_{1}(v_{1}, \ldots, v_{r}), \ldots, u_{k}(v_{1}, \ldots, v_{r})) = u_{1}(h_{1}(\x_{1}), \ldots, h_{r}(\x_{1})) \cdots u_{k}(h_{1}(\x_{k}), \ldots, h_{r}(\x_{k})). \]
If now $C^\infty\left({\cal H}_\omega^*\right)$ admits the Casimir  invariant $C=C(v_1,\dots,v_r)$,\footnote{Recall that Casimir invariants correspond to elements $C$ that Poisson-commute with any $v_{\alpha}$ with respect to the Poisson bracket (\ref{ca})} the functions constructed by 
\begin{equation}\label{cj}
F:= D(C),\qquad F^{(k)}(h_1,\dots,h_r) := D^{(k)}\left[\Delta^{(k)} \left({C( v_1,\dots,v_r)} \right) \right], \quad 2 \leq k \leq s+1,
\end{equation}
are  $t$-independent constants of the motion for the diagonal prolongation
$\widetilde {\bf X}$  of the LH system ${\bf X}$  to the product  manifold $N^{s+1}\subset M^{s+1}$ (see \cite{Ballesteros2013} for details). 
Each $F^{(k)}$ can thus  be interpreted as an element in $C^\infty(N^{s+1})$ for $k\leq s+1$.  Assuming that the $F^{(k)}$ do not reduce to constants, they form a set of  $s$ functionally independent functions in involution. Further constants of the motion can be derived using the permutation $S_{ij}$ of the variables ${\bf x}_{i}\leftrightarrow {\bf x}_j$:
\begin{equation}\label{ck}
F_{ij}^{(k)}=S_{ij} \left( F^{(k)}   \right) , \quad 1\le  i<j\le  s+1.
 \end{equation}
The sets (\ref{cj}) and (\ref{ck}) can be used, after some heavy algebraic manipulations, to determine a superposition principle for the LH system ${\bf X}$ in a direct and systematic way \cite{Ballesteros2013}.

\section{Lie systems from representations} \label{section:representations}

 Let $\g$ be an $r$-dimensional real Lie algebra with generators $X_{1}, \ldots, X_{r}$, and consider an $n$-dimensio\-nal  faithful representation  $\Gamma: \g \to \gl(n, \R)$ of $\g$. The representation naturally induces a (linear) realization $\Phi_{\Gamma}: \g \to \cv(\R^{n})$ by vector fields on $\R^{n}$ given by (see e.g. \cite{Campoamor2018,Campoamor2018a})
 \begin{equation}
 \X_{\alpha} := \Phi_{\Gamma}(X_{\alpha}):= x^{i} \Gamma(X_{\alpha})_{i}^{j} \pdv{x^{j}}, \qquad 1 \leq \alpha \leq r,
 \label{eq:Reduction:vfrealization}
 \end{equation}
 where $(x^{1}, \ldots, x^{n})$ are global coordinates in $\R^{n}$. The $t$-dependent vector field
 \begin{equation}
  \X := \sum_{\alpha = 1}^{r} b_{\alpha}(t) \X_{\alpha},
 \label{eq:Reduction:tvf}
 \end{equation}
 where $b_{\alpha}(t) \in C^{\infty}(\R)$ is an arbitrary $t$-dependent function $(1 \leq \alpha \leq r)$, determines a Lie system on $\R^{n}$ whose VG Lie algebra $V^{X}$, generated by $\X_{1}, \ldots, \X_{r}$, is isomorphic to the Lie algebra $\g$. The corresponding first-order ODEs system is given by 
 \begin{equation}\label{sislin}
 \dv{x^{j}}{t}= \sum_{\alpha=1}^{r} \sum_{i=1}^{n}b_\alpha(t) x^{i} \Gamma(X_{\alpha})_{i}^{j}, \qquad 1 \leq j \leq n.
\end{equation}
As a system of ODEs, \eqref{sislin} is a linear system, and hence always admits a fundamental system of solutions. Even in this case, to find such a fundamental system is usually a difficult problem, unless the coefficient matrix ${\bf A}$ with entries 
\begin{equation*}
\left({\bf A}\right)_{i}^{j}= \sum_{\alpha=1}^{r} b_\alpha(t)  \Gamma(X_{\alpha})_{i}^{j}, \qquad 1 \leq i, j \leq n
\end{equation*}
satisfies the constraint 
\begin{equation*}
\left[ {\bf A}, \int {\bf A} {\rm d}t\right]=0,
\end{equation*}
in which case the problem can be solved with exponentials \cite{Hur}. Otherwise, the knowledge of a group of symmetries allows to reduce the number of equations of the system. In particular, if the symmetry group is solvable and $r$-dimensional, then the system can be solved by quadratures alone \cite{OLV}. 
 
\medskip
In any case, this procedure shows that linear realizations of Lie algebras by vector fields, and hence higher-dimensional Lie systems (viz. non-autonomous linear systems) are naturally associated to faithful representations of Lie algebras. This approach can be used, assuming that the coefficient matrix ${\bf A}$ is not of maximal rank, as a starting point to construct lower-dimensional Lie systems whose associated first-order systems are no more linear, by means of the so-called reduction by invariants \cite{Campoamor2018}. Prior to the systematization of the method, we analyze the case of low-dimensional LH systems based on $\mathfrak{sl}(2,\mathbb{R})$ as an illustration.

\subsection{Reduction of $\mathfrak{sl}(2,\mathbb{R})$-systems}
\label{subsection:sl2}
Among Lie systems, those with a VG algebra isomorphic to $\frak{sl}(2,\mathbb{R})$ correspond to the most studied class. In this paragraph we show how to derive the $\frak{sl}(2,\mathbb{R})$-Lie systems in two dimensions from the three-dimensional Lie system associated to the adjoint representation, and how the latter can be reduced to systems on suitable (integral) submanifolds. We consider $\sl(2, \R)$ on a basis $\left\{h,e_{\pm}\right\}$ with commutation relations 
 \begin{equation*}
[h,e_{\pm}] = \pm e_{\pm}, \qquad [e_{-}, e_{+}] = 2h.
 \end{equation*}
The realization $\Phi_{\ad}: \sl(2, \R) \to \cv(\R^{3})$ induced by the adjoint representation $\ad: \sl(2, \R) \to \gl(3, \R)$ is given by the three vector fields 
 \begin{equation}
 \begin{split}
& \X_{1} := \Phi_{\ad} (e_{-}) = x^{1} \pdv{x^{2}} + 2x^{2} \pdv{x^{3}}, \qquad \X_{2} := \Phi_{\ad}(h) = - x^{1} \pdv{x^{1}} + x^{3} \pdv{x^{3}},\\
 & \X_{3} := \Phi_{\ad}(e_{+}) = - 2 x^{2} \pdv{x^{1}} - x^{3} \pdv{x^{2}},
 \end{split}
 \label{eq:Reduction:realizationsl2}
  \end{equation}
  expressed in the global coordinates $(x^{1}, x^{2}, x^{3})$ of $\R^{3}$. The $t$-dependent vector field 
  \begin{equation*}
  \X := \sum_{\alpha = 1}^{3} b_{\alpha}(t) \X_{\alpha},\quad b_{\alpha} \in C^{\infty}(\R); \qquad 1 \leq \alpha \leq 3
  \end{equation*}
 defines a Lie system in $\R^{3}$ with VG Lie algebra isomorphic to $\sl(2, \R)$ and associated first-order system of ODEs
\begin{equation}
  \dv{x^{1}}{t} = - b_{2}(t) x^{1} - 2 b_{3}(t) x^{2}, \qquad \dv{x^{2}}{t} =b_{1}(t) x^{1} - b_{3}(t) x^{3}, \qquad \dv{x^{3}}{t} = 2b_{1}(t) x^{2} + b_{2}(t) x^{3}.
\label{eq:Reduction:systemsl2}
\end{equation}
As system in $\mathbb{R}^3$, (\ref{eq:Reduction:systemsl2}) is linear with non-constant coefficients, and can thus be (formally) solved using standard procedures \cite{OLV}. In particular, if the coefficient functions $b_{\alpha}$ satisfy the dependence relation   $\Delta=2b_1(t)b_2(t)b_3(t)+\dv{b_3(t)}{t}b_1(t)-b_3(t)\dv{b_1(t)}{t}=0$, the system is easily seen to be solvable by quadratures. If $\Delta\neq 0$, then a straightforward manipulation shows that 
\begin{equation*}
\begin{split}
x^{1}(t)=\frac{1}{\Delta}\left(-b_3(t)\frac{{\rm d}^2x^{2}}{{\rm d}t^2}+\left(\frac{{\rm d}b_3}{{\rm d}t}+b_2(t)b_3(t)\right)\frac{{\rm d}x^{2}}{{\rm d}t} -4b_1(t)b_3(t)^2x^{2}\right),\\
x^{3}(t)=\frac{1}{\Delta}\left(-b_1(t)\frac{{\rm d}^2x^{2}}{{\rm d}t^2}+\left(\frac{{\rm d}b_1}{{\rm d}t}-b_1(t)b_2(t)\right)\frac{{\rm d}x^{2}}{{\rm d}t} -4b_1(t)^2 b_3(t)x^{2}\right),
\end{split}
\end{equation*}
while $x^2(t)$ is a solution to the third-order linear equation
\begin{equation*}
\frac{{\rm d}^3\xi}{{\rm d}t^3}- \frac{1}{\Delta}\frac{{\rm d}\Delta}{{\rm d}t}\frac{{\rm d}^2\xi}{{\rm d}t^2}+A_2(t)\frac{{\rm d}\xi}{{\rm d}t}+A_{3}(t)\xi=0,
\end{equation*}
where 
\begin{equation*}
\begin{split}
A_2(t)=&\frac{1}{\Delta}\left(\frac{{\rm d}^2b_1}{{\rm d}t^2}\frac{{\rm d}b_3}{{\rm d}t}-\frac{{\rm d}b_1}{{\rm d}t}\frac{{\rm d}^2b_3}{{\rm d}t^2} +4b_1(t)b_3(t)\left(\frac{{\rm d}b_1}{{\rm d}t}b_3(t)-b_1(t)\frac{{\rm d}b_3}{{\rm d}t}\right)-4b_1(t)b_3(t)b_2(t)^3 \right. \\
& \left. -8b_1(t)^2b_3(t)^2+6\frac{{\rm d}b_1}{{\rm d}t}\frac{{\rm d}b_3}{{\rm d}t}+\left(\frac{{\rm d}b_2}{{\rm d}t}-b_2(t)\right)\left(\frac{{\rm d}b_1}{{\rm d}t}b_3(t)+b_1(t)\frac{{\rm d}b_3}{{\rm d}t}\right)-3b_2(t)^2\Delta  \right),\\
A_3(t)=& -4\frac{b_1(t)b_3(t)}{\Delta}\frac{{\rm d}\Delta}{{\rm d}t}+6\left(\frac{{\rm d}b_1}{{\rm d}t}b_3(t)+b_1(t)\frac{{\rm d}b_3}{{\rm d}t}\right).
\end{split}
 \label{eq:IntC1}
\end{equation*}
The coefficients $A_2(t),A_3(t)$ eventually provide additional integrability conditions that enable to solve the system.

\begin{figure}[t]
\centering
\includegraphics[width=\textwidth]{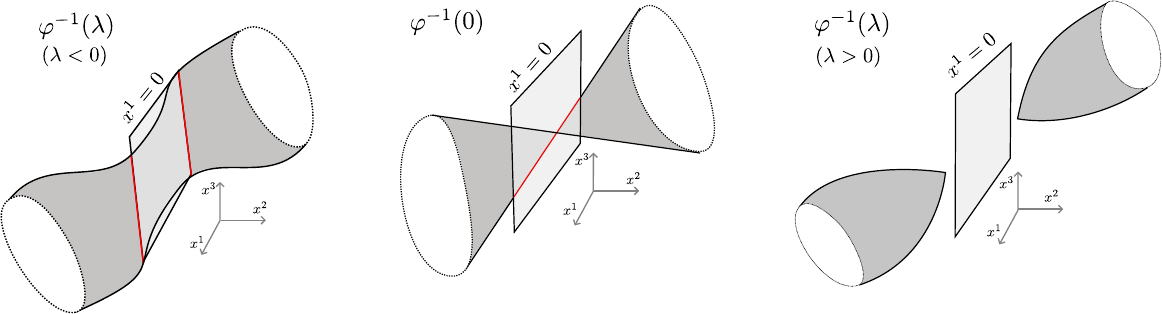}
\caption{The regular surfaces $\varphi^{-1}(\lambda) \subset U$ corresponding to the integral submanifolds of $\cD^{X}$: for $\lambda  <0$ we have a one-sheeted hyperboloid without two lines; for $\lambda = 0$ a cone with one of its generating lines removed, while the two-sheeted hyperboloid appears when $\lambda > 0$.}
\label{fig:surfaces}
\end{figure}

\bigskip
We can, however, proceed differently, and consider the vector fields \eqref{eq:Reduction:realizationsl2} as the generators of a (generalized) smooth distribution $\cD^{X} \subset T \R^{3}$ which is no longer integrable in the classical sense of Fr\"obenius, but in the Stefan--Sussman sense \cite{Sussman1973,Stefan1974}. Nevertheless, if we consider the restriction  $\cD^{X} \vert_{U}$  of $\cD^{X}$ to the open subset $U := \set{(x^{1}, x^{2}, x^{3}) \in \R^{3}: x^{1} \neq 0} \subset \R^{3}$, we see that $\cD^{X} \vert_{U} \subset TU$ is an involutive distribution (in the sense of Fr\"obenius) of constant rank $2$. Its annihilator $\Ann (\cD^{X} \vert_{U}) \subset \Omega(U)$ is locally generated by the exact $1$-form $\omega={\rm d}\varphi \in \Omega^{1}(U)$, where 
  \begin{equation}
  \varphi = x^{1} x^{3} - (x^{2})^{2} \in C^{\infty}(U) 
  \label{eq:Reduction:invariant}
  \end{equation}
  corresponds to an invariant of the realization \eqref{eq:Reduction:realizationsl2}. With respect to the ODEs system \eqref{eq:Reduction:systemsl2}, the constraint $\dot{\varphi} = 0$ implies that $y^{3} := \varphi$ can be taken as a cyclic coordinate of the local coordinate system   $(y^{1} := x^{1}, y^{2} := x^{2}, y^{3} = \varphi)$ on $U$. Moreover, the coordinate $x^{3}$ can be expressed on $U$ as 
  \begin{equation*}
  x^{3} = \frac{\lambda + (x^{2})^{2}}{x^{1}} =  \frac{\lambda + (y^{2})^{2}}{y^{1}}, 
  \label{eq:Reduction:constraint}
  \end{equation*}
  for a certain regular value $\lambda \in \R$ of the invariant \eqref{eq:Reduction:invariant}, so that $\varphi^{-1}(\lambda) \subset U$ is a regular surface (see Figure~\ref{fig:surfaces}).
Therefore, the realization \eqref{eq:Reduction:realizationsl2} can be reduced to a realization on $\R^{2}_{y^{1} \neq 0}$ spanned by the three vector fields 
  \begin{equation}
  \X^{\red}_{1} = y^{1} \pdv{y^{2}}, \qquad \X^{\red}_{2} = - y^{1} \pdv{y^{1}}, \qquad \X^{\red}_{3} = - 2y^{2} \pdv{y^{1}} - \left( \frac{\lambda + (y^{2})^{2}}{y^{1}} \right) \pdv{y^{2}},
  \label{eq:Reduction:reducedrealization}
  \end{equation}
 i.e., the Lie system \eqref{eq:Reduction:realizationsl2} on $\R^{3}$ reduces to the Lie system $\X^{\red}$ on $\R^{2}_{y^{1} \neq 0}$ associated to the realization \eqref{eq:Reduction:reducedrealization} \cite{Campoamor2018}. 
 Considering the change of coordinates given by 
  \begin{equation*}
  z^{1} := \frac{y^{2}}{y^{1}}, \qquad z^{2} := \frac{\beta}{y^{1}}, \qquad    y^{1} = \frac{\beta}{z^{2}}, \qquad y^{2} = \frac{\beta z^{1}}{z^{2}}, \qquad \beta \neq 0
  \end{equation*}
 the vector fields \eqref{eq:Reduction:reducedrealization}  on $\R^{2}_{y^{1} \neq 0}$ are transformed onto the vector fields 
  \begin{equation}
  \X^{\red}_{1} = \pdv{z^{1}}, \qquad \X^{\red}_{2} = z^{1} \pdv{z^{1}} + z^{2} \pdv{z^{2}}, \qquad \X^{\red}_{3} = \left( (z^{1})^{2} - c (z^{2})^{2} \right) \pdv{z^{1}} + 2 z^{1} z^{2} \pdv{z^{2}},
  \label{eq:Reduction:reducedrealization_standar}
  \end{equation}
  on $\R^{2}_{z^{2} \neq 0}$, where $c := \lambda \beta^{-2}$ is a constant labelling the realization of $\sl(2, \R)$ \cite{Blasco2015}, from which the three non-equivalent classes of realizations of $\sl(2, \R)$ on $\R^{2}$ for the values $c = 0, \pm 1$ are obtained \cite{GonzalezLopez1992}. We conclude that the Lie systems  $\X^{\red}$ on $\R^{2}_{z^{2} \neq 0}$ and $\X$ on the quadratic surface $\varphi^{-1}(\lambda) \subset U$ are  diffeomorphic via the  diffeomorphism 
\begin{equation*}
\phi: \R^{2}_{z^{2} \neq 0} \to \varphi^{-1}(\lambda) \subset U, \qquad (z^{1}, z^{2}) \mapsto \left( \frac{\beta}{z^{2}}, \frac{\beta z^{1}}{z^{2}}, \frac{\beta^{2} (z^{1})^{2} + \lambda (z^{2})^{2}}{\beta z^{2}} \right).
\label{eq:Reduction:localdiffeo}
\end{equation*}

\medskip
  In particular, for the value $c = + 1$, the vector fields \eqref{eq:Reduction:reducedrealization_standar} are conformal symmetries of the Euclidean plane $\E^{2} := \R^{2}$ equipped with the Riemannian metric $\dd s^{2} = \dd (z^{1})^{2} + \dd (z^{2})^{2}$,  interpreted as a translation, a dilation and a specific conformal transformation, respectively.  The reduced Lie system $\X^{\red}$ on $\R^{2}_{z^{2} \neq 0}$ is thus associated to the $t$-dependent vector field
  \begin{equation*}
  \X^{\red} := \sum_{\alpha = 1}^{3} b_{\alpha}(t) \X^{\red}_{\alpha},
  \end{equation*}
  and the corresponding first-order ODEs system is given by 
  \begin{equation*}
  \dv{z^{1}}{t} = b_{1}(t) + b_{2}(t) z^{1} + b_{3}(t)  \left( (z^{1})^{2} - c(z^{2})^{2} \right), \qquad \dv{z^{2}}{t} = b_{2}(t) z^{2} + 2b_{3}(t) z^{1} z^{2}.
  \label{eq:Reduction:systemstandar}
 \end{equation*}
We observe that, while the second equation is linear in $z^2$, the first equation is always a generalized Riccati equation for $b_3(t)\neq 0$, showing that the reduced system does not necessarily preserve the integrability properties of the original system.

\subsection{Differential forms reducing Lie systems} Using the ideas of the previous paragraph, we now discuss the reduction procedure developed in \cite{Campoamor2018}  for arbitrary Lie algebras and faithful representations, generalizing the result in Theorem~\ref{thm:Reduction}. Let $\X$ be a Lie system on $\R^{n}$, whose associated first-order system of ODEs is of the form 
\begin{equation}
\dv{x^{j}}{t} = \sum_{\alpha = 1}^{r} b_{\alpha}(t) \xi^{j}_{\alpha}(\x), \qquad 1 \leq j \leq n, \qquad \x := (x^{1}, \ldots, x^{n}),
\label{eq:Reduction:gen:system}
\end{equation} 
where the vector fields 
\begin{equation}
\X_{\alpha} := \xi_{\alpha}^{j}(\x) \pdv{x^{j}}, \qquad  1 \leq \alpha \leq r
\label{eq:Reduction:gen:vf}
\end{equation}
span a VG Lie algebra $V^{X}$ of the Lie system $\X$ isomorphic to an $r$-dimensional real Lie algebra $\g$. In particular, if 
the Lie system $\X$ is induced by a faithful representation $\Gamma: \g \to \gl(n, \R)$ as in \eqref{eq:Reduction:tvf}, the component functions are given in terms of the matrix elements of the representation (see \eqref{eq:Reduction:vfrealization}): 
\begin{equation*}
\xi_{\alpha}^{j}(\x) = x^{i} \Gamma(X_{\alpha})_{i}^{j}, \qquad 1 \leq j \leq n,\quad 1 \leq \alpha \leq r.
\end{equation*}

\medskip
 Suppose now that there exists some index $1 \leq j_{0} \leq n$ such that the components $\xi_{\alpha}^{j_{0}}$ of the system \eqref{eq:Reduction:gen:system} vanish identically, which is equivalent to $ \dd x^{j_{0}}(\X_{\alpha}) = \X_{\alpha}(x^{j_{0}}) = 0$ for all $1 \leq \alpha \leq r$. With respect to the system \eqref{eq:Reduction:gen:system}, the constraint $\dot{x}^{j_{0}} = 0$ implies that $x^{j_{0}} = \lambda_{0}$ is constant, hence it can be regarded as a cyclic coordinate. This allows us to reduce the system \eqref{eq:Reduction:gen:system} by neglecting the $x^{j_{0}}$ coordinate, yielding to a Lie  a system $\X^{\red}$ on $\R^{n-1}$ given by 
\begin{equation*}
\dv{x^{j}}{t} = \sum_{\alpha = 1}^{r} b_{\alpha}(t)\xi_{\alpha}^{j}(\widetilde{\x}), \qquad j  \neq j_{0}, \qquad \widetilde{\x} := (x^{1}, \ldots, x^{j_{0}-1}, \lambda_{0}, x^{j_{0}+1}, \ldots, x^{n}). 
\end{equation*}
The reduced vector fields coming from \eqref{eq:Reduction:gen:vf} read as 
\begin{equation*}
\X^{\red}_{\alpha} := \sum_{j = 1}^{j_{0}-1} \xi_{\alpha}^{j}(\widetilde{\x}) \pdv{x^{j}} + \sum_{ j =j_{0} +1}^{n}  \xi_{\alpha}^{j}(\widetilde{\x}) \pdv{x^{j}} \in \cv(\R^{n-1}), \qquad 1 \leq \alpha \leq r,
\end{equation*}
and they still span a VG Lie algebra isomorphic to $\g$. In the general case, i.e., in absence of cyclic coordinates, we assume that $\varphi \in C^{\infty}(\R^{n})$ is an \textit{invariant of the realization} \eqref{eq:Reduction:gen:vf} of the Lie algebra $\g$, i.e. $\dd \varphi (\X_{\alpha}) = \X_{\alpha}(\varphi) = 0$ for all $1 \leq \alpha \leq r$. With respect to the system \eqref{eq:Reduction:gen:system}, this implies the following constraint:
\begin{equation}
\dot{\varphi} = \dot{x}^{j} \pdv{\varphi}{x^{j}} = \sum_{\alpha= 1}^{r} b_{\alpha}(t) \xi_{\alpha}^{j} (\x) \pdv{\varphi}{x^{j}} = 0.
\label{eq:Reduction:gen:constraint}
\end{equation}
If we now consider  (possibly after a reordering of the variables) the (local) coordinate system $(y^{1}:= x^{1}, \ldots, y^{n-1}:= x^{n-1}, y^{n} := \varphi)$, the constraint \eqref{eq:Reduction:gen:constraint} implies that $y^{n}$ can be interpreted as a cyclic coordinate, allowing us to reduce the Lie system $\X$ on $\R^{n}$ by one degree of freedom and thus obtaining a  Lie system  $\X^{\red}$ on $\R^{n-1}$. The main feature of this procedure is that if $\lambda \in \R$ is a regular value of the invariant $\varphi$, then $\varphi^{-1}(\lambda) \subset \R^{n}$ is an $(n-1)$-dimensional regular submanifold of  $\R^{n}$, and we can find a local diffeomorphism 
\begin{equation*}
\phi: W' \subset \R^{n-1} \to W \subset \varphi^{-1}(\lambda) \subset \R^{n},
\end{equation*}
where $W' \subset \R^{n-1}$ and $W \subset \varphi^{-1}(\lambda)$ are open sets, mapping the vector fields of the VG Lie algebra of the reduced Lie system $\X^{\red}$ on $\R^{n-1}$ onto the vector fields \eqref{eq:Reduction:gen:vf} spanning the VG Lie algebra of the Lie system $\X$ on $\R^{n}$. 

A natural question that arises in this context is how many functionally independent invariants can be found for a realization \eqref{eq:Reduction:gen:vf}, and how to compute them effectively. Our approach is based on the analysis of the generalized distribution associated to a given Lie system and its annihilator, which actually provides an interpretation of the resulting integral submanifold as a (local) intersection of hypersurfaces in  $\R^{n}$. 

The \textit{generalized distribution} $\cD^{X} \subset T\R^{n}$ associated to the Lie system $\X$ \eqref{eq:Reduction:gen:system} on $\R^{n}$ is given by 
\begin{equation*}
\cD^{X}_{\xi} := \set{\mathbf{Y} \vert_{\xi}: \mathbf{Y} \in V^{X}} \subset T_{\xi} \R^{n}, \qquad \xi \in \R^{n},
\end{equation*}
so $\xi_{0} \in \R^{n}$ is called a \textit{generic point} of the distribution $\cD^{X}$ if the rank $r^{X} (\xi) := \dim \cD^{X}_{\xi}$ is locally constant around $\xi_{0}$. Then, the \textit{domain} $U \subset \R^{n}$ of $\cD^{X}$ is just the set of generic points, which is an open dense subset of $\R^{n}$. In general, the distribution $\cD^{X}$ is not completely integrable in the whole $\R^{n}$ in the classical sense of Fr\"obenius, but in the sense of Stefan--Sussman \cite{Sussman1973,Stefan1974}. Nevertheless, replacing $U$ by its subset where the rank of the distribution is maximal, we can assume that $U$ is open and that the rank of the distribution is constant on $U$, so $\cD^{X} \vert_{U} \subset TU$ is a completely integrable distribution in the sense of Fr\"obenius (also called regular).  This means that if we consider the \textit{annihilator}  $\Ann (\cD^{X} \vert_{U})$ of $\cD^{X} \vert_{U}$, given by the differential forms $\omega \in \Omega(U)$ annihilating the vector fields of $\cD^{X} \vert_{U} $, if $k = r^{X} \vert_{U}$ is the (constant) rank of the distribution on $U$, then $\Ann (\cD^{X} \vert_{U}) \subset \Omega(U)$ is locally generated by $\chi_{0} := n - k$ functionally independent $1$-forms $\omega^{1}, \ldots, \omega^{\chi_{0}}$, so the complete integrability of $\cD^{X} \vert_{U}$ is equivalent to $\Ann(\cD^{X} \vert_{U}) \subset \Omega (U)$ being a differential ideal, i.e. $\mathrm{d} ( \Ann (\cD^{X} \vert_{U})) \subset \Ann (\cD^{X} \vert_{U})$ \cite{Warner1983}. The main advantage of this reformulation is that $\varphi \in C^{\infty}(\R^{n})$ is a so-called invariant  of the realization \eqref{eq:Reduction:gen:vf} associated to the Lie system \eqref{eq:Reduction:gen:system} if and only if $\dd \varphi \in \Ann (\cD^{X})$.

Let us inspect the elements $\theta \in \Ann(\cD^{X} \vert_{U})$ more closely.  If $\theta$ is an exact form, then $\theta = \dd \varphi$ for a certain function $\varphi \in C^{\infty}(U)$, implying that $\varphi$ is an invariant of the realization. In the general case, although it is possible that $\theta$ may not be an exact form, we can always find (under certain local topological assumptions) two differentiable functions $f$ and $\varphi$ such that $f|_{U}\neq 0$ and $\theta = f \dd \varphi$ (which again implies that $\varphi$ is an invariant of the realization) provided that the condition 
\begin{equation*}
\theta \wedge \dd \theta = 0
\end{equation*}
is satisfied. This can be generalized to a local basis $\omega^{1}, \ldots, \omega^{\chi_{0}}$ of $\Ann(\cD^{X} \vert_{U})$ if the so-called Cartan's integrability condition (see  \cite[p. 46]{Cartan1945})
\begin{equation*}
\dd \omega^{i} \wedge \Omega = 0, \qquad 1 \leq i \leq \chi_{0},
\end{equation*}
is satisfied, where $\Omega := \omega^{1} \wedge \cdots \wedge \omega^{\chi_{0}}$. This approach provides a total amount of $\chi_{0}$ functionally independent invariants $\varphi_{1}, \ldots, \varphi_{\chi_{0}}$ of the realization \eqref{eq:Reduction:gen:vf}, yielding to the following result with respect to the reduction of the Lie system \eqref{eq:Reduction:gen:system} by these invariants, hence generalizing the criterion developed in \cite{Campoamor2018}:
 \begin{thm}
 \label{thm:Reduction}
Under the previous assumptions, let  $r^{X} \vert_{U} = k < n$ and $V^{X}$ be the VG Lie algebra of the Lie system $\bold{X}$, isomorphic to a real Lie algebra $\g$. The following properties hold: 
\begin{itemize}
\item[(1)] There exist  
\begin{equation*}
\chi_{0} = n - k 
\end{equation*}
functionally independent invariants $\varphi_{1}, \ldots, \varphi_{\chi_{0}} \in C^{\infty}(U)$ of the realization \eqref{eq:Reduction:gen:vf} that form part of a local coordinate system $(y^{1}, \ldots, y^{n})$ on $U$ given by  (after a reordering of the indices if necessary)
\begin{equation*}
y^{1} = x^{1}, \ldots, y^{k} = x^{k} 
\end{equation*}
and $\chi_{0}$ cyclic coordinates 
\[ y^{k+1} = \varphi_{1}, \ldots, y^{n} = \varphi_{\chi_{0}}. \]
\item[(2)] Consider now the differentiable mapping 
 \[ \psi := (\varphi_{1}, \ldots, \varphi_{\chi_{0}}): U \subset \R^{n} \to \R^{\chi_{0}}, \] 
and let $\lambda \in \R^{\chi_{0}}$ be a regular value of $\psi$, so that $\psi^{-1}(\lambda) \subset U$ is a $k$-dimensional submanifold. There exists a diffeomorphism 
\[ \phi : W' \subset \R^{k} \to W \subset \psi^{-1}(\lambda), \]
such that 
\begin{equation} 
(x^{1}, \ldots, x^{n}) = \phi(x^{1}, \ldots, x^{k}) 
\label{eq:cyclic_th}
\end{equation}
for every point $(x^{1}, \ldots, x^{n}) \in W \subset \psi^{-1}(\lambda)$, where $W \subset \psi^{-1}(\lambda)$ and $W' \subset \R^{k}$ are open subsets. In particular, $(\phi^{-1})_{*}(V^{X}) \simeq \g$ is the VG Lie algebra of a Lie system $\X^{\red}$ on $W' \subset \R^{k}$ called a reduction of $\X$ by invariants. 
\end{itemize}
\end{thm}
 
 \newcommand{\y}{\mathbf{y}}

 \begin{proof}
The first part was proved in \cite{Campoamor2018}, so we only prove here the second one. Let $\lambda \in \R^{\chi_{0}}$ be a regular value of $\psi$ (whose existence is guaranteed by  the Morse--Sard's Theorem \cite[p. 69]{Hirsch1976}), so $\psi^{-1}(\lambda) \subset U$ is a $k$-dimensional submanifold of $\R^{n}$. As $\lambda \in \R^{\chi_{0}}$ is a regular value of $\psi$, there exist open subsets $W \subset \psi^{-1}(\lambda)$ and $W' \subset \R^{k}$ and a diffeomorphism $\phi: W' \to W$ such that the equality \eqref{eq:cyclic_th} holds for every point $(x^{1}, \ldots, x^{n}) \in W$.\newline
\indent Considering the generators \eqref{eq:Reduction:gen:vf} of the VG Lie algebra $V^{X}$ of the Lie system $\X$ on $\R^{n}$, the images by the differential map $(\phi^{-1})_{*}(\X_{1}), \ldots, (\phi^{-1})_{*}(\X_{r})$ are the generators of a Lie algebra of vector fields $(\phi^{-1})_{*}(V^{X}) \simeq \g$ which is, indeed, a VG Lie algebra for the reduced Lie system $\X^{\red}$ on $W' \subset \R^{k}$ related to the $t$-dependent vector field $\X^{\red} = \sum_{\alpha = 1}^{r} b_{\alpha}(t) (\phi^{-1})_{*}(\X_{\alpha}) $. 
 \end{proof}
 The geometrical meaning of Theorem~\ref{thm:Reduction} is that the Lie system \eqref{eq:Reduction:gen:system} on $\R^{n}$ can be reduced by invariants to $\R^{n - \chi_{0}}$ if its associated generalized distribution $\cD^{X}$ possesses an $(n - \chi_{0})$-dimensional integral submanifold, which in our case is locally expressed as $\varphi^{-1}(\lambda) \subset \R^{n}$. The second part says that the local coordinates of the reduced Lie system on $\R^{n - \chi_{0}}$ just come from a suitable local chart of the latter submanifold $\varphi^{-1}(\lambda) \subset \R^{n}$. We emphasize also the importance of the chosen regular value $\lambda \in \R^{\chi_{0}}$, as different regular values may produce non-diffeomorphic reduced Lie systems (see Subsection~\ref{subsection:sl2} and Figure~\ref{fig:surfaces}). 
 
\begin{remark}
If $\set{X_{1}, \ldots, X_{r}}$ denotes a basis of the Lie algebra $\g$ with structure constants $[X_{\alpha}, X_{\beta}] = C_{\alpha \beta}^{\gamma} X_{\gamma}$, the realization $\Phi_{\ad}: \g \to \cv(\R^{r})$  induced by the adjoint representation $\ad: \g \to \gl(r, \R)$ of $\g$ can be naturally identified with the coadjoint representation of $\g$ (see e.g. \cite{Racah1951}) in terms of the first-order differential operators 
\begin{equation}
\widehat{\X}_{\alpha} := \Phi_{\ad}(X_{\alpha}) = x^{\gamma} C_{\alpha \beta}^{\gamma} \pdv{x^{\beta}}, \quad 1 \leq \alpha \leq r, 
\label{eq:Reduction:Realizationcoadj}
\end{equation}
where $(x^{1}, \ldots, x^{r})$ correspond to the coordinates on the dual basis of $\set{X_{1}, \ldots, X_{r}}$ on $\g^{*} \simeq \R^{r}$. In this context, an invariant of $\g$ is an operator $F(X_{1}, \ldots, X_{r})$ that is a solution of the following system of PDEs: 
\begin{equation}
\widehat{\X}_{\alpha} F(x^{1}, \ldots, x^{r}) = 0, \quad 1 \leq \alpha \leq r.
\label{eq:Reduction:BeltramettiBlasivf}
\end{equation} 
It is well known (see \cite{Beltrametti1966}) that the maximum number $\cN(\g)$ of functionally independent solutions of the system \eqref{eq:Reduction:BeltramettiBlasivf} is given by 
\begin{equation}
\cN(\g) = \dim \g - \mathrm{r}(\g),
\label{eq:BeltramettiBlasi}
\end{equation}
where $\mathrm{r}(\g)$ is the maximal (generic) rank of the functional matrix 
\begin{equation*}
A(\g) := \left( x^{\gamma} C_{\alpha \beta}^{\gamma}  \right)_{1 \leq \alpha, \beta \leq r}
\end{equation*} 
associated to the realization \eqref{eq:Reduction:Realizationcoadj}. We recall that an alternative reformulation of formula \eqref{eq:BeltramettiBlasi} can be obtained by means of differential forms \cite{Campoamor2004}, in analogy to the argument previously used for an arbitrary realization \eqref{eq:Reduction:gen:vf} of $\g$. Moreover, polynomial invariants of $\g$ correspond, after symmetrization, to elements belonging to the centre of the universal enveloping algebra $\mathcal{U}(\g)$, i.e., they correspond to the Casimir operators of $\g$. In Subsections~\ref{subsection:constants_h6}, \ref{subsection:constants_Lorentz} and \ref{subsection:sp4_superpositionrule}, Casimir operators will play an important role in order to derive superposition rules for LH systems. 
\end{remark}

 \subsection{Construction of new Lie systems}
 
 The preceding results enable us to determine whether Lie systems defined in a different number of variables can be seen as locally equivalent systems. 
  
\begin{defi} \label{def:Reduction:locdiffeo}
Let $\X$ be a Lie system on $\R^{n}$ and $\Y$ be a Lie system on $\R^{m}$, with $m < n$. We say that $\Y$ is \textit{locally diffeomorphic} to $\X$ if the Lie system $\X$ can be reduced by invariants to a Lie system $\X^{\red}$ on $\R^{m}$ which is locally diffeomorphic to $\Y$. 
\end{defi}

The following result is a direct consequence of the reduction procedure established in Theorem~\ref{thm:Reduction} and   Definition~\ref{def:Reduction:locdiffeo}: 

\begin{prop}
\label{prop:Reduction:criterio}
Let $\X$ be a Lie system on $\R^{n}$ and $\Y$ be a Lie system on $\R^{m}$, with $m < n$, and consider the (maximum and constant) ranks $r^{X} \vert_{U}$ and $r^{Y} \vert_{U'}$ of their associated generalized distributions $\cD^{X} \subset T \R^{n}$ and $\cD^{Y} \subset T \R^{m}$ on certain open subsets $U \subset \R^{n}$ and $U' \subset \R^{m}$, respectively. If $r^{X} \vert_{U} \neq r^{Y} \vert_{U'}$, then $\X$ and $\Y$ are not  locally diffeomorphic Lie systems. 
\end{prop}

The combination of the reduction method described previously and Proposition~\ref{prop:Reduction:criterio}  suggests how to construct higher-dimensional Lie systems which are not locally diffeomorphic to lower-dimensional Lie systems. Specifically, given a faithful representation $\Gamma: \g \to \gl(n, \R)$ of an $r$-dimensional real Lie algebra $\g$, with $n \geq 3$, consider the Lie system $\X$ on $\R^{n}$ (see \eqref{eq:Reduction:tvf}) induced by the realization \eqref{eq:Reduction:vfrealization}. Provided that the (maximum) rank of the generalized distribution $\cD^{X} \subset T \R^{n}$ associated to the latter Lie system is $r^{X} \vert_{U}  \geq n$ on a certain open subset $U \subset \R^{n}$, the Lie system  $\X$ on $\R^{n}$ cannot be locally diffeomorphic to a lower-dimensional Lie system, as a consequence of Proposition~\ref{prop:Reduction:criterio}. 

This property, however, does not exclude the possibility that the Lie system $\X$ on $\R^{n}$ is locally diffeomorphic to a `sum of copies' of lower dimensional Lie systems, i.e., that it can be described as a coupling of Lie systems.  

\begin{defi}
Let $\Y$ and $\mathbf{Z}$ be Lie systems on $\R^{p}$ and $\R^{q}$ such that their VG Lie algebras are both isomorphic to an $r$-dimensional real Lie algebra $\g$, spanned by the vector fields $\Y_{1}, \ldots, \Y_{r} \in \cv(\R^{p})$ and $\mathbf{Z}_{1}, \ldots, \mathbf{Z}_{r} \in \cv(\R^{q})$, respectively. Under the zero sections  
\begin{equation*}
\sigma_{p}: \R^{p} \hookrightarrow \R^{p+q},  \qquad \sigma_{q}: \R^{q} \hookrightarrow \R^{p+q}, 
\end{equation*}
of the bundles $\R^{p+q} \to  \R^{p}$ and $\R^{p+q} \to \R^{q}$, 
the vector fields 
\begin{equation*}
\sigma_{p_{*}}(\Y_{\alpha}) + \sigma_{q_{*}}(\mathbf{Z}_{\alpha}) \in \cv(\R^{p+q}), \qquad 1 \leq \alpha \leq r
\end{equation*}
span the VG Lie algebra, isomorphic to $\g$, of a Lie system on $\R^{p+q}$, called the \textit{uncoupled sum} of $\Y$ and $\mathbf{Z}$, and denoted by $\Y +_{{\rm u}} \mathbf{Z}$. 
\end{defi}
Similarly, by induction we can define the uncoupled sum of  $k \geq 1$  Lie systems. If $\X^{1}, \ldots, \X^{k}$ are Lie systems on $\R^{n_{1}}, \ldots, \R^{n_{k}}$, respectively, we denote their uncoupled sum by $\X^{1}+_{\mathrm{u}} \cdots +_{\mathrm{u}} \X^{k}$, which is a Lie system on $\R^{n_{1} + \cdots + n_{k}}$. 

\begin{example}
Consider the $n$-dimensional ($n$D in short) Smorodinsky--Winternitz (SW) oscillator with a $t$-dependent frequency $\Omega(t)$ on $T^{*}\R_{0}^{n}$, where $\R_{0}^{n}:= \R^{n} - \set{0}$,  and given by the $t$-dependent vector field $\X = \X_{3} + \Omega^{2}(t) \X_{1}$, where the vector fields 
\begin{equation*}
\X_{1} :=- \sum_{j = 1}^{n} q_{j} \pdv{p_{j}}, \qquad \X_{2} := \frac{1}{2} \sum_{j = 1}^{n} \left( p_{j} \pdv{p_{j}} - q_{j} \pdv{q_{j}} \right), \qquad \X_{3} :=  \sum_{j = 1}^{n} \left( p_{j} \pdv{q_{j}} + \frac{c}{q_{j}^{3}} \pdv{p_{j}} \right), \quad c \in \R 
\end{equation*} 
expressed in the global coordinates $(q_{j}, p_{j})$ of $T^{*}\R_{0}^{n}$, have commutation relations 
\begin{equation*}
[\X_{1}, \X_{2}] = \X_{1}, \qquad [\X_{1}, \X_{3}] = 2 \X_{2}, \qquad [\X_{2}, \X_{3}] = \X_{3},
\end{equation*}
and therefore span a VG Lie algebra isomorphic to $\sl(2, \R)$. The associated first-order system of ODEs reads
\begin{equation*}
\dv{q_{j}}{t} = p_{j}, \qquad \dv{p_{j}}{t} = - \Omega^{2}(t) q_{j} + \frac{c}{q_{j}^{3}}, \qquad    1\leq j \leq n
\end{equation*}
showing that the $n$D SW oscillator is the sum of $n$ uncoupled 1D SW oscillators. Indeed, if $\Y = \Y_{3} + \Omega^{2}(t) \Y_{1}$ is the $t$-dependent vector field associated to the 1D SW oscillator on $T^{*} \R_{0}$, with 
\[ \Y_{1} = -q \pdv{p}, \qquad \Y_{2} = \frac{1}{2} \left( p \pdv{p} - q \pdv{q} \right), \qquad \Y_{3} = p \pdv{q} + \frac{c}{q^{3}} \pdv{p}, \]
expressed in the global coordinates $(q, p)$ of $T^{*} \R$, the $n$D SW oscillator corresponds to the uncoupled sum 
\begin{equation*}
\X = \Y +_{{\rm u}} \overset{n}{\cdots} +_{{\rm u}}  \Y
\end{equation*}
of $n$ one-dimensional SW oscillators. 
\end{example}

\medskip
Let $\Gamma: \g \to \gl(n, \R)$ be a (faithful) completely reducible representation of an $r$-dimensional real Lie algebra $\g$,\footnote{This is ensured e.g. for reductive Lie algebras.} such that $\Gamma$ decomposes as 
\begin{equation*}
\Gamma =  k \Gamma_{0} \oplus  \Gamma_{1} \oplus \cdots \oplus \Gamma_{s} ,
\end{equation*}
 where  $\Gamma_{0}$ denotes the trivial representation, $k \Gamma_{0} := \Gamma_{0} \oplus \overset{k}{\cdots} \oplus \Gamma_{0}$ and each $\Gamma_{i}: \g \to \gl(n_{i}, \R)$  a (faithful) irreducible representation of $\g$ for $1 \leq i \leq s$. As the multiplicity $k = \mathrm{mult}_{\Gamma_{0}}(\Gamma)$ of the trivial representation $\Gamma_{0}$ in $\Gamma$ provides the number of independent linear invariants of the realization $\Phi_{\Gamma}: \g \to \cv(\R^{n})$ induced by $\Gamma$ (see \cite{Campoamor2018a}), the Lie system  $\X$  on $\R^{n}$ associated to this realization can be reduced by $k$ linear invariants to a Lie system $\X^{\red}$ on $\R^{n- k}$. If $k = 0$ and $\X^{i}$ denotes the Lie system on $\R^{n_{i}}$ induced by the representation $\Gamma_{i}$, then the Lie system $\X$ corresponds to  the uncoupled sum 
\begin{equation*}
\X = \X^{1} +_{{\rm u}} \overset{n}{\cdots} +_{{\rm u}} \X^{k},
\end{equation*}
showing that the Lie system $\X$ on $\R^{n}$ comes from lower-dimensional Lie systems in both cases. 

\medskip

Taking into account these properties, the following prescription can be established to construct
 higher-dimensional Lie systems:

\begin{itemize}
\item[1.] Given an $r$-dimensional real Lie algebra $\g$, take $\Gamma : \g \to \gl(n, \R)$ an indecomposable faithful representation of $\g$ (recall that if $\g$ is semisimple this is equivalent to  $\Gamma$ being irreducible), with $n \geq 3$.
\item[2.] Consider the Lie system $\X$ on $\R^{n}$, given by \eqref{eq:Reduction:tvf}, whose VG Lie algebra  is spanned by the vector fields \eqref{eq:Reduction:vfrealization} of the realization $\Phi_{\Gamma}: \g \to \cv(\R^{n})$ induced by the representation $\Gamma$. 
\item[3.] If the (maximum) rank of the generalized distribution $\cD^{X} \subset T \R^{n}$ associated to $\X$ is $r^{X} \vert_{U} \geq n$ on a certain open subset $U \subset \R^{n}$, then $\X$ is not locally diffeomorphic to a Lie system on $\R$ nor to a Lie system on $\R^{2}$ by Proposition~\ref{prop:Reduction:criterio}. 
\end{itemize}

In the most general situation,  Proposition~\ref{prop:Reduction:criterio} may not always detect whether a given higher-dimensional Lie system is the uncoupled sum of lower-dimensional Lie systems (see Section~\ref{section:Lorentz}). Nevertheless, for the new higher-dimensional Lie systems that will be constructed in the following sections, it is possible to show that they are not locally diffeomorphic to an uncoupled sum of lower-dimensional Lie systems. We just recall here a feasible scenario occurring when dealing with LH systems on symplectic manifolds in which this problem may be solved:
\begin{prop} \label{prop:Reduction:localdiffeo}
Let $\X$ and $\Y$ be Lie systems on a symplectic manifold $(M, \omega)$. Furthermore, suppose that $\X$ is a LH system on $(M, \omega)$ and that $\X$ and $\Y$ are locally diffeomorphic. Then, $\Y$ is a LH system on $(M, \omega)$.
\end{prop}
We omit the proof, as it follows immediately using the properties of the pullback of a local diffeomorphism. In addition, in order to identify whether a higher-dimensional Lie system is not a LH system, the following result, the proof of which is also rather immediate, will be useful in the following:

\begin{prop} \label{prop:Reduction:symmetries_volume}
Let $\X$ be a LH system on an $n$-dimensional symplectic Riemannian manifold $(M, \omega, g)$, where $\omega \in \Omega^{2}(M)$ is the symplectic form and $g$ is the Riemannian metric. There always exists a nowhere zero  function $f \in C^{\infty}(M)$ such that $\Omega_{\omega} = f \Omega_{g}$, where $\Omega_{\omega}$  and $\Omega_{g}$ are the volume forms associated to the symplectic form and the Riemannian metric, respectively. Therefore, the generators  $\X_{1}, \ldots, \X_{r}$  of the VG Lie algebra of the LH system $\X$ are symmetries of the volume form $\Omega_{\omega}$, that is, they satisfy the constraint 
\begin{equation*}
f \: \div \X_{\alpha} + \dd f (\X_{\alpha}) = 0, \qquad 1 \leq \alpha \leq r
\end{equation*}
where $\div \X_{\alpha}$ is the divergence of $\X_{\alpha}$ with respect to the volume form $\Omega_{g}$. 
\end{prop}

\medskip
We observe that, in general, the VG Lie algebra of a Lie system $\mathbf{X}$ on $\mathbb{R}^{n}$, isomorphic to an $r$-dimensional real Lie algebra $\mathfrak{g}$, gives rise to a local action of the corresponding (connected and simply connected) Lie group $G$ on $\mathbb{R}^{n}$, by means of the local diffeomorphisms associated to the vector fields realizing $\mathfrak{g}$. The generalized distribution $\mathcal{D}^{X} \subset T \mathbb{R}^{n}$ associated to the Lie system $\mathbf{X}$ is then integrable in the Stefan--Sussman sense \cite{Sussman1973, Stefan1974}. This leads to a (generalized) foliation $\mathcal{F}$ of $\R^{n}$ into connected submanifolds which are locally described as the orbits of the $G$-action \cite{Palais1957}. 

\medskip
In the following, the method described above will be applied to the case of LH systems on $T^{*}\R^{2} \equiv \R^{4}$. Within this context, the (global) coordinates of $T^{*}\R^{2}$ in which the realization \eqref{eq:Reduction:gen:vf} will be expressed are the canonical coordinates $(q_{1}, q_{2}, p_{1}, p_{2})$.

 \section{The two-photon Lie algebra} \label{section:twophoton}
 
 The two-photon Lie algebra $\h_{6}$, first considered in \cite{Zhang1990} in the context of coherent states, is spanned by the six operators
 \begin{equation}
 N :=  a^{\dagger} a, \qquad A_{+} := a^{\dagger}, \qquad A_{-}:= a, \qquad B_{+} := a^{\dagger 2}, \qquad B_{-} := a^{ 2}, \qquad M:= \I, 
 \label{eq:Schrodinger:boson}
 \end{equation}
 where the annihilation and creation operators $a$ and $a^{\dagger}$ are the generators of the boson algebra $[a, a^{\dagger}] = \I$.  Over this basis, the commutation relations of $\h_{6}$ read  
\begin{equation*}
\begin{array}{lllll}
&[A_{\pm}, B_{\pm}] = 0, \qquad  &[A_{-}, A_{+}] = M, \qquad &[M, \cdot] = 0,  \qquad &[B_{-}, B_{+}] = 4N + 2M,\\
&[A_{\pm}, B_{\mp}] = \mp 2 A_{\mp}, \qquad &[N, A_{\pm}] = \pm A_{\pm}, \qquad &[N, B_{\pm}]  = \pm 2 B_{\pm}. 
\end{array}
\label{eq:Schrodinger:basis1}
\end{equation*}
By means of the change of basis (see \cite{Ballesteros1997})
\begin{equation*}
 G:= A_{-}, \qquad P := -A_{+}, \qquad D:= -\left(N + \frac{1}{2}M \right), \qquad K:= \frac{1}{2} B_{-}, \qquad H:= \frac{1}{2} B_{+}, 
\end{equation*}
keeping $M$ as a central generator, the transformed commutation relations are given by   
\begin{equation}
\begin{array}{lllll}
& [D, K] = 2K,\qquad &[D, H] = -2H,\qquad  &[D, G] = G, \qquad &[D, P] = -P, \\
& [K, H] = -D, \qquad &[K, G] = 0, \qquad &[K, P] = -G, \qquad &[H, G] = P, \\
& [H, P] = 0, \qquad &[G, P] = -M, \qquad &[M, \cdot] = 0,
\end{array}
\label{eq:Schrodinger:basis2}
\end{equation}
from which it is easily seen that the two-photon Lie algebra $\h_{6}$ is isomorphic to the (centrally extended) Schr\"odinger algebra $\widehat{\cS}(1)$ in $(1+1)$ dimensions \cite{Niederer1972,Hagen1972,Feinsilver2004, Campoamor2005, Campoamor2020, Alshammari2018}. Over the latter basis \eqref{eq:Schrodinger:basis2}, we easily get the Levi decomposition $\widehat{\cS}(1) = \sl(2, \R) \overrightarrow{\oplus}_{\Gamma_{\frac{1}{2}} \oplus \Gamma_{0}} \h_{1}$, where $\sl(2, \R)$ acts on the three-dimensional Heisenberg algebra $\h_{1}$ by means of the direct sum of the fundamental representation $\Gamma_{\frac{1}{2}}$ and the trivial multiplet $\Gamma_{0}$. It is trivial to verify that the number of functionally independent invariants of the Schrödinger algebra is given by $\cN(\widehat{\cS}(1)) = 2$,\footnote{As a general fact, for any semidirect product $\g \overrightarrow{\oplus}_{\Gamma} \h_{N}$ of a semisimple Lie algebra $\g$ and the $(2N+1)$-dimensional Heisenberg Lie algebra $\h_{N}$, we have $\cN(\g \overrightarrow{\oplus}_{\Gamma} \h_{N}) = \mathrm{rank}(\g) + 1$ \cite{Quesne1988}.} where $C_{0} := M$ is a first-order invariant spanning the center of $\widehat{\cS}(1)$. A noncentral Casimir operator is computed either using the so-called matrix method \cite{Campoamor2005} or the trace-formulae approach \cite{Campoamor2020}, providing the fourth-order Casimir invariant 
\begin{equation*}
C_{4} := (-2MH + P^{2})(2MK-G^{2}) + (MD - GP)^{2},
\label{eq:Schrodinger:C4}
\end{equation*}
which factorizes as the product $C_{0} C_{3} = C_{4}$ of the central charge and the cubic invariant 
\begin{equation}
C_{3} := M (D^{2} -4KH) + 2(HG^{2} + KP^{2}- DGP). 
\label{eq:Schrodinger:C3}
\end{equation}

We now consider the following faithful (reducible) representation $\Gamma: \widehat{\cS}(1) \to \gl(4, \R)$ of $\widehat{\cS}(1)$ given by the  matrix 
\begin{equation*}
A_{\Gamma} := \begin{pmatrix}
 - D & 0 & H & P \\
 -G & 0 & P & 2M \\
 - K & 0 & D & G \\
 0 & 0 & 0 & 0 
\end{pmatrix}.
\end{equation*}
The generators of the associated realization $\Phi_{\Gamma}: \widehat{\cS}(1) \to \cv(T^{*} \R^{2})$ read as follows:
\begin{align}
&\X_{1} := \Phi_{\Gamma}(D) = - q_{1} \pdv{q_{1}} + p_{1} \pdv{p_{1}},  & &\X_{2}:= \Phi_{\Gamma}(K) = - p_{1} \pdv{q_{1}}, \nonumber   \\
&\X_{3} := \Phi_{\Gamma}(H) = q_{1} \pdv{p_{1}},  & &\X_{4} := \Phi_{\Gamma}(G) = - q_{2} \pdv{q_{1}} + p_{1} \pdv{p_{2}}, \label{eq:Schrodinger:vf} \\
& \X_{5} := \Phi_{\Gamma}(P) = q_{2} \pdv{p_{1}} + q_{1} \pdv{p_{2}}, & &\X_{6} := \Phi_{\Gamma}(M) = 2 q_{2} \pdv{p_{2}}. \nonumber
\end{align}

The Lie system on $T^{*} \R^{2}$  induced by the latter realization is determined the $t$-dependent vector field 
\begin{equation*}
\X := \sum_{\alpha = 1}^{6} b_{\alpha}(t) \X_{\alpha},
\label{eq:Schrodinger:tvf}
\end{equation*}
where $b_{\alpha} \in C^{\infty}(\R)$ is an arbitrary $t$-dependent function $(1 \leq \alpha \leq 6)$. The corresponding first-order system of differential equations on $T^{*} \R^{2}$ is 
\begin{equation}
\begin{split}
\dv{q_{1}}{t} &= - b_{1}(t) q_{1} - b_{4}(t) q_{2}  - b_{2}(t) p_{1}, \\
\dv{p_{1}}{t} & =  b_{3}(t) q_{1} + b_{5}(t) q_{2} + b_{1}(t) p_{1},
\end{split}
\qquad 
\begin{split}
\dv{q_{2}}{t} & = 0, \\
\dv{p_{2}}{t} & =b_{5}(t) q_{1} + 2b_{6}(t) q_{2} + b_{4}(t) p_{1}, 
\end{split}
\label{eq:Schrodinger:system}
\end{equation}
where $q_2$ is obviously a cyclic coordinate because of the reducibility of the representation. If
\begin{equation}
 \omega = \dd q_{1} \wedge \dd p_{1} + \dd q_{2} \wedge \dd p_{2} \in \Omega^{2}(T^{*}\R^{2})
\label{eq:symplecticform}
\end{equation}
 denotes the canonical symplectic form of $T^{*} \R^{2}$, it is straightforward to verify that the invariance condition $\cL_{\X_{\alpha}} \omega = 0$ is satisfied for all $1 \leq \alpha \leq 6$, showing that the vector fields \eqref{eq:Schrodinger:vf} are indeed Hamiltonian vector fields relative to $\omega$ on $T^{*}\R^{2}$ and, therefore, that $\X$ inherits the structure of a LH system on $T^{*}\R^{2}$. The Hamiltonian functions associated to the vector fields \eqref{eq:Schrodinger:vf}, determined by the condition $\iota_{\X_{\alpha}} \omega = \dd h_{\alpha}$, read 
\begin{equation}
h_{1} = - q_{1}p_{1}, \qquad h_{2} = - \frac{1}{2} p_{1}^{2}, \qquad h_{3} = - \frac{1}{2} q_{1}^{2}, \qquad h_{4} = - q_{2}p_{1}, \qquad h_{5} = - q_{1} q_{2}, \qquad h_{6} = -q_{2}^{2}.
\label{eq:Schrodinger:ham}
\end{equation}
The commutation relations of the Hamiltonian functions \eqref{eq:Schrodinger:ham} with respect to the Poisson bracket $\set{\cdot, \cdot}_{\omega}$ induced by the canonical symplectic form \eqref{eq:symplecticform} on $T^{*} \R^{2}$ are 
\begin{equation}
\begin{array}{lllll}
& \set{h_{1}, h_{2}}_{\omega} = - 2 h_{2}, \qquad & \set{h_{1}, h_{3}}_{\omega} = 2 h_{3}, \qquad &\set{h_{1}, h_{4}}_{\omega} = - h_{4}, \qquad &\set{h_{1}, h_{5}}_{\omega} = h_{5}, \\
& \set{h_{2}, h_{3}}_{\omega} = h_{1}, \qquad & \set{h_{2}, h_{4}}_{\omega} = 0, \qquad &\set{h_{2}, h_{5}}_{\omega} = h_{4}, \qquad &\set{h_{3}, h_{4}}_{\omega} = - h_{5}, \\
&\set{h_{3}, h_{5}}_{\omega} = 0, \qquad &\set{h_{4}, h_{5}}_{\omega} = h_{6}, \qquad &\set{h_{6}, \cdot}_{\omega} = 0,
\end{array}
\label{eq:Schrodinger:ham_cr}
\end{equation}
spanning a LH algebra $\cH_{\omega} \simeq \widehat{\cS}(1)$. 

It must be emphasized that the $\widehat{\cS}(1)$-LH system \eqref{eq:Schrodinger:system} on $T^{*}\R^{2}$ is intrinsic, as it is not locally diffeomorphic to an uncoupled sum of lower-dimensional LH systems, because no Lie system in $\R^{n}$ with $1 \leq n \leq 2$ and possesing a VG Lie algebra isomorphic to $\widehat{\cS}(1)$ can exist.

The ODEs system \eqref{eq:Schrodinger:system}, being associated to a representation, is linear, and directly integrable by quadratures for $b_3(t)=0$. For $b_3(t)\neq 0$, it can be shown after some computation that 
\begin{equation}\label{eqsm}
\begin{split}
q_2(t)=&\lambda_{1},\quad q_1(t)= -\frac{1}{b_3(t)}\left(\lambda_{1} b_5(t) +b_1(t)p_1-\dot{p}_1\right),\\
p_2(t)=&\int \frac{1}{b_3(t)}\left(2\lambda_{1}b_3(t)b_6(t)-\lambda_{1}b_5(t)^2+(b_3(t)b_4(t)-b_1(t)b_5(t))p_1 + b_{5}(t) \dot{p}_{1} \right) {\rm d}t +\lambda_{2},
\end{split}
\end{equation}
where $\lambda_{1}, \lambda_{2} \in \R$ are constants and $p_1$ is a solution to the second-order linear equation 
\begin{equation}\label{eqms1}
\begin{split}
\frac{{\rm d}^2\xi}{{\rm d}t^2} & -\frac{{\rm d}\ln b_3}{{\rm d}t}\frac{{\rm d}\xi}{{\rm d}t}+\left(b_1(t)\frac{{\rm d}\ln b_3}{{\rm d}t}+b_2(t)b_3(t)-b_1(t)^2-\frac{{\rm d}b_1}{{\rm d}t}\right)\xi\\
& -\lambda_{1}\left(b_1(t)b_5(t)-b_3(t)b_4(t)+\frac{{\rm d}b_5}{{\rm d}t}-\frac{{\rm d}\ln b_3}{{\rm d}t} \, b_5(t)\right)=0.
\end{split}
\end{equation}
This equation can always be reduced to the free equation $\ddot{z}(s)=0$ by means of a point transformation \cite{OLV}, allowing a priori to solve the system completely. If certain relations among the coefficients are given, the previous equation may further simplify, providing a solution by quadratures of the system. This holds e.g. if $b_2(t)$ and $b_4(t)$ are such that the coefficient in $\xi$ and the inhomogeneous term of (\ref{eqms1}) vanish. 

As $q_{2}$ is a cyclic coordinate of the system \eqref{eq:Schrodinger:system}, the latter can directly be seen as a three-dimensional system. With the choice $q_{2} = 1$ for the regular value, the reduction procedure of Theorem~\ref{thm:Reduction} yields   a reduced Lie system $\X^{\red}$ on $T^{*} \R \times \R \equiv \R^{3}$ whose VG Lie algebra $V^{X^{\red}} \simeq \widehat{\cS}(1)$ is spanned by the vector fields
\begin{equation}
\begin{split}
&\X_{1}^{\red}= - q \pdv{q} + p \pdv{p}, \qquad \X_{2}^{\red} = - p \pdv{q}, \qquad \X_{3}^{\red}  = q \pdv{p},\qquad  \X_{4}^{\red} =-\pdv{q} + p \pdv{s}, \\
&\X_{5}^{\red} = \pdv{p} + q \pdv{s}, \qquad \quad \: \X_{6}^{\red} = 2 \pdv{s}
\end{split}
\label{eq:Schrodinger:vf3d}
\end{equation}
expressed in the global coordinates $(q, p, s)$ of $T^{*}\R \times \R$, where $(q, p)$ are the coordinates of $T^{*} \R$, while $s$ denotes the coordinate of $\R$. Clearly, the reduced Lie system is equivalent to the Lie system determined by the three-dimensional representation with matrix elements
\begin{equation*}
B_{\widehat{\Gamma}}:= \left(A_{\Gamma}\right)_{(4,2)}
\end{equation*}
obtained after deletion of the fourth row and second column of the matrix $A_{\Gamma}$. The resulting system of differential equations on $T^{*} \R \times \R$ is given by  
\begin{equation}
\begin{split}
\dv{q}{t} &= - b_{1}(t) q - b_{4}(t)   - b_{2}(t) p, \qquad \dv{p}{t}  =  b_{3}(t) q + b_{5}(t)  + b_{1}(t) p, \\
\dv{s}{t} & =b_{5}(t) q + 2b_{6}(t)  + b_{4}(t) p, 
\end{split}
\label{eq:Schrodinger:system3d}
\end{equation}
and its solution can be read off from (\ref{eqsm})--(\ref{eqms1}) with $\lambda_{1}=1$, $p=p_1$, $q_1=q$ and $s=p_2$.  

Although the system \eqref{eq:Schrodinger:system3d} does not define a LH system, we observe that the Hamiltonian functions \eqref{eq:Schrodinger:ham}, when restricted to $\set{q_{2} = 1} \equiv T^{*}\R \times \R$, simplify to 
\begin{equation}
h_{1} = - qp, \qquad h_{2} = - \frac{1}{2} p^{2}, \qquad h_{3} = - \frac{1}{2} q^{2}, \qquad h_{4} = - p, \qquad h_{5} = - q, \qquad h_{6} = -1,
\label{eq:Schrodinger:ham3d}
\end{equation}
which are, after a change of basis, the Hamiltonian functions spanning the LH algebra of the so-called P$_{5}$-LH class on $T^{*}\R \equiv \R^{2}$ (see \cite{Ballesteros2015}). This suggests us to consider the projection 
\begin{equation}
\pr : T^{*} \R \times \R \to T^{*}\R, \quad (q, p, s) \mapsto (q, p) .
\label{eq:Schrodinger:projection}
\end{equation}
For the projection $\Y_{\alpha} :=  \pr_{*}(\X_{\alpha}^{\red})$  of the reduced vector fields  \eqref{eq:Schrodinger:vf3d}, we see in particular that $\Y_{6} = \pr_{*}(\X_{6}) = 0$, while the remaining vector fields
\begin{equation}
\Y_{1} = - q  \pdv{q}+ p\pdv{p}, \qquad \Y_{2} = -p \pdv{q}, \qquad \Y_{3} = q \pdv{p}, \qquad \Y_{4} = -\pdv{q}, \qquad \Y_{5} = \pdv{p}
\label{eq:Schrodinger:projectedvf}
\end{equation}
are the generators of the VG Lie algebra of the P$_{5}$-LH class on $T^{*}\R$. Their associated Hamiltonian functions with respect to the canonical symplectic form $\omega = \dd q \wedge \dd p$ on $T^{*}\R$ are given in \eqref{eq:Schrodinger:ham3d}, as they are invariant under the projection \eqref{eq:Schrodinger:projection}.  We conclude that the LH system of the P$_{5}$-LH class on $T^{*}\R$, with associated first-order system of ODEs
\begin{equation}
\dv{q}{t} = - b_{1}(t) q - b_{4}(t)   - b_{2}(t) p, \qquad \dv{p}{t}  =  b_{3}(t) q + b_{5}(t)  + b_{1}(t) p, 
\label{eq:Schrodinger:system2d}
\end{equation}
appears as the projection of the Lie system \eqref{eq:Schrodinger:system3d} on $T^{*} \R \times \R$, which in turn arises as the reduction of the LH system \eqref{eq:Schrodinger:system} on $T^{*}\R^{2}$. It follows that any solution  of (\ref{eqms1}) with $\lambda=1$, $p=p_1$, $q_1=q$ also provides the solution to the projected system \eqref{eq:Schrodinger:system2d}.

\subsection{Constants of the motion and superposition rules}
\label{subsection:constants_h6}

Besides a direct attempt to integrate the system \eqref{eq:Schrodinger:system}, which may be computationally quite demanding, in spite of the linearity of the equation \eqref{eqms1}, an alternative strategy is to find a superposition rule, the existence of which is ensured by the structure of 
LH (in particular Lie) system \cite{Lucas}. One of the advantages of a compatible symplectic structure for LH systems, when compared to classical Lie systems, is that the coalgebra formalism (see \cite{Ballesteros2013} and references therein for details) provides a systematic and algorithmic procedure to construct $t$-independent constants of the motion, using the Casimir invariants of the LH algebra, as explained in Section~\ref{subsection:coalgebra}. We now consider the LH algebra $\cH_{\omega}$ of the $\widehat{\cS}(1)$-LH system $\X$ on $T^{*} \R^{2}$, expressed on a basis $\set{v_{1}, \ldots, v_{6}}$  satisfying the brackets  \eqref{eq:Schrodinger:ham_cr}. It admits a cubic Casimir invariant (compare with \eqref{eq:Schrodinger:C3}) given by 
\begin{equation}
C_{3} = v_{6} (v_{1}^{2} - 4 v_{2} v_{3}) + 2(v_{3} v_{4}^{2} + v_{2} v_{5}^{2} - v_{1}v_{4}v_{5}), \qquad \set{C_{3}, \cdot} = 0.
\label{eq:Schrodinger:CasimirLH}
\end{equation}
In this case, the morphism of Lie algebras \eqref{corre} is given by 
\[ \phi: \cH_{\omega} \to C^{\infty}(T^{*}\R^{2}), \qquad v_{\alpha} \mapsto \phi(v_{\alpha}) = h_{\alpha}, \qquad 1 \leq \alpha \leq 6,\]
while the morphism of Poisson algebras \eqref{cg} $D: C^{\infty}(\cH_{\omega}^{*}) \to C^{\infty}(T^{*}\R^{2})$ is just 
\[ D(v_{\alpha}) = h_{\alpha}(q_{1}, p_{1}, q_{2},p_{2}), \qquad 1 \leq \alpha \leq 6. \]

Now, the smallest integer $s$ such that the diagonal prolongations of the vector fields \eqref{eq:Schrodinger:vf} to the product manifold $(T^{*}\R^{2})^{s}$ are linearly independent at a generic point is $s =3$, so we construct the following Hamiltonian functions $h_{\alpha}^{(k)} = D^{(k)}(\Delta^{(k)}(v_{\alpha})) \in C^{\infty}((T^{*} \R^{2})^{k})$ for $1 \leq k \leq 4 = s+1$ (see \eqref{cg}):
\begin{equation}
\begin{split}
&h_{1}^{(k)} = - \sum_{\ell = 1}^{k} q_{1}^{(\ell)} p_{1}^{(\ell)}, \qquad h_{2}^{(k)} =  - \frac{1}{2} \sum_{\ell = 1}^{k} \left( p_{1}^{(\ell)}\right)^{2}, \qquad h_{3}^{(k)} = - \frac{1}{2} \sum_{\ell = 1}^{k} \left( q_{1}^{(\ell)}\right)^{2}, \qquad \\
 &h_{4}^{(k)} = - \sum_{\ell = 1}^{k} q_{2}^{(\ell)} p_{1}^{(\ell)},\qquad h_{5}^{(k)} = - \sum_{\ell = 1}^{k} q^{(\ell)}_{1} q^{(\ell)}_{2}, \qquad h_{6}^{(k)}  = - \sum_{\ell = 1}^{k} \left( q_{2}^{(\ell)} \right)^{2},
 \end{split}
 \label{eq:Schrodinger:ham_pro}
\end{equation}
where $(q_{1}^{(\ell)}, q_{2}^{(\ell)}, p_{1}^{(\ell)}, p_{2}^{(\ell)})$ denote the coordinates in the $\ell^{\textup{th}}$-copy of $T^{*} \R^{2}$. Each set of functions in \eqref{eq:Schrodinger:ham_pro} satisfies the relations  \eqref{eq:Schrodinger:ham_cr} with respect to the Poisson bracket induced in the product  $(T^{*}\R^{2})^{4}$  by the symplectic form
\begin{equation*}
\omega^{[4]} := \sum_{\ell = 1}^{4} \dd q_{1}^{(\ell)} \wedge \dd p_{1}^{(\ell)} + \dd q_{2}^{(\ell)} \wedge \dd p_{2}^{(\ell)}. 
\end{equation*} 
Using the Casimir \eqref{eq:Schrodinger:CasimirLH}, we obtain the constants of the motion  $F^{(k)} = D^{(k)}(\Delta^{(k)} (C_{3}))$ for the diagonal prolongation $\widetilde{\X}^{4}$ of $\X$ to $(T^{*} \R^{2})^{4}$ (see \eqref{cj}). As expected, for the first two values we get $F^{(1)} = F^{(2)} = 0$, while  
\begin{equation}
F^{(3)} = \left( q_{2}^{(1)} \left( p_{1}^{(2)} q^{(3)}_{1} - p_{1}^{(3)} q^{(2)}_{1} \right) +  q_{2}^{(2)} \left( q_{1}^{(1)} p_{1}^{(3)} - q_{1}^{(3)} p_{1}^{(1)} \right) + q_{2}^{(3)} \left( p_{1}^{(1)} q^{(2)}_{1} - p_{1}^{(2)} q^{(1)}_{1} \right) \right)^{2}.
\label{eq:Schrodinger:3rdconstant}
\end{equation}
This constant of the motion gives rise to three additional constants (see \eqref{ck}) through the permutation $S_{ij}$ of the variables $(q_{1}^{(i)}, q_{2}^{(i)}, p_{1}^{(i)}, p_{2}^{(i)}) \leftrightarrow (q_{1}^{(j)}, q_{2}^{(j)}, p_{1}^{(j)}, p_{2}^{(j)})$; specifically 
\begin{equation}{\small
\begin{split}
&F_{34}^{(3)} = S_{34}(F^{(3)}) = \left( q_{2}^{(1)} \left( p_{1}^{(2)} q^{(4)}_{1} - p_{1}^{(4)} q^{(2)}_{1} \right) +  q_{2}^{(2)} \left( q_{1}^{(1)} p_{1}^{(4)} - q_{1}^{(4)} p_{1}^{(1)} \right) + q_{2}^{(4)} \left( p_{1}^{(1)} q^{(2)}_{1} - p_{1}^{(2)} q^{(1)}_{1} \right) \right)^{2},  \\
& F_{24}^{(3)} = S_{24}(F^{(3)}) = \left( q_{2}^{(1)} \left( p_{1}^{(4)} q^{(3)}_{1} - p_{1}^{(3)} q^{(4)}_{1} \right) +  q_{2}^{(4)} \left( q_{1}^{(1)} p_{1}^{(3)} - q_{1}^{(3)} p_{1}^{(1)} \right) + q_{2}^{(3)} \left( p_{1}^{(1)} q^{(4)}_{1} - p_{1}^{(4)} q^{(1)}_{1} \right) \right)^{2}, \\
& F_{14}^{(3)} = S_{14} (F^{(3)}) = \left( q_{2}^{(4)} \left( p_{1}^{(2)} q^{(3)}_{1} - p_{1}^{(3)} q^{(2)}_{1} \right) +  q_{2}^{(2)} \left( q_{1}^{(4)} p_{1}^{(3)} - q_{1}^{(3)} p_{1}^{(4)} \right) + q_{2}^{(3)} \left( p_{1}^{(4)} q^{(2)}_{1} - p_{1}^{(2)} q^{(4)}_{1} \right) \right)^{2}.
\end{split} }
\label{eq:Schrodinger:3rdconstant_permutations}
\end{equation}
The fourth-order constant of the motion $F^{(4)}$, on the other hand, does not provide a new constant, as it can be expressed in terms of the  constants of the motion \eqref{eq:Schrodinger:3rdconstant}--\eqref{eq:Schrodinger:3rdconstant_permutations} as 
\begin{equation*}
F^{(4)} = F^{(3)} + F_{34}^{(3)} + F_{24}^{(3)} + F_{14}^{(3)}.
\end{equation*}
In order to derive a superposition rule for the Lie system \eqref{eq:Schrodinger:system}, we need $n=4$ functionally independent first integrals $I_{1}, \ldots, I_{4}$ common to all diagonal prolongations of the vector fields \eqref{eq:Schrodinger:vf}, i.e., such that the Jacobian condition 
\begin{equation}
\pdv{(I_{1}, \ldots, I_{4})}{(q_{1}^{(1)}, q_{2}^{(1)}, p_{1}^{(1)}, p_{2}^{(1)})} \neq 0
\label{eq:Schrodinger:reg_super}
\end{equation}
is satisfied. The constants of the motion $F^{(3)}$, $F_{34}^{(3)}$, $F_{24}^{(3)}$ and $F_{14}^{(3)}$ clearly do not satisfy this requirement, as they are independent on $p_{2}^{(1)}$. However, the $\widehat{\cS}(1)$-LH system admits an additional constant of the motion that does not arise from the coalgebra method,\footnote{In Subsection~\ref{subsection:sp4_superpositionrule}, we explain the origin of this constant of the motion (see \eqref{eq:sp4:2dconstant}).} and given by 
\begin{equation}
G_{2} := - \left( p_{1}^{(2)}q^{(1)}_{1} - p_{1}^{(1)} q^{(2)}_{1} + p_{2}^{(2)} q^{(1)}_{2} - p_{2}^{(1)} q^{(2)}_{2} \right)^{2}. 
\label{eq:Schrodinger:G2}
\end{equation}
The set $\left\{F^{(3)}, F_{34}^{(3)}, F_{24}^{(3)}, G_2\right\}$ now fulfills the condition \eqref{eq:Schrodinger:reg_super}, and can thus be used for deriving a superposition rule. With four significative constants $F^{(3)} = k_{1}^{2}, F^{(3)}_{34} = k_{2}^2, F^{(3)}_{24} = k_{3}^{2}$ and $G_{2} = -k_{4}^{2}$, the general solution $(q_{1}(t), q_{2}(t), p_{1}(t), p_{2}(t))$ can be obtained by solving the latter system of algebraic equations with respect to three particular solutions $(q_{1}^{(\ell)}(t), q_{2}^{(\ell)}(t), p_{1}^{(\ell)}(t), p_{2}^{(\ell)}(t))$, $1 \leq \ell \leq 3$. After some cumbersome but routine computation, it follows that
\begin{equation}
\begin{split}
&q_{i}(t) = \frac{q_{i}^{(1)} - k_{2} q_{i}^{(2)} + k_{1} q_{i}^{(3)}}{k_3} \quad (1 \leq i \leq 2), \qquad p_{1}(t) = \frac{p_{1}^{(1)} - k_{2}p_{1}^{(2)} + k_{1}p_{1}^{(3)}}{k_3}, \\
&p_{2}(t) = \frac{p_{2}^{(1)} q_{2}^{(1)} + k_{2}(p_{1}^{(2)} q_{1}^{(1)} - p_{1}^{(1)} q_{1}^{(2)} -p_{2}^{(1)} q_{2}^{(2)}) + k_{1}(-p_{1}^{(3)} q_{1}^{(1)} + p_{1}^{(1)} q_{1}^{(3)} + p_{2}^{(1)} q_{2}^{(3)}) - k_3 k_{4} }{k_3q_{2}^{(1)} }.
\end{split}
\label{eq:Schrodinger:superposition}
\end{equation}
In particular, the preceding expression can also be used to derive a superposition rule for the reduced Lie system \eqref{eq:Schrodinger:system3d} on $T^{*}\R \times \R$, setting  $q_{2}^{(\ell)} = 1$ for $1 \leq \ell \leq 3$:  
\begin{equation}
\begin{split}
&q(t) = \frac{q^{(1)} - k_{2} q^{(2)} + k_{1}q^{(3)}}{k_3}, \qquad p(t) = \frac{p^{(1)} - k_{2}p^{(2)} + k_{1}p^{(3)}}{k_{3}},\\
& s(t) =  \frac{s^{(1)}  + k_{2}(p^{(2)} q^{(1)} - p^{(1)} q^{(2)} -s^{(1)}) + k_{1}(-p^{(3)} q^{(1)} + p^{(1)} q^{(3)} + s^{(1)}) -k_3 k_{4}}{k_{3}}.
\end{split}
\label{eq:Schrodinger:superpositionred}
\end{equation} 
Furthermore, after the projection to $T^{*}\R$ of the superposition rule \eqref{eq:Schrodinger:superpositionred} via \eqref{eq:Schrodinger:projection}, we recover the superposition rule of the P$_{5}$-LH class \eqref{eq:Schrodinger:system2d} on $T^{*}\R$ in terms of three particular solutions $(q^{(\ell)}, p^{(\ell)})$ and three significative constants (see \cite{Blasco2015}): 
\begin{equation*}
q(t) = \frac{q^{(1)} - k_{2} q^{(2)} + k_{1}q^{(3)}}{k_{3}}, \qquad p(t) = \frac{p^{(1)} - k_{2}p^{(2)} + k_{1}p^{(3)}}{k_{3}}. 
\end{equation*}
LH systems associated to the P$_{5}$-LH class arise, among other applications, when studying dissipative or damped harmonic oscillators, particles under the action of specific electric fields, quadratic Hamiltonians, second-order Riccati equations or time-dependent epidemic models (see \cite{Blasco2015,Campoamor2023} and references therein).

\section{The Lorentz Lie algebra} \label{section:Lorentz}

From either the mathematical or physical point of view, the Lorentz algebra $\so(1, 3)$ has turned out to be one of the most relevant Lie algebras, due various branches where it appears, such as constant curvature spacetimes \cite{Bacry1968,Herranz1999}, coherent states representations \cite{Campoamor2013} or noncommutative spacetimes \cite{Cerchiai1998,Ballesteros2021}, among others. 
As real Lie algebra, $\so(1, 3)$ is isomorphic to $\sl(2, \C)$, hence corresponding to a real form generated by outer involutive automorphisms of a compact Lie algebra with semisimple and non-simple complexification. As a basis of $\so(1,3)$ we consider the six generators  $J$, $P_{i}$, $K_{i}$ and $H$ $(1 \leq i \leq 2)$, interpreted as infinitesimal generators of spatial rotations, space translations, boosts and time translations, respectively. The commutation relations are given by ($1 \leq i \leq 2$)
\begin{equation*}
\begin{array}{llllll}
& [J, P_{i}] = \epsilon_{ij}P_{j}, \qquad &[J, K_{i}] = \epsilon_{ij} K_{j}, \qquad &[J, H] = 0, \qquad & [P_{1}, P_{2}] = - J, & \\
& [K_{1}, K_{2}] = - J, \qquad &[P_{i}, K_{j}] = - \delta_{ij} H, \qquad &[H, P_{i}] = - K_{i}, \qquad & [H, K_{i}] = - P_{i}.
\end{array}
\end{equation*}
As $\so(1, 3)$ has rank two, it has two quadratic Casimir invariants,\footnote{The fact that both invariants are quadratic is a direct consequence of the structure of $\so(1, 3)$ as a real form.} which can be obtained by standard methods:  
\begin{equation*}
C_{2} :=J^{2} + P_{1}^{2} + P_{2}^{2} - H^{2} - K_{1}^{2} - K_{2}^{2}, \qquad C'_{2} := - JH - P_{1}K_{2} + P_{2} K_{1},
\label{eq:Lorentz:Casimirs}
\end{equation*}
where $C_{2}$ is related to the Killing--Cartan form of $\so(1, 3)$, while $C_{2}'$ is related to the Pauli--Lubanski vector. 

Consider now the faithful and irreducible representation $\Gamma: \so(1, 3) \to \gl(4, \R)$ given by the matrix 
\begin{equation*}
A_{\Gamma} := \frac{1}{2} \begin{pmatrix}
K_{1} & P_{2} & - H - P_{1} & -J + K_{2} \\
- P_{2} & K_{1} & - J + K_{2} & H + P_{1} \\
- H + P_{1} & J + K_{2} & - K_{1} & P_{2} \\
J + K_{2} & H - P_{1} & - P_{2} & - K_{1}
\end{pmatrix}.
\end{equation*}
The realization $\Phi_{\Gamma}: \so(1, 3) \to \cv(T^{*} \R^{2})$ induced by $\Gamma$ is spanned by the vector fields 
\begin{equation}
\begin{split}
&\X_{1} := \Phi_{\Gamma}(J) = \frac{1}{2} \left( p_{2} \pdv{q_{1}} + p_{1} \pdv{q_{2}} - q_{2} \pdv{p_{1}} - q_{1} \pdv{p_{2}} \right), \\
& \X_{2} := \Phi_{\Gamma} (P_{1}) = \frac{1}{2} \left( p_{1} \pdv{q_{1}} - p_{2} \pdv{q_{2}} - q_{1} \pdv{p_{1}} + q_{2} \pdv{p_{2}} \right), \\
& \X_{3} := \Phi_{\Gamma} (P_{2} ) = \frac{1}{2} \left( - q_{2} \pdv{q_{1}} + q_{1} \pdv{q_{2}} - p_{2} \pdv{p_{1}} + p_{1} \pdv{p_{2}} \right), \\
 & \X_{4} := \Phi_{\Gamma}(H ) = \frac{1}{2}  \left( -p_{1} \pdv{q_{1}} + p_{2} \pdv{q_{2}} - q_{1} \pdv{p_{1}} + q_{2} \pdv{p_{2}} \right), \\
  & \X_{5} := \Phi_{\Gamma} (K_{1}) = \frac{1}{2} \left( q_{1} \pdv{q_{1}} + q_{2} \pdv{q_{2}} - p_{1} \pdv{p_{1}} - p_{2} \pdv{p_{2}} \right), \\
   & \X_{6} := \Phi_{\Gamma} (K_{2} ) = \frac{1}{2} \left( p_{2} \pdv{q_{1}} + p_{1} \pdv{q_{2}} + q_{2} \pdv{p_{1}} + q_{1} \pdv{p_{2}} \right). 
\end{split}
\label{eq:Lorentz:vf}
\end{equation}
The Lie system induced by the realization $\Phi_{\Gamma}$ of $\so(1, 3)$ on $T^{*} \R^{2}$ is associated to the $t$-dependent vector field 
\begin{equation}
\X := \sum_{\alpha = 1}^{6} b_{\alpha}(t) \X_{\alpha},\quad b_{\alpha} \in C^{\infty}(\R);\quad 1 \leq \alpha \leq 6.
\label{eq:Lorentz:tvf}
\end{equation}
The associated system of first-order ODEs on $T^{*} \R^{2}$ is given by 
\begin{equation}
\begin{split}
\dv{q_{1}}{t} &= \frac{1}{2} \bigl(  b_{5}(t) q_{1}- b_{3}(t) q_{2}   +  (b_{2}(t) - b_{4}(t))p_{1} + (b_{1}(t) + b_{6}(t))p_{2}    \bigr),\\ 
\dv{q_{2}}{t} & = \frac{1}{2} \bigl( b_{3}(t) q_{1} + b_{5}(t) q_{2} +  (b_{1}(t) + b_{6}(t))p_{1} + (b_{4}(t) - b_{2}(t)) p_{2}   \bigr), \\
\dv{p_{1}}{t} & = \frac{1}{2} \bigl( -(b_{2}(t) + b_{4}(t)) q_{1} + (b_{6}(t) - b_{1}(t)) q_{2} - b_{5}(t) p_{1} - b_{3}(t) p_{2} \bigr), \\
\dv{p_{2}}{t} & = \frac{1}{2} \bigl( (b_{6}(t) - b_{1}(t)) q_{1} + (b_{2}(t) + b_{4}(t)) q_{2} + b_{3}(t) p_{1} - b_{5}(t) p_{2} \bigr).
\end{split}
\label{eq:Lorentz:system}
\end{equation}
As before, being a system associated to a matrix representation, it is linear and can be transformed in such manner that the solution is expressed in terms of one of the coordinates and its derivatives, e.g. $p_2$, while $p_2$ itself satisfies a linear fourth-order ODE. However, due to the fact that the $b_{\alpha}$ are arbitrary functions, the resulting ODE is not very useful, for which reason we skip the explicit formulae.

With respect to the canonical symplectic form \eqref{eq:symplecticform} on $T^{*} \R^{2}$, we have that $\cL_{\X_{\alpha}} \omega = 0$ for $1 \leq \alpha \leq 6$, showing that $\X$ actually defines a LH system on $T^{*} \R^{2}$. A short computation shows that the Hamiltonian functions associated to the vector fields are 
\begin{equation}
\begin{split}
&h_{1}  = \frac{1}{2} (q_{1}q_{2} + p_{1}p_{2}), \\
& h_{4} = \frac{1}{4} \left( q_{1}^{2} - q_{2}^{2} - p_{1}^{2} + p_{2}^{2} \right), 
\end{split}
\qquad 
\begin{split}
&h_{2} = \frac{1}{4} \left( q_{1}^{2} -q_{2}^{2} + p_{1}^{2} - p_{2}^{2} \right), \\
&h_{5} = \frac{1}{2}(q_{1}p_{1} + q_{2}p_{2} ),
\end{split}
\qquad
\begin{split}
&h_{3} = \frac{1}{2}(q_{1} p_{2} - q_{2} p_{1}),\\
&h_{6} = \frac{1}{2} (- q_{1} q_{2} + p_{1}p_{2}),
\end{split}
\label{eq:Lorentz:ham}
\end{equation}
with commutation relations (with respect to the Poisson bracket $\set{\cdot, \cdot}_{\omega}$ induced by the canonical symplectic form of $T^{*} \R^{2}$ on $C^{\infty} (T^{*} \R^{2})$) 
\begin{equation}
\begin{array}{lllll}
& \set{h_{1}, h_{2}}_{\omega} = -  h_{3}, \qquad & \set{h_{1}, h_{3}}_{\omega} = h_{2}, \qquad &\set{h_{1}, h_{4}}_{\omega} = 0, \qquad &\set{h_{1}, h_{5}}_{\omega} =- h_{6}, \\
& \set{h_{1}, h_{6}}_{\omega} = h_{5}, \qquad & \set{h_{2}, h_{3}}_{\omega} = -h_{1}, \qquad & \set{h_{2}, h_{4}}_{\omega} = -h_{5}, \qquad & \set{h_{2}, h_{5}}_{\omega} = h_{4}, \\
& \set{h_{2}, h_{6}}_{\omega} = 0, \qquad & \set{h_{3}, h_{4}}_{\omega} = -h_{6}, \qquad & \set{h_{3}, h_{5}}_{\omega} = h_{4}, \qquad & \set{h_{3}, h_{6}}_{\omega} = - h_{6}, \\
& \set{h_{4}, h_{5}}_{\omega} = h_{2}, \qquad & \set{h_{4}, h_{6}}_{\omega} = h_{3}, \qquad & \set{h_{5}, h_{6}}_{\omega} = h_{1},
\end{array}
\label{eq:Lorentz:ham_cr}
\end{equation}
spanning a LH algebra $\cH_{\omega} \simeq \so(1, 3)$. 

We first prove that the $\so(1, 3)$-LH system \eqref{eq:Lorentz:system} on $T^{*} \R^{2}$ is intrinsic. Recall that the so-called P$_{7} \simeq \so(1, 3)$  class on $T^{*} \R$ is spanned by the vector fields \cite{GonzalezLopez1992}
\begin{equation*}
\begin{split}
& \Y_{1} = \pdv{q}, \qquad \Y_{2} = \pdv{p}, \qquad \Y_{3} = q \pdv{q} + p \pdv{p}, \qquad \Y_{4} = p \pdv{q} - q \pdv{p},\\
& \Y_{5} = (q^{2} - p^{2}) \pdv{q} + 2qp \pdv{p}, \qquad \Y_{6} = 2qp \pdv{q} + (p^{2} - q^{2}) \pdv{p},
\end{split}
\label{eq:Lorentz:P7}
\end{equation*} 
expressed in the local coordinates $(q, p)$ of $T^{*}\R$. However, as proved in \cite{Ballesteros2015}, this does not define a LH system on $T^{*}\R$. 

\begin{prop}
The $\so(1, 3)$-LH system \eqref{eq:Lorentz:system} on $T^{*} \R^{2}$ is intrinsic, i.e., it is not locally diffeomorphic to either the $\mathrm{P}_{7}$ class on $T^{*}\R$ or to the uncoupled sum $\mathrm{P}_{7} +_{{\rm u}} \mathrm{P}_{7}$ on $T^{*} \R^{2}$.
\end{prop}

\begin{proof}
The maximum rank of the generalized distribution associated to the $\so(1, 3)$-LH system \eqref{eq:Lorentz:system} on $T^{*} \R^{2}$ is $4$, so that, as Lie system, it cannot be locally diffeomorphic to a lower-dimensional Lie system by Proposition~\ref{prop:Reduction:criterio}. In particular, it is not locally diffeomorphic to the $\mathrm{P}_{7}$ class on $T^{*}\R$.

The uncoupled sum $\mathrm{P}_{7} +_{{\rm u}} \mathrm{P}_{7}$ on $T^{*} \R^{2}$ is spanned by the vector fields 
\begin{equation}
\begin{split}
&\mathbf{Z}_{1} = \pdv{q_{1}} + \pdv{q_{2}}, \qquad \mathbf{Z}_{2} = \pdv{p_{1}} + \pdv{p_{2}}, \\
& \mathbf{Z}_{3} = q_{1} \pdv{q_{1}} + p_{1} \pdv{p_{1}} + q_{2} \pdv{q_{2}} + p_{2} \pdv{p_{2}}, \\
& \mathbf{Z}_{4} = p_{1} \pdv{q_{1}} - q_{1} \pdv{p_{1}} + p_{2} \pdv{q_{2}} - q_{2} \pdv{p_{2}},\\
& \mathbf{Z}_{5} = \left( q_{1}^{2} - p_{1}^{2} \right) \pdv{q_{1}} + 2q_{1} p_{1} \pdv{p_{1}}+ \left( q_{2}^{2} - p_{2}^{2} \right) \pdv{q_{2}} + 2q_{2} p_{2} \pdv{p_{2}}, \\
& \mathbf{Z}_{6}= 2q_{1} p_{1} \pdv{q_{1}} + \left( p_{1}^{2} - q_{1}^{2} \right) \pdv{p_{1}} + 2q_{2} p_{2} \pdv{q_{2}} + \left( p_{2}^{2} - q_{2}^{2} \right) \pdv{p_{2}}, 
\end{split}
\label{eq:Lorentz:P7+P7:vf}
\end{equation}
and the maximum rank of the associated generalized distribution is $r^{\mathrm{P}_{7} +_{{\rm u}} \mathrm{P}_{7}} = 4$. We show that the uncoupled sum $\mathrm{P}_{7} +_{{\rm u}} \mathrm{P}_{7}$ does not admit the structure of a LH system on $T^{*} \R^{2}$, hence implying, by Proposition~\ref{prop:Reduction:localdiffeo}, that it cannot be locally diffeomorphic to the system \eqref{eq:Lorentz:system}. Suppose that there exists a symplectic form $\omega \in \Omega^{2}(T^{*} \R^{2})$ such that the vector fields \eqref{eq:Lorentz:P7+P7:vf} satisfy the constraint $\cL_{\mathbf{Z}_{\alpha}} \omega = 0$ for all $1 \leq \alpha \leq 6$. Using Proposition~\ref{prop:Reduction:symmetries_volume}, there exists a nowhere zero smooth function $f \in C^{\infty}(T^{*} \R^{2})$ such that $\Omega_{\omega} = f (\dd q_{1} \wedge \dd p_{1} \wedge \dd q_{2} \wedge \dd p_{2})=  \omega \wedge \omega$, implying that the vector fields \eqref{eq:Lorentz:P7+P7:vf} are symmetries of the volume form $\Omega_{\omega}$. Therefore, 
\begin{equation}
f \div \mathbf{Z}_{\alpha} + \dd f (\mathbf{Z}_{\alpha}) = 0, \qquad 1 \leq \alpha \leq 6,
\label{eq:Lorentz:P7+P7:symmetries}
\end{equation}
where $\div \mathbf{Z}_{\alpha}$ is the divergence of the vector field $\mathbf{Z}_{\alpha}$ with respect to $f^{-1}
\Omega_{\omega}$. Now, condition \eqref{eq:Lorentz:P7+P7:symmetries} applied to $\mathbf{Z_{1}}$ and $\mathbf{Z_{2}}$, implies that $f = g_{1} (z_{1}, z_{2}) = g_{1} (q_{1}- q_{2}, p_{1}- p_{2})$ for a certain differentiable function $g_{1} \in C^{\infty}(\R^{2})$, where $z_{1} = z_{1}(q_{1}, q_{2}) := q_{1} - q_{2}$ and $z_{2} = z_{2}(p_{1}, p_{2}):= p_{1}- p_{2}$. On the other hand, applying the condition \eqref{eq:Lorentz:P7+P7:symmetries} to $\mathbf{Z}_{4}$ shows that 
\begin{equation*}
f(q_{1}, q_{2}, p_{1}, p_{2}) = g_{1}(z_{1}, z_{2}) = g_{2}(z),  \qquad z = z(z_{1}, z_{2}):= z_{1} z_{2}
\end{equation*}
for a certain smooth function $g_{2} \in C^{\infty}(\R)$. We conclude from this that $g_{2} = 0$, implying that $f = 0$, which is a contradiction with the assumption that $f$ is nowhere zero. 
\end{proof}

\subsection{Constants of the motion} \label{subsection:constants_Lorentz}
As done in Subsection~\ref{subsection:constants_h6} for $\widehat{\cS}(1)$, a superposition rule for the LH system $\X$ in \eqref{eq:Lorentz:tvf}
can be deduced by application of the coalgebra formalism, starting from one of the two quadratic Casimir invariants of the  LH algebra $\cH_{\omega} \simeq \so(1,3)$, given by  
\begin{equation}
\begin{split}
C_{2} =& \ v_{1}^{2} + v_{2}^{2} + v_{3}^{2} -v_{4}^{2} - v_{5}^{2} - v_{6}^{2},  \\
C_{2}' =& -v_{1}v_{4} - v_{2}v_{6} + v_{3}v_{5}
\label{eq:Lorentz:CasimirLH}
\end{split}
\end{equation}
over a basis  $\set{v_{1}, \ldots, v_{6}}$ satisfying the brackets \eqref{eq:Lorentz:ham_cr}. In this case, the smallest integer $s$ such that the diagonal prolongations of the vector fields \eqref{eq:Lorentz:vf} to the product manifold $(T^{*}\R^{2})^{s}$ are linearly independent at a generic point is $s = 4$, thus a superposition rule depending on four particular solutions and four significant constants can be constructed. However, in spite of the apparent simplicity of the invariants, the resulting superposition formulae are rather cumbersome from the computational point of view, for which reason we skip their explicit expression here. However, in Subsection~\ref{subsection:sp4_superpositionrule} we show how to derive a superposition rule for the system \eqref{eq:Lorentz:system} using an appropriate embedding of $\so(1, 3)$ into a larger simple Lie algebra, by restricting a more general superposition principle to a subsystem equivalent to \eqref{eq:Lorentz:system}. This indirect approach is motivated by the following observation. 
An inspection of the ODE system \eqref{eq:Lorentz:system} suggests to consider a relabeling of the coefficient functions as follows 
\begin{equation}
\begin{split}
a_{1} :=\frac 12 ( b_{1} + b_{6}) , \qquad a_{2} :=\frac 12 ( b_{2} - b_{4}), \qquad a_{3}:= \frac 12 b_{3}, \\[2pt]
 a_{4}:= \frac 12 (b_{2} + b_{4}), \qquad a_{5} := \frac 12 b_{5}, \qquad a_{6}:= \frac 12 (b_{1} - b_{6}). 
\label{eq:Lorentz:new_tdep}
\end{split}
\end{equation}
Writing the resulting system in matrix form, we obtain 
\begin{equation}\label{eq:Lorentz:system_mod}
 \begin{pmatrix}
 \dot{q}_{1} \\
 \dot{q}_{2} \\
 \dot{p}_{1} \\
 \dot{p}_{2}
\end{pmatrix}  = \left(
\begin{array}[c]{rrrr}
a_5(t) & -a_3(t) & a_2(t) & a_1(t)\\
a_3(t) & a_5(t) & a_1(t) & -a_2(t)\\
-a_4(t) & -a_6(t) & -a_5(t) & -a_3(t)\\
-a_6(t) & a_4(t) & a_3(t) & -a_5(t)\\
\end{array} \right) \begin{pmatrix}
q_{1} \\
q_{2} \\
p_{1} \\
p_{2}
\end{pmatrix}.
\end{equation}
It is obvious that for any value of $t$, the coefficient matrix $\mathbf{A}(t)$ is symplectic, i.e., satisfies the condition $J\mathbf{A}(t)+ \mathbf{A}^{T}(t)J=0$, where 
\begin{equation}\label{symp}
J=\left(
\begin{array}[c]{cc}
0 & {\rm Id}_2\\
-{\rm Id}_2 & 0
\end{array}\right).
\end{equation}
Hence, the \eqref{eq:Lorentz:system_mod} can be seen as a particular Lie system associated to the 4-dimensional fundamental representation of the symplectic Lie algebra $\mathfrak{sp}(4,\mathbb{R})$. As will be shown in Subsection~\ref{subsection:sp4_superpositionrule}, such Lie system is actually a LH system, and a superposition rule can be derived for the latter using the coalgebra formalism. It turns out that this rule, when restricted to the subalgebra $\mathfrak{so}(1,3)$, is computationally more manageable than those obtained using the Casimir invariants of the Lorentz algebra.   

\subsection{Applications to coupled systems} \label{subsection:Lorentz:applications}

Using the preceding reformulation \eqref{eq:Lorentz:new_tdep} of the coefficients we obtain the following change of basis on the LH algebra $\cH_{\omega}$ (see \eqref{eq:Lorentz:ham}): 
\begin{align*}
&h_{1}' := h_{1} + h_{6} = p_{1}p_{2}, & &h_{2}':= h_{2} - h_{4} = \frac{1}{2} \left( p_{1}^{2} - p_{2}^{2} \right), &  &
h_{3}' := 2 h_{3} =  q_{1}p_{2} - q_{2} p_{1}, \nonumber \\
&h_{4}':= h_{2} + h_{4} = \frac{1}{2} \left( q_{1}^{2} - q_{2}^{2} \right), & &
h_{5}':= 2 h_{5} =  q_{1}p_{1} + q_{2}p_{2},  & &h_{6}':= h_{1} - h_{6} = q_{1}q_{2}.
\label{eq:Lorentz:changeHam}
\end{align*}
Therefore, the system \eqref{eq:Lorentz:system_mod} corresponds to the Hamilton equations of the $t$-dependent Hamiltonian $h':= \sum_{\alpha=1}^6 a_\alpha(t) h'_\alpha$, namely
\begin{equation}
h':= a_{1}(t) p_{1}p_{2} + \frac{a_{2}(t)}{2}\bigl( p_{1}^{2} - p_{2}^{2} \bigr) +a_{3}(t)(q_{1}p_{2} - q_{2}p_{1})  + \frac{a_{4}(t)}{2} \bigl( q_{1}^{2} - q_{2}^{2} \bigr) + a_{5}(t)  (q_{1}p_{1} + q_{2}p_{2})  + a_{6}(t) q_{1}q_{2}.
\label{eq:Lorentz:tHam_mod}
\end{equation}
The commutation relations of the $h^{\prime}_{\alpha}$ are immediately obtained from \eqref{eq:Lorentz:ham_cr}, and, in particular, we observe that 
\begin{equation*}\label{eq:Lorentz:ham'_cr}
\set{h_{2}', h_{3}'}_{\omega} = - 2h_{1}', \qquad  \set{h_{2}', h_{4}'}_{\omega} = -h_{5}' , \qquad  \set{h_{3}', h_{4}'}_{\omega} = 2 h_{6}' ,
\end{equation*}
showing that the Hamiltonian functions $h_{2}', h_{3}'$ and $h_{4}'$ actually span the LH algebra. This implies that $\mathfrak{so}(1,3)$ is actually the \textit{minimal} LH algebra for the LH system \eqref{eq:Lorentz:system_mod}, in the sense of \cite{Ballesteros2015}. This suggests us to consider the choice $a_{\alpha}(t) \equiv 0$ for all $\alpha \neq 2, 3, 4$, so that the Hamiltonian \eqref{eq:Lorentz:tHam_mod} reduces to  
\begin{equation}
h' = \frac{a_{2}(t)}{2}\bigl(p_{1}^{2} - p_{2}^{2} \bigr) +a_{3}(t)(q_{1}p_{2} - q_{2}p_{1}) + \frac{a_{4}(t)}{2} \bigl(q_{1}^{2} - q_{2}^{2} \bigr).
\label{eq:Lorentz:tHam_modBateman}
\end{equation}
This Hamiltonian exhibits some remarkable properties. If we consider the (1+1)D Minkowskian spacetime $\M^{1+1}$, i.e., the real plane $\R^{2}$ equipped with the Lorentzian metric $\dd s^{2} = \dd q_{1}^{2} - \dd q_{2}^{2}$, the Hamiltonian function $h_{2}' = \frac{1}{2}(p_{1}^{2} -p_{2}^{2})$ can be interpreted as the Lorentzian kinetic energy coming from the free Lagrangian associated to the Lorentzian metric. Moreover, $h_{3}' = (q_{1}p_{2} - q_{2}p_{1})$ corresponds to the angular momentum, while $h_{4}' = \frac{1}{2}(q_{1}^{2} - q_{2}^{2})$ is a potential of the isotropic oscillator on $\M^{1+1}$. Thus, the $t$-dependent Hamiltonian \eqref{eq:Lorentz:tHam_modBateman} describes a $t$-dependent oscillator Hamiltonian system on the Minkowskian plane $\M^{1+1}$.

As pointed out in \cite{Blasone1996}, the Lorentzian metric $\dd s^{2} = \dd q_{1}^{2} - \dd q_{2}^{2}$ of the Minkowskian plane is necessary for the decomposition of a canonical (Hamiltonian) system into two subsystems, whenever these are noncanonical ones. In this context, the Hamiltonian \eqref{eq:Lorentz:tHam_modBateman} can be considered as a coupled system. First of all, let us consider the cotangent bundle $T^{*}\M^{1+1}$ as the Whitney sum $T^{*}\M^{1+1} = T^{*}\R \oplus T^{*}\R$ together with the projections
\begin{equation}
\begin{split}
&\pr_{1}: T^{*} \M^{1+1} \to T^{*}\R, \qquad (q_{1}, q_{2}, p_{1}, p_{2}) \mapsto (q_{1}, p_{1}), \\
&\pr_{2}: T^{*}\M^{1+1} \to T^{*}\R, \qquad  (q_{1}, q_{2}, p_{1}, p_{2}) \mapsto (q_{2}, p_{2}),
\end{split}
\label{eq:Lorentz:Whitneyproj}
\end{equation} 
onto its factors. The 1D $t$-dependent Hamiltonian $h'_{\mathrm{1D}}$ on $T^{*}\R$ given by 
\begin{equation}
h'_{\mathrm{1D}} := \frac{a_{2}(t)}{2}p^{2} + \frac{a_{4}(t)}{2} q^{2}
\label{eq:Lorentz:coupled1D}
\end{equation}
is a LH system on $T^{*}\R$ such that the $t$-dependent Hamiltonian \eqref{eq:Lorentz:tHam_modBateman} on $T^{*}\M^{1+1}$ can be expressed as\footnote{In general, if $H$ is a $t$-dependent Hamiltonian on $T^{*}\R$, we refer to $\widetilde{H} := \pr_{i}^{*}(H)$ as the $t$-dependent Hamiltonian  on $T^{*}\M^{1+1}$ such that $\widetilde{H}_{t}  = H_{t} \circ \pr_{i}$ for all $t \in \R$, with $1 \leq i \leq 2$. } 
\begin{equation*}
h' = \pr_{1}^{*}(h'_{\mathrm{1D}}) - \pr_{2}^{*}(h'_{\mathrm{1D}})+ a_{3}(t)  (q_{1}p_{2} - q_{2}p_{1}),
\end{equation*}
showing that the angular momentum  $h_{3}' =  (q_{1}p_{2} - q_{2}p_{1})$ acts as a `coupling' term between the Hamiltonian systems on $T^{*}\R$ given by \eqref{eq:Lorentz:coupled1D}. Moreover, the Hamilton equations of \eqref{eq:Lorentz:tHam_modBateman} on $T^{*}\M^{1+1}$ are  given by  (see  \eqref{eq:Lorentz:system_mod})
\begin{align}
\dv{q_{1}}{t} &=  - a_{3}(t) q_{2}   +  a_{2}(t)p_{1}   , \qquad
\dv{q_{2}}{t}  =  a_{3}(t) q_{1}  -a_{2}(t)p_{2}  , \nonumber \\
\dv{p_{1}}{t} & =  -a_{4}(t) q_{1}   - a_{3}(t) p_{2}  ,  \qquad 
\dv{p_{2}}{t}  = a_{4}(t) q_{2} + a_{3}(t) p_{1} .
\label{eq:Lorentz:system_coupled}
\end{align} 
Now, the first projection \eqref{eq:Lorentz:Whitneyproj} of the LH system \eqref{eq:Lorentz:system_coupled} onto $T^{*}\R$, yields the following ODEs system on $T^{*}\R$:
\begin{equation*} 
\dv{q}{t} =  {a_{2}(t)} p, \qquad \dv{p}{t} = -  {a_{4}(t)} q.
\label{eq:Lorentz:system1D}
\end{equation*}
These are, as it can be expected, the Hamilton equations of the $t$-dependent Hamiltonian \eqref{eq:Lorentz:coupled1D} on $T^{*}\R$ with respect to the canonical symplectic form $\omega = \dd q \wedge \dd p$, so the ODEs system \eqref{eq:Lorentz:system_coupled} is a coupling of two copies of this 1D system. 

\medskip

We observe that, even in the uncoupled case ($a_3(t)=0$), the solution of the system \eqref{eq:Lorentz:system_coupled} differs from being trivially deduced. For $a_2(t)a_4(t)\neq 0$, the system is equivalent to the system of homogeneous second-order equations 
\begin{equation}\label{red}
\frac{{\rm d}^2q_{i}}{{\rm d}t^2}-\frac{1}{a_2(t)}\dv{a_2}{t}\dv{q_{i}}{t}+ a_2(t)a_4(t)q_i=0,\quad 1 \leq i \leq 2.  
\end{equation}
By means of the transformation 
\begin{equation*}
\dv{q_{i}}{t}=u_i(t)q_i(t),\quad 1 \leq i \leq 2, 
\end{equation*}
the system \eqref{red} can be written as
\begin{equation*}
q_i(t)\left(\dv{u_{i}}{t}+u_i(t)^2-\frac{1}{a_2(t)}\dv{a_2}{t} \, u_i(t)+ a_2(t)a_4(t)\right)=0,\quad 1 \leq i \leq 2,  
\end{equation*}
leading to a generalized Riccati equation. 

In the following, we make particular choices of the $t$-dependent functions in \eqref{eq:Lorentz:tHam_modBateman}, obtaining some coupled systems on $T^{*}\M^{1+1}$ that generalize well-known systems to the non-autonomous case. 

\subsubsection{A $t$-dependent Bateman Hamiltonian}
Consider $m(t), k(t), \gamma(t) \in C^{\infty}(\R)$ such that
\begin{equation*}
m(t), k(t), \gamma(t) > 0, \qquad k(t) > \frac{\gamma^{2}(t)}{4m(t)}, \qquad t \in \R.
\end{equation*}
With the choice 
\begin{equation*}
a_{2}(t) = \frac{1}{m(t)}, \qquad  a_{3}(t) = \frac{\gamma(t)}{2m(t)}, \qquad a_{4}(t) = m(t) \Omega^{2}(t),  \qquad \Omega(t) := \sqrt{ \frac{1}{m(t)} \left( k(t) - \frac{\gamma^{2}(t)}{4m(t)} \right) },
\end{equation*}
the Hamiltonian $h^{\mathrm{B}}$ on $T^{*} \M^{1+1}$ obtained from  \eqref{eq:Lorentz:tHam_modBateman} reads 
\begin{equation}
h^{\mathrm{B}} = \left( \frac{1}{2 m(t)} p_{1}^{2} + \frac{1}{2} m(t) \Omega^{2}(t) q_{1}^{2} \right) - \left(\frac{1}{2m(t)}p_{2}^{2} + \frac{1}{2} m(t) \Omega^{2}(t) q_{2}^{2} \right) + \frac{\gamma(t)}{2m(t)}  ( q_{1}p_{2} - q_{2}p_{1}).
\label{eq:Lorentz:tHam_modBateman_st}
\end{equation}
If  $m(t) \equiv m_{0}, k(t) \equiv k_{0}, \gamma(t)  \equiv \gamma_{0}$ and $ \Omega(t)\equiv \Omega_{0}$ are positive constants,  $h^{\mathrm{B}}$ reduces to the Hamiltonian of the so-called Bateman oscillator model \cite{Blasone1996,Bateman1931,Deguchi2019}. Recall that, in this particular case, the ($t$-independent) Hamiltonian 
\begin{equation*}
h^{\mathrm{B}} = \left( \frac{1}{2 m_{0}} p_{1}^{2} + \frac{1}{2} m_{0} \Omega_{0}^{2} q_{1}^{2} \right) - \left(\frac{1}{2m_{0}}p_{2}^{2} + \frac{1}{2} m_{0} \Omega_{0}^{2}q_{2}^{2} \right) + \frac{\gamma_{0}}{2m_{0}}  ( q_{1}p_{2} -q_{2}p_{1}) ,
\end{equation*}
is induced by the Bateman Lagrangian (see \cite{Bateman1931}), which in certain coordinates $(Q_1,Q_2)$ can be expressed as
\begin{equation*}
L^{\mathrm{B}} = m_{0} \dot{Q}_{1} \dot{Q}_{2} + \gamma_{0} \bigl(Q_{1} \dot{Q}_{2} - \dot{Q}_{1} Q_{2} \bigr) - k_{0} Q_{1} Q_{2} .
\end{equation*}
Observe that it describes a double system formed by the (uncoupled) damped harmonic oscillator 
\begin{equation*}
m_{0} \ddot{Q}_{1} + \gamma_{0} \dot{Q}_{1} + k_{0} Q_{1} = 0
\end{equation*}
and its `time-reversed' image 
\begin{equation*}
m_{0} \ddot{Q}_{2} - \gamma_{0} \dot{Q}_{2} + k_{0} Q_{2} = 0,
\end{equation*}        
which is an amplified harmonic oscillator, being $m_{0}$ the mass, $\gamma_{0}$ the damping constant, $k_{0}$ the spring constant and $\Omega_{0}$ the frequency \cite{Blasone1996,Deguchi2019}. 

In the general case, when the functions $m(t), k(t), \gamma(t)$ (and consequently also $\Omega(t)$) are not constant, we call \eqref{eq:Lorentz:tHam_modBateman_st} the \textit{$t$-dependent Bateman oscillator Hamiltonian}, where the functions $m(t), \gamma(t), k(t)$ and $\Omega(t)$ can be regarded as a $t$-dependent mass, a $t$-dependent damping constant, a $t$-dependent spring constant and a $t$-dependent frequency, defining a Hamiltonian analogue of the generalization proposed in  \cite{Campoamor2017}.   The corresponding Hamilton equations on $T^{*} \M^{1+1}$ (coming from the system \eqref{eq:Lorentz:system_mod}) are 
\begin{align*}
 &\dv{q_{1}}{t} =   \frac{1}{2 m(t)} \left( - \gamma(t) q_{2}  +   2 p_{1}   \right),   & &\dv{q_{2}}{t}  =  \frac{1}{2 m(t)} \left(\gamma(t) q_{1}  -  2 p_{2}   \right), \nonumber \\
 &\dv{p_{1}}{t}  = - \frac{1}{2m(t)} \left(  2  m^{2}(t) \Omega^{2}(t) q_{1}   + \gamma(t) p_{2} \right),   & &\dv{p_{2}}{t}  =  \frac{1}{2m(t)} \left(  2 m^{2}(t) \Omega^{2}(t) q_{2} + \gamma(t) p_{1}  \right).
\end{align*}
 In contrast to what happens at the Lagrangian level in the $t$-independent case, the $t$-dependent Hamiltonian \eqref{eq:Lorentz:tHam_modBateman_st} is not formed by two uncoupled harmonic oscillators. Indeed, if we consider the Whitney sum $T^{*}\M^{1+1} = T^{*}\R \oplus T^{*}\R$ together with the projections \eqref{eq:Lorentz:Whitneyproj}, we see that 
 \begin{equation*}
h^{\mathrm{B}} = \pr_{1}^{*} \bigl(h^{\mathrm{HO}} \bigr) -  \pr_{2}^{*} \bigl(h^{\mathrm{HO}} \bigr) + \frac{\gamma(t)}{2m(t)} ( q_{1}p_{2} - q_{2}p_{1}),
\end{equation*}
where 
\begin{equation}
h^{\mathrm{HO}} =  \frac{1}{2 m(t)} p^{2} + \frac{1}{2} m(t) \Omega^{2}(t) q^{2}
\label{eq:Lorentz:bateman:proj}
\end{equation}
is the $t$-dependent Hamiltonian on $T^{*}\R$ describing a harmonic oscillator  with a varying mass $m(t)$ and a time-dependent frequency $\Omega(t)$. Hence, the $t$-dependent Bateman oscillator Hamiltonian \eqref{eq:Lorentz:tHam_modBateman_st} is a coupled system through the $\gamma(t)$-term formed by two $t$-dependent harmonic oscillators \eqref{eq:Lorentz:bateman:proj}. 

\subsubsection{Coupled Caldirola--Kanai Hamiltonian}
Let $m(t), k(t), \gamma(t) \in C^{\infty}(\R)$ be positive $t$-dependent functions, and consider the smooth functions
\begin{equation*}
\lambda(t):= \frac{\gamma(t)}{m(t)}, \qquad \Omega(t) := \sqrt{\frac{k(t)}{m(t)}}, \qquad  t \in \R. 
\end{equation*}
Making the choices 
\begin{equation*}
a_{2}(t) = \frac{1}{m(t)} \e^{- 2\int_{0}^{t} \lambda(s) \: \dd s}, \qquad a_{4}(t) = m(t) \Omega^{2}(t) \e^{2 \int_{0}^{t} \lambda(s) \: \dd s}, 
\end{equation*}
the Hamiltonian $h^{2\mathrm{CK}}$ obtained from  \eqref{eq:Lorentz:tHam_modBateman} is expressed as 
\begin{equation}
\begin{split}
h^{2\mathrm{CK}} &= \left( \frac{1}{2m(t)} \e^{- 2\int_{0}^{t} \lambda(s) \: \dd s}p_{1}^{2} + \frac{1}{2} m(t) \Omega^{2}(t) \e^{2 \int_{0}^{t} \lambda(s) \: \dd s} q_{1}^{2} \right)  \\
&- \left( \frac{1}{2m(t)} \e^{- 2\int_{0}^{t} \lambda(s) \: \dd s}p_{2}^{2} + \frac{1}{2} m(t) \Omega^{2}(t) \e^{2 \int_{0}^{t} \lambda(s) \: \dd s} q_{2}^{2} \right) +  a_{3}(t)(q_{1}p_{2} - q_{2}p_{1}).
\end{split}
\label{eq:Lorentz:CK}
\end{equation}
As in the case of the $t$-dependent Bateman oscillator Hamiltonian \eqref{eq:Lorentz:tHam_modBateman_st}, we see that the  Hamiltonian \eqref{eq:Lorentz:CK} does also correspond to a coupled system. Indeed, the $t$-dependent Hamiltonian 
\begin{equation}
h^{\mathrm{CK}} := \frac{1}{2m(t)} \e^{- 2\int_{0}^{t} \lambda(s) \: \dd s}p^{2} + \frac{1}{2} m(t) \Omega^{2}(t) \e^{2 \int_{0}^{t} \lambda(s) \: \dd s} q^{2},
\label{eq:Lorentz:CK1D}
\end{equation}
on $T^{*}\R$ is the well-known Caldirola--Kanai Hamiltonian \cite{Caldirola1941,Kanai1948}, describing the motion of a particle of $t$-dependent mass $m(t)$ tied to a spring with a $t$-dependent Hooke `constant' $k(t) = m(t) \Omega^{2}(t)$ and subjected to a frictional force with a $t$-dependent friction constant $\gamma(t) = m(t) \lambda(t)$, where  $\Omega(t)$ is the frequency \cite{Maamache2000}. In this case, its explicit solution can be obtained using the theory of Lie systems \cite{Carinena2009}. Now, considering the Whitney sum $T^{*}\M^{1+1} = T^{*}\R \oplus T^{*}\R$ together with the  projections \eqref{eq:Lorentz:Whitneyproj} onto its factors, we see that the $t$-dependent Hamiltonian \eqref{eq:Lorentz:CK} can be written as  
\begin{equation*}
h^{2 \mathrm{CK}} = \pr_{1}^{*}\bigl(h^{\mathrm{CK}}\bigr) - \pr_{2}^{*}\bigl(h^{\mathrm{CK}}\bigr)  +  a_{3}(t) (q_{1}p_{2} - q_{2}p_{1}).
\end{equation*}
Thus, we call the $t$-dependent Hamiltonian \eqref{eq:Lorentz:CK} a \textit{coupled Caldirola--Kanai Hamiltonian}. Its associated Hamilton equations on $T^{*}\M^{1+1}$ coming form the ODEs system \eqref{eq:Lorentz:system_mod} are 
\begin{align*}
 & \dv{q_{1}}{t} =   - a_{3}(t) q_{2} +  \frac{1}{m(t)} \e^{- 2\int_{0}^{t} \lambda(s) \: \dd s}p_{1}    , &  &  \dv{q_{2}}{t}  =   a_{3}(t) q_{1}  -\frac{1}{m(t)} \e^{- 2\int_{0}^{t} \lambda(s) \: \dd s}p_{2}  , \nonumber \\
& \dv{p_{1}}{t}  =   - m(t) \Omega^{2}(t) \e^{2 \int_{0}^{t} \lambda(s) \: \dd s} q_{1}   - a_{3}(t) p_{2} ,  	&  &  \dv{p_{2}}{t}  =   m(t) \Omega^{2}(t) \e^{2 \int_{0}^{t} \lambda(s) \: \dd s} q_{2} + a_{3}(t) p_{1}  .
\end{align*}
Several relevant choices can be made for the $t$-dependent coupling `constant' $a_{3}(t)$, as the LH formalism allows us to consider it as   depending on the mass $m(t)$, the Hooke `constant' $k(t)$ and the friction constant $\gamma(t)$.

\section{The symplectic Lie algebra $\sp(4, \R)$}
\label{section:symplectic}
The real symplectic Lie algebra $\sp(4, \R)$ is defined by the matrix equation  
\begin{equation*}
\sp(4, \R) = \set{X \in \gl(4, \R): X^{T}J + JX = 0},
\label{eq:sp4:constraint}
\end{equation*}
where $J$ is the matrix given in \eqref{symp}.
Instead of using the usual basis of $\sp(4, \R)$ described in terms of elementary matrices, we use the so-called boson basis of $\sp(4, \R)$, as done for the two-photon Lie algebra $\h_{6}$ in Section~\ref{section:twophoton} (see \eqref{eq:Schrodinger:boson}).
 
 The creation $a_{i}$ and annihilation $a_{i}^{\dagger}$ operators $(1 \leq i \leq 2)$ satisfy the commutation relations 
 \begin{equation*}
 [a_{i}, a_{j}^{\dagger}] = \delta_{ij}, \qquad [a_{i}, a_{j}] = [a_{i}^{\dagger}, a_{j}^{\dagger}] = 0,
 \end{equation*}
 so the symplectic real Lie algebra $\sp(4, \R)$ is generated by the operators $a_{i}^{\dagger}a_{j}$, $a_{i}^{\dagger} a_{j}^{\dagger}$ and $a_{i} a_{j}$, with this basis being appropriate for studying realizations \cite{Quesne1988,Campoamor2005,SoaresBarbosa2004}. For convenience we label the basis as 
\begin{equation*}
X_{i,j} := a_{i}^{\dagger} a_{j}, \qquad X_{-i,j} := a_{i}^{\dagger} a_{j}^{\dagger}, \qquad X_{i,-j} := a_{i} a_{j}
\end{equation*}
satisfying
\begin{equation*}
X_{i,j} + \varepsilon_{i} \varepsilon_{j} X_{-j,-i} = 0, \qquad \varepsilon_{i} := \sgn(i), \qquad \varepsilon_{j} := \sgn(j), \qquad  \quad -2 \leq i, j \leq 2.
\end{equation*}
Over this basis, the commutation relations of $\sp(4, \R)$ can be described with the single expression 
\begin{equation*}
[X_{i,j}, X_{k, \ell}] = \delta_{j,k} X_{i, \ell} - \delta_{i, \ell} X_{k,j} + \varepsilon_{i} \varepsilon_{j} \delta_{j, -\ell} X_{k,-i} - \varepsilon_{i} \varepsilon_{j} \delta_{i,-k} X_{-j,\ell}, \qquad -2 \leq i,j,k,\ell \leq 2.
\end{equation*}
As the rank is two, the Lie algebra $\sp(4, \R)$ possesses  $\cN(\sp(4, \R)) = 2$ functionally independent polynomial invariants $C_{2}$ and $C_{4}$ of degrees $2$ and $4$, given by  
\begin{equation}
\begin{split}
C_{2} &:= X_{1,1}^2 + X_{2,2}^2-4 X_{-1,1} X_{1,-1} -4 X_{-2,2} X_{2,-2} -2 X_{-1,2} X_{1,-2}+2 X_{1,2} X_{2,1}, \\[2pt]
C_{4} &:= X_{1,-2}^2 X_{-1,2}^2-4 X_{1,-1} X_{2,-2} X_{-1,2}^2-2 X_{1,-2} X_{1,2} X_{2,1} X_{-1,2}+4 X_{1,-1} X_{1,2} X_{2,2} X_{-1,2} \\
&\qquad  -4 X_{-2,2} X_{-1,1} X_{1,-2}^2-4 X_{-2,2} X_{1,-1} X_{1,2}^2+X_{1,2}^2 X_{2,1}^2-4 X_{-1,1} X_{2,-2} X_{2,1}^2\\
&\qquad -4 X_{-1,1} X_{1,-1} X_{2,2}^2+16 X_{-2,2} X_{-1,1} X_{1,-1} X_{2,-2}+4 X_{-1,1} X_{1,-2} X_{2,1} X_{2,2}\\
&\qquad -2 X_{1,1} (-2 X_{-2,2} X_{1,-2} X_{1,2}+X_{2,1} X_{2,2} X_{1,2}-2 X_{-1,2} X_{2,-2} X_{2,1}+X_{-1,2} X_{1,-2} X_{2,2}) \\
&\qquad +X_{1,1}^2 (X_{2,2}^2-4 X_{-2,2} X_{2,-2}).
\end{split}
\label{eq:sp4:Casimirs}
\end{equation}
Consider now the fundamental representation $\Gamma_{\omega_{1}}: \sp(4, \R) \to \gl(4, \R)$ of $\sp(4, \R)$  given by  
\begin{equation*}
A_{\Gamma_{\omega_{1}}} := \begin{pmatrix}
X_{1,1} & X_{1,2} & -X_{-1,1} & -X_{-1,2} \\
X_{2,1} & X_{2,2} & -X_{-1,2} & - X_{-2,2} \\
X_{1,-1} & X_{1,-2} & -X_{1,1} & -X_{2,1} \\
X_{1,-2} & X_{2,-2} & - X_{1,2} & - X_{2,2}
\end{pmatrix}.
\end{equation*}
The realization $\Phi_{\Gamma_{\omega_{1}}}: \sp(4, \R) \to \cv(T^{*} \R^{2})$ induced by $\Gamma_{\omega_{1}}$ is spanned by the vector fields
\begin{align}
 & \X_{1} := \Phi_{\Gamma_{\omega_{1}}}(X_{1,1}) = q_{1} \pdv{q_{1}} - p_{1} \pdv{p_{1}}, & & \X_{2} := \Phi_{\Gamma_{\omega_{1}}} (X_{1,2}) = q_{1} \pdv{q_{2}} - p_{2} \pdv{p_{1}}, \nonumber  \\
  &  \X_{3} := \Phi_{\Gamma_{\omega_{1}}} (X_{2,1}) = q_{2} \pdv{q_{1}} - p_{1} \pdv{p_{2}},& & \X_{4} := \Phi_{\Gamma_{\omega_{1}}} (X_{2,2}) = q_{2} \pdv{q_{2}} - p_{2} \pdv{p_{2}}, \nonumber \\
  & \X_{5} := \Phi_{\Gamma_{\omega_{1}}} (X_{-1,1}) = -q_{1} \pdv{p_{1}}, & &  \X_{6} := \Phi_{\Gamma_{\omega_{1}}} (X_{-1,2}) = - q_{2} \pdv{p_{1}} - q_{1} \pdv{p_{2}},\label{eq:sp4:vf} \\
  & \X_{7} := \Phi_{\Gamma_{\omega_{1}}} (X_{-2,2}) = - q_{2} \pdv{p_{2}}, & & \X_{8} := \Phi_{\Gamma_{\omega_{1}}}(X_{1,-1}) = p_{1} \pdv{q_{1}}, \nonumber \\
  & \X_{9} := \Phi_{\Gamma_{\omega_{1}}} (X_{1,-2}) = p_{2} \pdv{q_{1}} + p_{1}\pdv{q_{2}} , & &  \X_{10} := \Phi_{\Gamma_{\omega_{1}}} (X_{2,-2}) = p_{2} \pdv{q_{2}}. \nonumber
\end{align}
Thus, the Lie system induced by the latter realization $\Phi_{\Gamma_{\omega_{1}}}$ of $\sp(4, \R)$ on $T^{*} \R^{2}$ has a VG Lie algebra isomorphic to $\sp(4, \R)$ and the associated $t$-dependent vector field reads
\begin{equation*}
\X := \sum_{\alpha= 1}^{10} b_{\alpha}(t) \X_{\alpha},\quad b_{\alpha} \in C^{\infty}(\R);\quad 1 \leq \alpha \leq 10.
\end{equation*}
The first-order system of differential equations on $T^{*} \R^{2}$ reads 
\begin{equation}\label{eq:sp4:system}
 \left( \begin{array}[c]{c} {\dot{q}_1}\\{\dot{q}_2}\\{\dot{p}_1}\\{\dot{p}_2}\end{array} \right)  = \left(
\begin{array}[c]{rrrr}
b_1(t) & b_3(t) & b_8(t) & b_9(t)\\
b_2(t) & b_4(t) & b_9(t) & b_{10}(t)\\
-b_5(t) & -b_6(t) & -b_1(t) & -b_2(t)\\
-b_6(t) & -b_7(t) & -b_3(t) & -b_4(t)\\
\end{array} \right)  \left( \begin{array}[c]{c} {{q}_1}\\{{q}_2}\\{{p}_1}\\{{p}_2}\end{array} \right). 
\end{equation}
Now, with respect to the canonical symplectic form \eqref{eq:symplecticform} on $T^{*} \R^{2}$, we have that $\cL_{\X_{\alpha}} \omega = 0$ for every $1 \leq\alpha \leq 10$, showing that the $\sp(4, \R)$-Lie system \eqref{eq:sp4:system} is a LH system on $T^{*} \R^{2}$. The Hamiltonian functions associated with the vector fields \eqref{eq:sp4:vf} spanning the VG Lie algebra  are
\begin{equation}
\begin{split}
&h_{1} = q_{1} p_{1}, \qquad h_{2} = q_{1} p_{2}, \qquad h_{3} = q_{2}p_{1}, \qquad h_{4} = q_{2} p_{2}, \qquad h_{5} =  \frac{1}{2} q_{1}^{2} , \\
& h_{6} = q_{1} q_{2}, \qquad h_{7} = \frac{1}{2} q_{2}^{2}, \qquad h_{8} = \frac{1}{2} p_{1}^{2}, \qquad h_{9} = p_{1}p_{2}, \qquad h_{10} = \frac{1}{2}p_{2}^{2},
\end{split}
\label{eq:sp4:ham}
\end{equation} 
and it is immediately verified that, with respect to the Poisson bracket $\set{\cdot, \cdot}_{\omega}$ on $T^{*} \R^{2}$, they span a LH algebra isomorphic to $\sp(4, \R)$. We mention that the $\sp(4, \R)$-LH system \eqref{eq:sp4:system} on $T^{*} \R^{2}$ is intrinsic in the sense that it is not locally diffeomorphic to a Lie system on $\R^{n}$ ($1 \leq n \leq 3$). This fact follows by applying Proposition~\ref{prop:Reduction:criterio}, as the (maximum) rank of its generalized distribution $\cD^{X}$ spanned by the vector fields \eqref{eq:sp4:vf} is $4$. Moreover, it is not locally diffeomorphic to an uncoupled sum of Lie systems on the real plane $\R^{2}$ or on the real line $\R$, as there are no Lie systems on $\R$ or $\R^{2}$ admitting a VG Lie algebra isomorphic to $\sp(4, \R)$. 

\subsection{Constants of the motion and superposition rules} \label{subsection:sp4_superpositionrule}

In this Section we determine a superposition rule for the LH system \eqref{eq:sp4:system} on $T^{*} \R^{2}$ by considering the $\sp(4, \R)$-LH algebra $\cH_{\omega}$ over a  basis $\set{v_{1}, \ldots, v_{10}}$ satisfying the same brackets as the Hamiltonian functions \eqref{eq:sp4:ham}. Then, from \eqref{corre} we have a morphism of Lie algebras 
\[ \phi: \cH_{\omega} \to C^{\infty}(T^{*}\R^{2}), \qquad v_{\alpha} \mapsto \phi(v_{\alpha}) = h_{\alpha}, \qquad 1 \leq \alpha \leq 10,\]
 while the morphism of Poisson algebras $D: C^{\infty}(\cH_{\omega}^{*}) \to C^{\infty}(T^{*}\R^{2})$ (see \eqref{cg})  in this case is just 
\[  D(v_{\alpha}) = h_{\alpha}(q_{1}, p_{1}, q_{2}, p_{2}), \qquad 1 \leq \alpha \leq 10. \]
Now,  the  $\sp(4, \R)$-LH algebra admits a quadratic Casimir invariant $C_2$ (see \eqref{eq:sp4:Casimirs}) given by 
\begin{equation}
C_{2} =   v_{1}^{2} + v_{4}^{2} + 2 v_{2}v_{3} - 2v_{6}v_{9} -4v_{5}v_{8} - 4 v_{7}v_{10}. 
\label{eq:sp4:CasimirLH}
\end{equation}
The smallest integer $s$ such that the diagonal prolongations of the vector fields \eqref{eq:sp4:vf} are linearly independent at a generic point of $(T^{*}\R^{2})^{s}$ is $s = 4$, so that we construct the following Hamiltonian functions $h_{\alpha}^{(k)} =D^{(k)}(\Delta^{(k)}(v_{\alpha})) \in C^{\infty}((T^{*} \R^{2})^{k})$ for $1 \leq k \leq  5 = s+1$ (see \eqref{cg}):   
\begin{align*}
&h_{1}^{(k)} = \sum_{\ell = 1}^{k} q_{1}^{(\ell)} p_{1}^{(\ell)}, &  &h_{2}^{(k)} = \sum_{\ell = 1}^{k} q^{(\ell)}_{1} p_{2}^{(\ell)}, & &h_{3}^{(k)} = \sum_{\ell = 1}^{k} q^{(\ell)}_{2} p_{1}^{(\ell)}, & &h_{4}^{(k)} = \sum_{\ell = 1}^{k}q^{(\ell)}_{2} p_{2}^{(\ell)}, \nonumber\\
& h_{5}^{(k)} =  \frac{1}{2}  \sum_{\ell = 1}^{k} \left(q^{(\ell)}_{1}\right)^{2},    & &h_{6}^{(k)} =  \sum_{\ell = 1}^{k} q^{(\ell)}_{1} q^{(\ell)}_{2},  & &h_{7}^{(k)} =  \frac{1}{2} \sum_{\ell = 1}^{k} \left(q^{(\ell)}_{2} \right)^{2},  & &h_{8}^{(k)} = \frac{1}{2} \sum_{\ell = 1}^{k} \left(p^{(\ell)}_{1} \right)^{2}, \\
& h_{9}^{(k)} = \sum_{\ell = 1}^{k} p_{1}^{(\ell)}p_{2}^{(\ell)}, &  &h_{10}^{(k)} = \frac{1}{2} \sum_{\ell = 1}^{k} \left(p^{(\ell)}_{2} \right)^{2}. & &  \nonumber
\end{align*}
Using the Casimir \eqref{eq:sp4:CasimirLH} we obtain the constants of motion $F^{(k)} :=  D^{(k)} (\Delta^{(k)} (C_{2}))$ for the diagonal prolongation $\widetilde{\X}^{5}$ of $\X$ to $(T^{*} \R^{2})^{5}$ (see \eqref{cj}). In particular, we find that $F^{(1)} =0$, while 
\begin{equation}
F^{(2)} = - \left( p_{1}^{(2)}q^{(1)}_{1} - p_{1}^{(1)} q^{(2)}_{1} + p_{2}^{(2)} q^{(1)}_{2} - p_{2}^{(1)} q^{(2)}_{2} \right)^{2}.
\label{eq:sp4:2dconstant}
\end{equation}
The second-order constant of the motion \eqref{eq:sp4:2dconstant} gives rise to the following six constants of motion (see \eqref{ck}) through the permutation $S_{ij}$ of the variables $(q_{1}^{(i)}, q_{2}^{(i)}, p_{1}^{(i)}, p_{2}^{(i)}) \leftrightarrow (q_{1}^{(j)}, q_{2}^{(j)}, p_{1}^{(j)}, p_{2}^{(j)})$ : 
\begin{equation*}
\begin{split}
&F^{(2)}_{13} = S_{13}(F^{(2)}), \qquad F^{(2)}_{14} = S_{14}(F^{(2)}), \qquad F_{15}^{(2)} =  S_{15}(F^{(2)}), \qquad  \\
&F^{(2)}_{23} = S_{23}(F^{(2)}), \qquad F^{(2)}_{24} = S_{24}(F^{(2)}), \qquad F^{(2)}_{25} = S_{25}(F^{(2)}).
\end{split}
\label{eq:sp4:2dconstant_permutations}
\end{equation*}
A short computation shows that the third-order constant of the motion $F^{(3)}$ is expressed in terms of the quadratic  invariants as 
\begin{equation*}
F^{(3)} = F^{(2)} + F_{13}^{(2)} + F_{23}^{(2)}.
\end{equation*}
 After a long but routine computation, a superposition rule for the $\sp(4, \R)$-LH system \eqref{eq:sp4:system} on $T^{*}\R^{2}$ can be derived by taking into account the four significative constants $F^{(2)} = - k_{1}^{2}, F^{(2)}_{23} = - k_{2}^{2}, F_{24}^{(2)} = - k_{3}^{2}$ and $F^{(2)}_{25} = - k_{4}^{2}$ and four particular solutions $(q_{1}^{(\ell)}, q_{2}^{(\ell)}, p_{1}^{(\ell)}, p_{2}^{(\ell)})$, with $1 \leq \ell \leq 4$. In order to simplify the resulting expression, let us define the generalized angular momenta 
 \begin{equation}\label{fakt}
L_{\alpha,\beta}^{(i,j)}=p^{(i)}_\alpha q^{(j)}_\beta-q^{(i)}_\alpha p^{(j)}_\beta,\quad 1 \leq \alpha,\beta \leq 2;\quad 1\leq i,j\leq 4. 
\end{equation}
 Then the general solution $(q_{1}(t), q_{2}(t), p_{1}(t), p_{2}(t))$ can be described as follows:
\begin{equation}
\begin{split}
q_{1}(t) =  \frac{1}{\Xi}& \left[k_1\left(q_1^{(2)}L_{2,2}^{(4,3)}+q_1^{(3)}L_{2,2}^{(2,4)}+q_1^{(4)}L_{2,2}^{(3,2)}\right)+ k_2\left(q_1^{(1)}L_{2,2}^{(3,4)}+q_1^{(3)}L_{2,2}^{(4,1)}+q_1^{(4)}L_{2,2}^{(1,3)}\right) \right.\\
& \left.\   +k_3\left(q_1^{(1)}L_{2,2}^{(2,4)}+q_1^{(2)}L_{2,2}^{(4,1)}+q_1^{(4)}L_{2,2}^{(1,2)}\right)+k_4\left(q_1^{(1)}L_{2,2}^{(3,2)}+q_1^{(2)}L_{2,2}^{(1,3)}+q_1^{(3)}L_{2,2}^{(2,1)}\right) \right],\\
q_{2}(t) =  \frac{1}{\Xi}& \left[k_1\left(q_2^{(2)}L_{1,1}^{(3,4)}+q_2^{(3)}L_{1,1}^{(4,2)}+q_2^{(4)}L_{1,1}^{(2,3)}\right)+ k_2\left(q_2^{(1)}L_{1,1}^{(4,3)}+q_2^{(3)}L_{1,1}^{(1,4)}+q_2^{(4)}L_{1,1}^{(3,1)}\right) \right.\\
& \left. \   +k_3\left(q_2^{(1)}L_{1,1}^{(4,2)}+q_2^{(2)}L_{1,1}^{(1,4)}+q_2^{(4)}L_{1,1}^{(2,1)}\right)+k_4\left(q_2^{(1)}L_{1,1}^{(2,3)}+q_2^{(2)}L_{1,1}^{(3,1)}+q_2^{(3)}L_{1,1}^{(1,2)}\right) \right],\\
p_{1}(t) =  \frac{1}{\Xi}& \left[k_1\left(p_1^{(2)}L_{2,2}^{(4,3)}+p_1^{(3)}L_{2,2}^{(2,4)}+p_1^{(4)}L_{2,2}^{(3,2)}\right)+ k_2\left(p_1^{(1)}L_{2,2}^{(3,4)}+p_1^{(3)}L_{2,2}^{(4,1)}+p_1^{(4)}L_{2,2}^{(1,3)}\right) \right.\\
& \left.  \   +k_3\left(p_1^{(1)}L_{2,2}^{(2,4)}+p_1^{(2)}L_{2,2}^{(4,1)}+p_1^{(4)}L_{2,2}^{(1,2)}\right)+k_4\left(p_1^{(1)}L_{2,2}^{(3,2)}+p_1^{(2)}L_{2,2}^{(1,3)}+p_1^{(3)}L_{2,2}^{(2,1)}\right) \right],\\
p_{2}(t) =  \frac{1}{\Xi}& \left[k_1\left(p_2^{(2)}L_{1,1}^{(3,4)}+p_2^{(3)}L_{1,1}^{(4,2)}+p_2^{(4)}L_{1,1}^{(2,3)}\right)+ k_2\left(p_2^{(1)}L_{1,1}^{(4,3)}+p_2^{(3)}L_{1,1}^{(1,4)}+p_2^{(4)}L_{1,1}^{(3,1)}\right) \right.\\
& \left. \   +k_3\left(p_2^{(1)}L_{1,1}^{(4,2)}+p_2^{(2)}L_{1,1}^{(1,4)}+p_2^{(4)}L_{1,1}^{(2,1)}\right)+k_4\left(p_2^{(1)}L_{1,1}^{(2,3)}+p_2^{(2)}L_{1,1}^{(3,1)}+p_2^{(3)}L_{1,1}^{(1,2)}\right) \right],
\end{split}
\label{eq:sp4:superpositionrule}
\end{equation} 
where
\begin{equation*}
\begin{split}
\Xi &= L_{1,1}^{(1,3)}L_{2,2}^{(4,2)}+L_{1,1}^{(4,2)}L_{2,2}^{(1,3)}+L_{1,1}^{(2,1)}L_{2,2}^{(4,3)}+L_{1,1}^{(4,3)}L_{2,2}^{(2,1)}+L_{1,1}^{(3,2)}L_{2,2}^{(4,1)}+L_{1,1}^{(4,1)}L_{2,2}^{(3,2)}.
\end{split}
\end{equation*}
The superposition rule exhibits some interesting symmetry properties. In particular, the expressions for   $q_\alpha(t)$ and   $p_\alpha(t)$ are symmetric with respect to the substitution $q^{(i)}_{\alpha} L_{\beta,\beta}^{(j,k)}\mapsto   q^{(i)}_{\beta}L_{\alpha,\alpha}^{(k,j)}$ and $p^{(i)}_{\alpha}L_{\beta,\beta}^{(j,k)}\mapsto   p^{(i)}_{\beta}L_{\alpha,\alpha}^{(k,j)}$, respectively, while $\Xi$ is symmetric with respect to the transformation $L_{\alpha,\alpha}^{(k,j)}\mapsto L_{\beta,\beta}^{(k,j)}$ (see \eqref{fakt}). Moreover, we observe that for each term $q^{(i)}_{\alpha}L_{\beta,\beta}^{(j,k)}$ appearing in the expression of $q_{\alpha}(t)$, a corresponding term $p^{(i)}_{\alpha}L_{\beta,\beta}^{(j,k)}$ appears in the expression of $p_{\alpha}(t)$, associated to the same significant constant. 

\medskip
In Subsection~\ref{subsection:constants_Lorentz} we pointed out that the derivation of a superposition rule for the $\so(1,3)$-LH system \eqref{eq:Lorentz:system} on $T^{*}\R^{2}$ using the constants of the motion obtained through the Casimir invariants   \eqref{eq:Lorentz:CasimirLH} is computationally cumbersome, and that an improved rule can be derived considering an appropriate embedding. A routine computation shows  that the vector fields $\left\{{\bf X}_1,\cdots ,{\bf X}_6\right\}$ in \eqref{eq:Lorentz:vf} are obtained as the following linear combination of the generators \eqref{eq:sp4:vf} here denoted $\bigl\{\tilde{{\bf X}}_1,\cdots ,\tilde{{\bf X}}_{10}\bigr\}$: 
\begin{equation*}
\begin{split}
&{\bf X}_1=\frac 12 \bigl( \tilde{{\bf X}}_6+\tilde{{\bf X}}_9 \bigr),\qquad {\bf X}_2=\frac 12 \bigl(\tilde{{\bf X}}_5-\tilde{{\bf X}}_7+\tilde{{\bf X}}_8-\tilde{{\bf X}}_{10}  \bigr),\qquad  {\bf X}_3=\frac 12 \bigl(\tilde{{\bf X}}_2-\tilde{{\bf X}}_3  \bigr),\\[2pt]
&{\bf X}_4=\frac 12 \bigl(\tilde{{\bf X}}_5-\tilde{{\bf X}}_7-\tilde{{\bf X}}_8+\tilde{{\bf X}}_{10}  \bigr),\qquad {\bf X}_5=\frac 12 \bigl(\tilde{{\bf X}}_1+\tilde{{\bf X}}_4  \bigr),\qquad {\bf X}_6=\frac 12 \bigl(\tilde{{\bf X}}_9-\tilde{{\bf X}}_6 \bigr).
\end{split}
\end{equation*}
Therefore, the $\mathfrak{so}(1,3)$-LH determined by the vector field ${\bf X}=\sum_{\alpha=1}^6 b_{\alpha}(t){\bf X}_{\alpha}$ corresponds to a particular case of the LH system \eqref{eq:sp4:system} determined by $\tilde{{\bf X}}=\sum_{\alpha=1}^{10} \tilde{ b}_{\alpha}(t)\tilde{{\bf X}}_{\alpha}$, for the values of $\tilde{b}_{\alpha}$ given by 
\begin{equation*}
\begin{split}
\tilde{b}_1(t) &=\frac 12 b_5(t),\qquad \tilde{b}_2(t)=\frac 12 b_3(t),\qquad \tilde{b}_3(t)=-\frac 12 b_3(t),\qquad \tilde{b}_4(t)=\frac 12 b_5(t) ,\\[2pt]
\tilde{b}_5(t)&=\frac 12 \bigl( b_2(t)+b_4(t) \bigr) , \qquad \tilde{b}_6(t) =\frac 12 \bigl( b_1(t)-b_6(t)  \bigr),\qquad \tilde{b}_7(t)=-\frac 12 \bigl( b_2(t)+b_4(t)  \bigr), \\[2pt]
 \tilde{b}_8(t)&=\frac 12 \bigl( b_2(t)-b_4(t)  \bigr),\qquad \tilde{b}_9(t) =\frac 12 \bigl( b_1(t)+b_6(t) \bigr),\qquad \tilde{b}_{10} =\frac 12 \bigl( b_4(t)-b_2(t) \bigr).
\end{split}
\end{equation*}
As a consequence, the constants of the motion of the diagonal prolongations of the $\sp(4, \R)$-LH system obtained applying the coalgebra formalism are also constants of the motion for the system  \eqref{eq:Lorentz:system}, and thus the superposition rule \eqref{eq:sp4:superpositionrule}, applied to particular solutions of \eqref{eq:Lorentz:system}, provides an admissible superposition principle.  

\medskip
A similar argument applies to the second-order constant of the motion \eqref{eq:Schrodinger:G2} used for the computation of the superposition rule \eqref{eq:Schrodinger:superposition} for the $\widehat{\cS}(1)$-LH system \eqref{eq:Schrodinger:system}. It is straightforward to verify that, among the Hamiltonian functions \eqref{eq:sp4:ham}, the subset formed by $\left\{h_1,h_3,h_5,h_6,h_7,h_{8}\right\}$ generates $\widehat{\cS}(1)$ (see (\ref{eq:Schrodinger:ham})). In this context, the function $G_2$ in \eqref{eq:Schrodinger:G2} arises as a common invariant of the prolonged vector fields, i.e., as a differential invariant of the prolongation. This justifies why it cannot be recovered by the coalgebra formalism, as it has no counterpart within the $\widehat{\cS}(1)$-invariants.  

\medskip
As a general observation, the procedure can be applied to any subalgebra of $\mathfrak{sp}(4, \R)$, once the generating vector fields have been expressed as linear combinations of the generators \eqref{eq:sp4:vf}. The corresponding Lie system, which identifies with a particular case of \eqref{eq:sp4:system}, can hence be solved using the superposition principle \eqref{eq:sp4:superpositionrule} for appropriate particular solutions of the subalgebra system. We can thus state the following general fact on  LH systems on $T^{*}\mathbb{R}^{2}$ associated to the various subalgebras of $\mathfrak{sp}(4, \R)$: 

\begin{thm}
For each subalgebra $\mathfrak{g} \subset \mathfrak{sp}(4, \R)$, there exists a LH system on $T^{*}\mathbb{R}^{2}$ with respect to the canonical symplectic form \eqref{eq:symplecticform} possessing a VG Lie algebra isomorphic to $\mathfrak{g}$. 
\end{thm}

\subsection{Applications to coupled systems}

As observed, the Hamiltonian $h$ obtained from the Hamiltonian functions \eqref{eq:sp4:ham} is simply 
\begin{equation*}
h = \sum_{\alpha = 1}^{10} b_{\alpha}(t) h_{\alpha},
\end{equation*}
with the system \eqref{eq:sp4:system} corresponding to the Hamilton equations with respect to the canonical symplectic form \eqref{eq:symplecticform} on $T^{*}\R^{2}$. In the most general case, when all the $b_{\alpha}(t)$ are nonzero, the minimal LH algebra for this system is $\cH_{\omega} \simeq \sp(4, \R)$. Some of the Hamiltonian functions allow a natural mechanical interpretation. Actually, $h_{5} = \frac{1}{2} q_{1}^{2}$, $h_{6} = q_{1}q_{2}$ and $h_{7} = \frac{1}{2} q_{2}^{2}$  can be regarded as quadratic potentials, while $h_{8} = \frac{1}{2} p_{1}^{2}$, $h_{9} = p_{1}p_{2}$ and $h_{10} = \frac{1}{2} p_{2}^{2}$ can be interpreted as kinetic energy terms. The `mixed' terms $h_{1} = q_{1}p_{1}$, $h_{2} = q_{1}p_{2}$, $h_{3} = q_{2}p_{1}$ and $h_{4} = q_{2}p_{2}$ are required to guarantee that the system \eqref{eq:sp4:system} is not equivalent to an uncoupled sum, thus leading to intrinsic systems in $T^{*}\R^{2}$. On the other hand, these functions are also  the coupling terms of two 1D $t$-dependent Hamiltonians (see Subsection~\ref{subsubsection:sp4:coupling} below). 

 In order to find relevant applications, we first look for a minimal set of generators of the LH algebra $\cH_{\omega} \simeq \sp(4, \R)$. It can be easily verified that the Hamiltonian functions  
\begin{equation*}
h_{2} = q_{1}p_{2}, \qquad h_{3} = q_{2}p_{1}, \qquad h_{5} = \frac{1}{2} q_{1}^{2}, \qquad h_{7} = \frac{1}{2} q_{2}^{2}, \qquad h_{8} = \frac{1}{2} p_{1}^{2}, \qquad h_{10} = \frac{1}{2} p_{2}^{2}
\end{equation*}
can be chosen as a set of generators of the LH algebra $\cH_{\omega}$. The Hamiltonian $h$ corresponding to this choice is given by  
\begin{equation}
h = b_{2}(t) q_{1}p_{2} + b_{3}(t)q_{2}p_{1} + \frac{b_{5}(t)}{2} q_{1}^{2}+ \frac{b_{7}(t)}{2} q_{2}^{2} + \frac{b_{8}(t)}{2} p_{1}^{2}+ \frac{b_{10}(t)}{2} p_{2}^{2},
\label{eq:sp4:tHam_mod}
\end{equation}
while the associated Hamilton equations (see \eqref{eq:sp4:system}), assuming that $b_{1}(t), b_{4}(t), b_{6}(t), b_{9}(t) \equiv 0$, are given by
\begin{equation}
\begin{split}
\dv{q_{1}}{t} &= b_{3}(t) q_{2} + b_{8}(t) p_{1}, \\
\dv{p_{1}}{t} & = -b_{2}(t) p_{2} - b_{5}(t) q_{1},
\end{split}
\qquad 
\begin{split}
\dv{q_{2}}{t} &=b_{2}(t) q_{1} + b_{10}(t) p_{2}, \\
\dv{p_{2}}{t} & =-b_{3}(t) p_{1} - b_{7}(t) q_{2}.
\end{split}
\label{eq:sp4:system_mod}
\end{equation}
We observe that, as first-order ODEs system, it has the same generic form as \eqref{eq:Lorentz:system_coupled},\footnote{The system \eqref{eq:Lorentz:system_coupled} is actually a special case of \eqref{eq:sp4:system_mod}.} and thus its explicit integration for arbitrary functions is subjected to the same technical difficulties. For $b_2(t)=b_3(t)=0$, the system decouples, as the Hamiltonian \eqref{eq:sp4:tHam_mod} splits into the sum of two 1D Hamiltonians. Therefore, in order to have an intrinsic LH system, either $b_2(t)$ or $b_3(t)$ cannot vanish. A special case is given for $b_5(t)=b_7(t)=0$ and $b_2(t)b_3(t)\neq 0$, corresponding to the vanishing of the quadratic potential. In this situation, the solution of the system can be expressed as  
\begin{equation}
\begin{split}
q_{1}(t) & = -Z_3(t)-\frac{1}{b_2(t)b_3(t)}\left(b_2(t)b_3(t)b_{8}(t)-b_2(t)\dv{b_{3}}{t}b_8(t)+b_2(t)b_3(t)\dv{b_{8}}{t}+b_3(t)^2b_{10}(t)\right)Z_2(t),   \\
p_{1}(t) & = -\frac{ \dv{Z_{2}}{t}}{b_2(t)},\qquad q_2(t)=\frac{-a_8(t)Z_2(t)+ \dv{Z_{2}}{t}}{b_3(t)},\qquad  p_{2}(t) =-\frac{ \dv{Z_{2}}{t}}{b_3(t)},
\end{split}
\label{eq:sp4:system_mod2}
\end{equation}
where $Z_\alpha(t)$ is the general solution of the second-order ODE 
\begin{equation}\label{sec}
\frac{{\rm d}^2 Z_\alpha}{{\rm d}t^2}-\frac{1}{b_\alpha(t)}\dv{b_\alpha}{t}\dv{Z_\alpha}{t}-b_2(t)b_3(t)Z_\alpha=0,\quad 2\leq \alpha\leq 3.
\end{equation}
The latter equation can further be reduced by a transformation (compare with \eqref{red}), essentially reducing the analysis of \eqref{eq:sp4:system_mod} to a pair or Riccati equations. In particular, if $b_2(t)=b_3(t)$, the equation (\ref{sec}) can be solved directly, providing 
\begin{equation}\label{sol1}
Z_\alpha(t)= \lambda_{1}\sinh\left(\int b_3(t) {\rm d}t\right)+\lambda_{2}\cosh\left(\int b_3(t) {\rm d}t\right),\quad 2\leq \alpha\leq 3,
\end{equation}
where $\lambda_{1}, \lambda_{2} \in \R$ are constants. 
In the following, we make particular choices of the coefficient functions in \eqref{eq:sp4:tHam_mod}, obtaining natural generalizations of the systems presented in Subsection~\ref{subsection:Lorentz:applications}, as well as some new types.
 
\subsubsection{A $t$-dependent electromagnetic field} Let us consider $t$-dependent functions $m_{1}(t), m_{2}(t), e_{1}(t), e_{2}(t), \gamma(t) \in C^{\infty}(\R)$ such that 
\begin{equation}
m_{1}(t),\; m_{2}(t) > 0, \qquad \ddot{\gamma}(t) \neq 0, \qquad t \in \R. 
\label{eq:sp4:electro_choices_const}
\end{equation}
Define the following $t$-dependent vector potential $\mathbf{A}$  on $\R^{3}$:
\begin{equation*}
A_{1} := - \frac{1}{2} q_{2} \gamma(t), \qquad A_{2} := \frac{1}{2} q_{1} \gamma(t), \qquad A_{3} := 0,
\end{equation*}
as well as the scalar potential 
\begin{equation*}
\phi:= \phi_{1} + \phi_{2}, \qquad 
\phi_{1} := \frac{1}{2} q_{1}^{2}, \qquad \phi_{2}:= \frac{1}{2} q_{2}^{2}.
\end{equation*}
With the choice 
\begin{equation*}\label{eq:sp4:electro_choices}
\begin{array}{lll}
\displaystyle b_{2}(t) = - \frac{\gamma(t) e_{2}(t)}{2m_{2}(t)},  & \displaystyle b_{3}(t)  = \frac{\gamma(t) e_{1}(t)}{2m_{1}(t)},  & \displaystyle b_{5}(t) = e_{1}(t) + \frac{\gamma^{2}(t) e_{2}^{2}(t)}{4m_{2}(t)}  ,  \\[10pt]
\displaystyle b_{7}(t) =  e_{2}(t) + \frac{\gamma^{2}(t) e_{1}^{2}(t)}{4 m_{1}(t)},
  & \displaystyle b_{8}(t) = \frac{1}{m_{1}(t)},  &\displaystyle  b_{10}(t) = \frac{1}{m_{2}(t)},  
\end{array}
\end{equation*}
the $t$-dependent Hamiltonian $h^{\mathrm{E}}$ obtained from \eqref{eq:sp4:tHam_mod} reads 
\begin{equation*}
h^{\mathrm{E}} = \frac{1}{2m_{1}(t)} \left(p_{1} - e_{1}(t) A_{1}\right)^{2} + e_{1}(t) \phi_{1} + \frac{1}{2m_{2}(t)} \left(p_{2} - e_{2}(t)A_{2}\right)^{2} + e_{2}(t) \phi_{2}.
\end{equation*}
This Hamiltonian describes, in a natural way, the motion of two particles with time-dependent masses $m_{1}(t), m_{2}(t)$ and time-dependent electric charges $e_{1}(t), e_{2}(t)$, respectively, on a hyperplane in $\R^{3}$, under the action of a $t$-dependent electromagnetic field $\mathbf{B}$. Specifically, the latter is given by 
\begin{equation*}
\mathbf{B} = \nabla \times \mathbf{A} = (0, 0, \gamma(t)), 
\end{equation*}
while the electric field is 
\begin{equation*}
\mathbf{E} = - \nabla \phi - \pdv{\mathbf{A}}{t} =-\frac{1}{2} \left(  2 q_{1} - q_{2} \dot{\gamma}(t),  2q_{2} + q_{1} \dot{\gamma}(t), 0 \right). 
\end{equation*}
Note that the condition $\ddot{\gamma}(t) \neq 0$ in \eqref{eq:sp4:electro_choices_const} implies that the magnetic field $\mathbf{B}$ and the electric fields $\mathbf{E}$ are not constant. The resulting equations of the motion on $T^{*}\R^{2}$, for within this choice of the parameter functions  are 
\begin{equation*}
\begin{split}
\dv{q_{1}}{t} &=\frac{1}{m_{1}(t)} \left(  \frac{\gamma(t) e_{1}(t)}{2} q_{2} +  p_{1} \right), \\
\dv{q_{2}}{t} &= \frac{1}{m_{2}(t)} \left( - \frac{\gamma(t) e_{2}(t)}{2} q_{1} +  p_{2} \right), 
\end{split}
\qquad 
\begin{split}
\dv{p_{1}}{t} & = \frac{\gamma(t) e_{2}(t)}{2m_{2}(t)} p_{2} -\left( e_{1}(t) + \frac{\gamma^{2}(t) e_{2}^{2}(t)}{4m_{2}(t)} \right) q_{1}, \\
\dv{p_{2}}{t} & =-\frac{\gamma(t) e_{1}(t)}{2m_{1}(t)}p_{1} - \left( e_{2}(t) + \frac{\gamma^{2}(t) e_{1}^{2}(t)}{4m_{1}(t)} \right) q_{2}.
\end{split}
\end{equation*}
In particular, if 
\begin{equation*}
e_{1}(t) + \frac{\gamma^{2}(t) e_{2}^{2}(t)}{4m_{2}(t)}=0, \qquad e_{2}(t) + \frac{\gamma^{2}(t) e_{1}^{2}(t)}{4m_{1}(t)}=0
\end{equation*}
holds, i.e., when 
\begin{equation*}
e_{1}(t) =  -\frac{4m_1(t)^{\frac{2}{3}}m_2(t)^{\frac{1}{3}}}{\gamma^{2}(t)},\quad e_{2}(t) = - \frac{4m_1(t)^{\frac{1}{3}}m_2(t)^{\frac{2}{3}}}{\gamma^{2}(t)},
\end{equation*} 
the solution can be written in the form \eqref{eq:sp4:system_mod2}--\eqref{sec}.

\subsubsection{Generalized coupled oscillators}
\label{subsubsection:sp4:coupling}
In Subsection~\ref{subsection:Lorentz:applications}, we obtained the $t$-dependent Bateman oscillator Hamiltonian \eqref{eq:Lorentz:tHam_modBateman_st} and the coupled Caldirola--Kanai Hamiltonian \eqref{eq:Lorentz:CK} on $T^{*}\M^{1+1}$ as a coupling of two 1D oscillators with the same $t$-dependent mass and frequency, through a $t$-dependent angular momentum term. We show next that these systems can be generalized to a coupling of two 1D oscillators of the same type, again through a $t$-dependent angular momentum, but with different $t$-dependent masses and frequencies. 
First, we consider coefficient functions $a_{2}, a_{5}, a_{7}, a_{8}, a_{10} \in C^{\infty}(\R)$ given by 
\begin{equation*}
a_{2}:= b_{2} = -b_{3}, \qquad a_{5}:= b_{5}, \qquad  a_{7}:= b_{7}, \qquad a_{8}:= b_{8}, \qquad a_{10} := b_{10}, 
\label{eq:sp4:choices}
\end{equation*} 
so that the modified Hamiltonian $h'$ obtained from \eqref{eq:sp4:tHam_mod} reads 
\begin{equation}
h' =  \left( \frac{a_{8}(t)}{2} p_{1}^{2} + \frac{a_{5}(t)}{2} q_{1}^{2} \right) + \left( \frac{a_{10}(t)}{2} p_{2}^{2} + \frac{a_{7}(t)}{2} q_{2}^{2} \right) + a_{2}(t) (q_{1}p_{2} - q_{2}p_{1}). 
\label{eq:sp4:tham'}
\end{equation}
With the latter choice, it is convenient to consider the following change of basis on the LH algebra $\cH_{\omega}$ spanned by the Hamiltonian functions \eqref{eq:sp4:ham}: 
\begin{equation*}
h_{2}' := h_{2} - h_{3} = q_{1}p_{2} - q_{2}p_{1}, \qquad h_{\alpha}':= h_{\alpha}, \qquad \alpha \neq 2.  
\label{eq:sp4:ham_cr'}
\end{equation*}
It is immediate to verify that the Hamiltonian functions 
\begin{equation*}
h_{2}' = q_{1}p_{2} - q_{2}p_{1}, \qquad h_{5}' = \frac{1}{2} q_{1}^{2}, \qquad h_{7}' = \frac{1}{2} q_{2}^{2}, \qquad h_{8}' = \frac{1}{2} p_{1}^{2}, \qquad h_{10}' = \frac{1}{2}p_{2}^{2},
\label{eq:sp4:ham_basis}
\end{equation*}
span the LH algebra $\cH_{\omega} \simeq \sp(4, \R)$, showing that it is the minimal LH algebra for the $t$-dependent Hamiltonian \eqref{eq:sp4:tham'}. 
In particular, $h_{3}' = q_{1}p_{2} - q_{2}p_{1}$ can be identified with the angular momentum. The Hamilton equations of $h^{\prime}$ are 
 \begin{equation}
\begin{split}
\dv{q_{1}}{t} &= -a_{2}(t) q_{2} +a_{8}(t)p_{1}, \\
\dv{p_{1}}{t} & = -a_{2}(t)p_{2} - a_{5}(t)q_{1},
\end{split}
\qquad 
\begin{split}
\dv{q_{2}}{t} &=a_{2}(t) q_{1} +a_{10}(t) p_{2}, \\
\dv{p_{2}}{t} & =a_{2}(t) p_{1} - a_{7}(t)q_{2}.
\end{split}
\label{eq:sp4:system_mod'}
\end{equation}
We observe that, for the particular cases $a_8(t)=a_{10}(t)=0$ and $a_5(t)=a_7(t)=0$, the system can be explicitly solved by quadratures. However, due to the length of the resulting expressions, we skip its detailed description. If $a_8(t)=a_{10}(t)=a_5(t)=a_7(t)$, the system can be solved using matrix exponentials \cite{Hur}, while for $a_8(t)=a_{10}(t)=a_5(t)=a_7(t)=0$, the solution reduces to the form given by \eqref{sol1}. 

\smallskip
A remarkable fact is that, within the Whitney sum $T^{*}\R^{2} = T^{*}\R \oplus T^{*}\R$, together with the projections 
\begin{equation*}
\begin{split}
&\pr_{1}: T^{*} \R^{2} \to T^{*}\R, \qquad (q_{1}, q_{2}, p_{1}, p_{2}) \mapsto (q_{1}, p_{1}), \\
&\pr_{2}: T^{*}\R^{2} \to T^{*}\R, \qquad  (q_{1}, q_{2}, p_{1}, p_{2}) \mapsto (q_{2}, p_{2}),
\end{split}
\label{eq:sp4:Whitneyproj}
\end{equation*} 
the $t$-dependent Hamiltonian \eqref{eq:sp4:tham'} can be expressed as 
\begin{equation*}
h' = \pr_{1}^{*}(h'_{\mathrm{1D_{1}}}) + \pr_{2}^{*}(h'_{\mathrm{1D_{2}}}) + a_{2}(t) (q_{1}p_{2} - q_{2}p_{1})
\end{equation*}
where $h'_{\mathrm{1D_{1}}}$ and $h'_{\mathrm{1D_{2}}}$ are the 1D $t$-dependent Hamiltonians given by 
\begin{equation*}
h'_{\mathrm{1D_{1}}} :=  \frac{a_{8}(t)}{2} p_1^{2} + \frac{a_{5}(t)}{2} q_1^{2}, \qquad h'_{\mathrm{1D_{2}}} :=  \frac{a_{10}(t)}{2} p_2^{2} + \frac{a_{7}(t)}{2} q_2^{2},
\end{equation*}
respectively. This shows, as pointed out before, that the $t$-dependent Hamiltonian \eqref{eq:sp4:tham'} is a coupling of two 1D $t$-dependent Hamiltonians through a $t$-dependent angular momentum term. 

\medskip
We finally consider some particular choices for the coefficient functions appearing in \eqref{eq:sp4:tham'}, which generalize the coupled systems studied in Subsection~\ref{subsection:Lorentz:applications}. Consider positive functions $m_{i}(t)$, $k_{i}(t)$, $\omega_{i}(t) \in C^{\infty}(\R)$  for $1 \leq i \leq 2$, as follows: 
\begin{itemize}
\item[•] \textit{A $t$-dependent coupled harmonic oscillators Hamiltonian}. Suppose that 
\begin{equation*}
k_{i}(t) > \frac{\gamma_{i}^{2}(t)}{4 m_{i}(t)}, \qquad t \in \R, \qquad 1 \leq i \leq 2,
\end{equation*}
and define 
\begin{equation*}
\Omega_{i}(t) := \sqrt{ \frac{1}{m_{i}(t)} \left( k_{i}(t) - \frac{\gamma_{i}^{2}(t)}{4m_i(t)} \right) }, \qquad t \in \R, \qquad 1 \leq i \leq 2.
\end{equation*}
The $t$-dependent Hamiltonian $h^{\mathrm{CHO}}$ obtained from \eqref{eq:sp4:tham'} for the choice 
\begin{equation*}
a_{8}(t) = \frac{1}{m_{1}(t)}, \qquad a_{5}(t) = m_{1}(t) \Omega_{1}^{2}(t), \qquad a_{10}(t) = \frac{1}{m_{2}(t)}, \qquad a_{7} = m_{2}(t) \Omega_{2}^{2}(t),
\end{equation*}
is given by 
\begin{equation}
h^{\mathrm{CHO}} =  \left( \frac{1}{2m_{1}(t)} p_{1}^{2} + \frac{1}{2}m_{1}(t) \Omega_{1}^{2}(t) q_1^2\right) +  \left( \frac{1}{2m_{2}(t)} p_{2}^{2} + \frac{1}{2}m_{2}(t) \Omega_{2}^{2}(t)q_2^2 \right) + a_{2}(t) (q_{1}p_{2} - q_{2}p_{1}), 
\label{eq:sp4:cho}
\end{equation}
and corresponds to a coupling of two 1D harmonic oscillators with $t$-dependent masses $m_{i}(t)$ and frequencies $\Omega_{i}(t)$ $(1 \leq i \leq 2)$ of the type \eqref{eq:Lorentz:bateman:proj}. This generalizes the coupling of the $t$-dependent Bateman oscillator Hamiltonian \eqref{eq:Lorentz:tHam_modBateman_st}. The Hamilton equations on $T^{*}\R^{2}$ coming from \eqref{eq:sp4:system_mod'} read
 \begin{equation*}
\begin{split}
\dv{q_{1}}{t} &= -a_{2}(t) q_{2} + \frac{1}{m_{1}(t)} p_{1}, \\
\dv{p_{1}}{t} & = -a_{2}(t)p_{2} - m_{1}(t) \Omega_{1}^{2}(t) q_{1},
\end{split}
\qquad 
\begin{split}
\dv{q_{2}}{t} &=a_{2}(t) q_{1} + \frac{1}{m_{2}(t)} p_{2}, \\
\dv{p_{2}}{t} & =a_{2}(t) p_{1} -m_{2}(t) \Omega_{2}^{2}(t)q_{2}.
\end{split}
\end{equation*}
We call \eqref{eq:sp4:cho} the \textit{$t$-dependent coupled harmonic oscillator Hamiltonian}. 

\item[•] \textit{A generalized coupled Caldirola--Kanai Hamiltonian}. If we now define 
\begin{equation*}
\lambda_{i} := \frac{\gamma_{i}(t)}{m_{i}(t)}, \qquad \Omega_{i}(t) := \sqrt{\frac{k_{i}(t)}{m_{i}(t)}}, \qquad t \in \R, \qquad 1 \leq i \leq 2,
\end{equation*}
the  $t$-dependent Hamiltonian $h^{\mathrm{CCK}}$ obtained from \eqref{eq:sp4:tham'} for the choice 
\begin{equation*}
\begin{split}
&a_{8}(t) = \frac{1}{m_{1}(t)} \e^{- 2\int_{0}^{t} \lambda_{1}(s) \: \dd s}, \qquad a_{5}(t) = m_{1}(t) \Omega_{1}^{2}(t) \e^{2 \int_{0}^{t} \lambda_{1}(s) \: \dd s}, \\
&a_{10}(t) =  \frac{1}{m_{2}(t)} \e^{- 2\int_{0}^{t} \lambda_{2}(s) \: \dd s}, \qquad a_{7}(t) = m_{2}(t) \Omega_{2}^{2}(t) \e^{2 \int_{0}^{t} \lambda_{2}(s) \: \dd s}, 
\end{split}
\end{equation*}
reads  
\begin{equation}
\begin{split}
h^{\mathrm{CCK}} &= \left( \frac{1}{2m_{1}(t)} \e^{- 2\int_{0}^{t} \lambda_{1}(s) \: \dd s}p_{1}^{2} + \frac{1}{2} m_{1}(t) \Omega_{1}^{2}(t) \e^{2 \int_{0}^{t} \lambda_{1}(s) \: \dd s} q_{1}^{2} \right)  \\
&\quad + \left( \frac{1}{2m_{2}(t)} \e^{- 2\int_{0}^{t} \lambda_{2}(s) \: \dd s}p_{2}^{2} + \frac{1}{2} m_{2}(t) \Omega_{2}^{2}(t) \e^{2 \int_{0}^{t} \lambda_{2}(s) \: \dd s} q_{2}^{2} \right) \\
& \quad + a_{2}(t) (q_{1}p_{2} - q_{2}p_{1}).
\label{eq:sp4:CCK}
\end{split}
\end{equation}
We call \eqref{eq:sp4:CCK} the \textit{generalized coupled Caldirola--Kanai Hamiltonian}, as it corresponds to the coupling of two 1D Caldirola--Kanai Hamiltonians with $t$-dependent masses $m_{i}(t)$ and frequencies $\Omega_{i}(t)$ of the type \eqref{eq:Lorentz:CK1D}. This  generalizes  the coupled Caldirola--Kanai Hamiltonian \eqref{eq:Lorentz:CK}, with the associated Hamilton equations  given by 
\begin{equation*}
\begin{split}
\dv{q_{1}}{t} &= -a_{2}(t) q_{2} + \frac{1}{m_{1}(t)} \e^{- 2\int_{0}^{t} \lambda_{1}(s) \: \dd s}p_{1}, \\
\dv{p_{1}}{t} & = -a_{2}(t)p_{2} -m_{1}(t) \Omega_{1}^{2}(t) \e^{2 \int_{0}^{t} \lambda_{1}(s) \: \dd s}q_{1},
\end{split}
\qquad 
\begin{split}
\dv{q_{2}}{t} &=a_{2}(t) q_{1} +\frac{1}{m_{2}(t)} \e^{- 2\int_{0}^{t} \lambda_{2}(s) \: \dd s} p_{2}, \\
\dv{p_{2}}{t} & =a_{2}(t) p_{1} - m_{2}(t) \Omega_{2}^{2}(t) e^{2 \int_{0}^{t} \lambda_{2}(s) \: \dd s}  q_{2}.
\end{split}
\end{equation*}
\end{itemize}
Note that  multiple possible choices can be made for the $t$-dependent coupling `constant' $a_{2}(t)$ in \eqref{eq:sp4:cho} and  \eqref{eq:sp4:CCK} as, for example, its dependence on the masses $m_{i}(t)$ or the frequencies $\Omega_{i}(t)$. 

\section{Concluding remarks} \label{section:concluding}

In this work, the representation theoretical approach to Lie systems proposed in \cite{Campoamor2018,Campoamor2018a} has been refined and extended to the case of LH systems, whenever the underlying faithful representation is compatible with a symplectic structure. From the perspective of differential equations, the first-order time-dependent systems of ODEs associated to a linear realization by vector fields of a Lie algebra are linear systems with non-constants coefficients, and thus always possess a fundamental system of solutions, regardless of the fact that they are Lie systems \cite{Hur}. However, up to certain special types of the coefficient matrices, such fundamental solutions are quite difficult to compute, and the structure of LH systems constitutes a powerful tool, as it provides an algorithmic construction of time-independent constants of the motion, from which a superposition principle can be constructed \cite{Blasco2015}. LH systems associated to representations of Lie algebras provide other systems by means of the so-called reduction by invariants, giving rise to Lie (or LH, depending on the dimension) systems that are generally no more related to a representation, but that admit a natural interpretation in terms of generalized distributions \cite{FLAN,Stefan1974}. An important question that has not been addressed to in detail in the literature is whether a given LH system is intrinsic, i.e., if the dimension of the phase space where it is defined is minimal, in the sense that the system cannot be locally diffeomorphic to a lower dimensional LH system. A criterion that enables us to determine whether a LH system has this property has been given, also leading to the notion of the uncoupled sum of LH systems. 

The approach has been applied to construct new (intrinsic) LH systems based on the simple Lie algebra $\mathfrak{sp}(4,\mathbb{R})$, as well as on some relevant subalgebras, such as the (centrally extended) Schr\"odinger algebra $\widehat{\cS}(1)$ and the Lorentz algebra $\mathfrak{so}(1,3)$. It has been shown that, although for each of the subalgebras a superposition rule can be derived, by means of the coalgebra formalism \cite{Rag}, it may be computationally more efficient to derive it by restriction of an $\mathfrak{sp}(4,\mathbb{R})$-superposition principle. Several time-dependent LH systems associated to $\mathfrak{sp}(4,\mathbb{R})$ and the subalgebras have been obtained, generalizing relevant physical systems as the Bateman or Caldirola--Kanai Hamiltonians, as well as LH systems associated to time-dependent electromagnetic fields. Besides the use of the superposition rules, some of these systems can be reduced, for appropriate choices of the coefficient functions, to a system essentially consisting of two Riccati equations.  

We further observe that most of the above systems are related, to some extent, to various well-known (super)integrable systems, a feature that is not uncommon within the theory of LH systems. In this context, it should be observed that, by construction, such LH systems  are non-reducible, and as the constants of the motion do not necessarily Poisson commute, the systems are generally not integrable in the sense of Liouville. An open question in this direction that deserves further analysis is whether from LH systems associated to (super)integrable systems, additional constraints on the parameter functions can be imposed in order to derive Liouville-integrable subsystems. 

There are several general questions that emerge from this analysis that deserve further inspection. For instance, the reduction by invariants of the $\widehat{\cS}(1)$-LH system \eqref{eq:Schrodinger:system} on $T^{*} \R^{2}$ to the $\widehat{\cS}(1)$-Lie system \eqref{eq:Schrodinger:system3d} on $T^{*}\R \times \R$ shows that the system is, intrinsically, three-dimensional. In this context, it is natural to ask whether there is any geometric structure on $T^{*}\R \times \R$ compatible with the $\widehat{\cS}(1)$-Lie system \eqref{eq:Schrodinger:system3d}. When considering the embedding 
\begin{equation*}
i: T^{*} \R \times \R \hookrightarrow T^{*} \R^{2}, \qquad (q, p, s) \mapsto (q, p, 1, s),
\end{equation*} 
we have identified $T^{*}\R \times \R$ with the hypersurface $\set{q_{2} = 1} \subset T^{*} \R^{2}$. Now
\begin{equation*}
\mathbf{\Delta} := \frac{1}{2} \left( q_{1} \pdv{q_{1}} + q_{2} \pdv{q_{2}} + p_{1} \pdv{p_{1}} + p_{2} \pdv{p_{2}} \right)
\end{equation*}
is a Liouville vector field for  $T^{*} \R^{2}$ with respect to the canonical symplectic form $\omega$ (i.e., $\cL_{\mathbf{\Delta}} \omega = \omega$, so $\omega = \dd \iota_{\mathbf{\Delta}} \omega$ is an exact symplectic form) which, in addition,  is transverse to the hypersurface $i(T^{*}\R \times \R) \equiv \set{q_{2} = 1} \subset T^{*} \R^{2}$. This implies that the identity  
\begin{equation*}
\eta := i^{*}(\iota_{\mathbf{\Delta}} \omega)  = \frac{1}{2} \left( \dd s - p \dd q + q \dd p \right),
\label{eq:Schrodinger:contactform}
\end{equation*}
defines a contact form on $T^{*} \R \times \R$, turning $(T^{*}\R \times \R, \eta)$ into a (co-orientable) contact manifold \cite[Theorem 5.9, p. 136]{Godbillon1969}. The Reeb vector field $\cR \in \cv(T^{*} \R \times \R)$, determined by the conditions 
\begin{equation*}
\iota_{\cR} \eta = 1, \qquad \iota_{\cR} \dd \eta = 0,
\end{equation*}
is the vector field $\cR = \X^{\red}_{6} = 2 \partial_{s}$  belonging to the VG Lie algebra spanned by the vector fields \eqref{eq:Schrodinger:vf3d}. It is worthy to be observed that, if we consider the differentiable functions $h_{\alpha} \in C^{\infty}(T^{*} \R \times \R)$ given in \eqref{eq:Schrodinger:ham3d} (which are the restrictions  to $T^{*} \R \times \R$ of the Hamiltonian functions \eqref{eq:Schrodinger:ham}), the following relations hold: 
 \begin{equation}
\iota_{\X_{\alpha}^{\red}} \eta = - h_{\alpha}, \qquad \iota_{\X_{\alpha}^{\red}} \dd \eta = \dd h_{\alpha} - (\cL_{\cR} h_{\alpha}) \eta, \qquad 1 \leq \alpha \leq 6. 
\label{eq:Schrodinger:contactvf}
 \end{equation}
 From \eqref{eq:Schrodinger:contactvf} we see that  $\X_{\alpha}^{\red}$ is a Hamiltonian vector field with respect to the contact form $\eta$ whose associated Hamiltonian function is $h_{\alpha}$, for all $1 \leq \alpha \leq 6$. This shows that the $\widehat{\cS}(1)$-Lie system \eqref{eq:Schrodinger:system3d} on $T^{*} \R \times \R$ is a contact Lie system with respect to the contact form $\eta$ \cite{deLucas2023}. Moreover, as every Hamiltonian function \eqref{eq:Schrodinger:ham3d} is a first integral of the Reeb vector field $\cR = \X^{\red}_{6} = 2 \pdv{s}$, the contact Lie system \eqref{eq:Schrodinger:system3d} is of Liouville type in the terminology of \cite{deLucas2023}. Therefore, using \cite[Prop. 3.9]{deLucas2023}, the contact Lie system \eqref{eq:Schrodinger:system3d} on $T^{*} \R \times \R$ can be projected onto a LH system on the space $(T^{*}\R \times \R) / \cR \simeq T^{*}\R$ of integral curves of the Reeb vector field $\cR$ with respect to the symplectic form $\omega \in T^{*} \R$ given by $\pr^{*} \omega = \dd \eta$, with $\pr: T^{*} \R \times \R \to T^{*}\R$ being the projection \eqref{eq:Schrodinger:projection}. More precisely, we have that $\omega = \dd q \wedge \dd p$ is the canonical symplectic form of $T^{*} \R$, and that the projected vector fields are, precisely, those of \eqref{eq:Schrodinger:projectedvf}, so the projected LH system is just the system \eqref{eq:Schrodinger:system2d} of the P$_{5}$-LH class on $T^{*}\R$ \cite{Ballesteros2015}. 
 We also recall that, the constants of motion of the diagonal prolongations  of the  $\widehat{\cS}(1)$-contact Lie system \eqref{eq:Schrodinger:system3d}, which are precisely those of  \eqref{eq:Schrodinger:3rdconstant} and \eqref{eq:Schrodinger:3rdconstant_permutations}, can be obtained by applying the coalgebra method for Jacobi--Lie systems formed by \textit{good Hamiltonian functions} \cite[Prop.~6.7]{deLucas2023}.
 
 Summarizing, this alternative representation-theoretical approach to LH systems can also be considered, in some cases, in order to obtain Lie systems compatible with other geometric structures. Along these lines, it would be interesting to analyze the existence of compatible geometric structures for realizations coming from representations of the $N$-dimensional centrally extended Schrödinger algebra $\widehat{\cS}(N) = \sl(2, \R) \overrightarrow{\bigoplus}_{\Gamma_{\frac{j}{2}} \oplus \Gamma_{0}}  \h_{\frac{2j+1}{2}}$, with $N = 2j$, with the contact geometry setting being a natural and suitable candidate, as the central generator of $\widehat{\cS}(N)$  can be realized as a Reeb vector field. Another natural extension of this work concerns the construction of LH systems on $T^{*}\R^{n}$ associated to the symplectic real Lie algebra $\sp(2n, \R)$ and the defining representation, generalizing the construction carried out in Section~\ref{section:symplectic} for the $n = 2$ case. It is worthy to be observed that this generalization has been shown to be compatible with some \textit{scaling symmetries} \cite{Bravetti2023}, allowing us to reduce LH system on $\R^{2n} \equiv T^{*}\R^{n}$ to a contact Lie system on the $(2n-1)$-dimensional sphere $\S^{2n-1}$, by making use of the so-called \textit{Kirillov structures} \cite{Kirillov1976}. In this context,  when taking into account a \textit{Sasakian structure} \cite{Blair2010} compatible with the contact structure obtained on $\S^{2n-1}$ after the reduction, Liouville-type contact Lie systems enable us to provide an elegant and precise geometrical description, as they are associated to isometries of the associated Riemannian structure. Work along these various lines is currently in progress.  


\section*{Acknowledgements}

\phantomsection
\addcontentsline{toc}{section}{Acknowledgments}

This work has been supported by Agencia Estatal de Investigaci\'on (Spain)   under  the grant PID2023-148373NB-I00 funded by MCIN/AEI/10.13039/501100011033/FEDER, UE.
F.J.H.~acknowledges support  by the  Q-CAYLE Project  funded by the Regional Government of Castilla y Le\'on (Junta de Castilla y Le\'on, Spain) and by the Spanish Ministry of Science and Innovation (MCIN) through the European Union funds NextGenerationEU (PRTR C17.I1).
O.C. acknowledges a fellowship (grant C15/23) supported by  Universidad Complutense de Madrid and Banco Santander. The authors also acknowledge the contribution of  RED2022-134301-T  funded by MCIN/AEI/10.13039/501100011033 (Spain).


\appendix

\section{Some algebraic structures within the coalgebra method} \label{section:app}
  
In this appendix we briefly recall the main algebraic structures of the coalgebra method for LH systems described in Section~\ref{subsection:coalgebra}. For additional details the reader
is referred to \cite{Ballesteros2013,Lucas,Vaisman1994,Dixmier1974} and references therein.

A {\it Poisson algebra} over $\R$  is an $\R$-vector space $A$ equipped with a multiplication $A \times A \ni (a_{1}, a_{2}) \mapsto a_{1}a_{2} \in A$ inducing an associative $\R$-algebra structure on $A$, such that $(A, \set{\cdot, \cdot})$ is a Lie algebra for which the Leibniz rule 
\[ \set{b  c, a} = b  \set{c, a}  + \set{b, a}  c, \qquad a, b, c \in A\]
is satisfied. A {\it Casimir element} of $A$ is an element $C \in A$ such that $\set{C, a} = 0$ for all $a \in A$. If $A$ and $B$ are Poisson algebras, their tensor product $A \otimes B$ is also a Poisson algebra with the Poisson bracket 
\[ \set{a_{1} \otimes b_{1}, a_{2} \otimes b_{2}}_{A \otimes B} := \set{a_{1}, a_{2}}_{A} \otimes b_{1}b_{2} + a_{1}a_{2} \otimes \set{b_{1}, b_{2}}_{B}, \qquad a_{1},a_{2} \in A, \quad b_{1}, b_{2} \in B, \]
being $\set{\cdot, \cdot}_{A}$ and $\set{\cdot, \cdot}_{B}$ the Poisson brackets of $A$ and $B$, respectively. A {\it morphism of Poisson algebras} is a morphism $\varphi: A \to B$ of $\R$-algebras preserving the Poisson brackets:
\[ \varphi (\set{a, b}_{A}) = \set{\varphi(a), \varphi(b)}_{B}, \qquad a, b \in A. \] 
A {\it Poisson coalgebra} is a Poisson algebra $(A, \set{\cdot, \cdot})$ equipped with 
a morphism of Poisson algebras $\Delta: A \to A \otimes A$, the so-called {\it coproduct}, which is coassociative: $(\Delta \otimes \mathrm{Id}) \circ \Delta = ( \mathrm{Id} \otimes \Delta) \circ \Delta$.

Let $\g$ be an $r$-dimensional real Lie algebra with generators $X_{1}, \ldots, X_{r}$ and commutation rules $[X_{\alpha}, X_{\beta}] = C_{\alpha\beta}^{\gamma} X_{\gamma}$. Consider now the tensor algebra $T(\g) := \oplus_{k \geq 0} \g^{\otimes k}$ of $\g$, where $\g^{\otimes 0} := \R$ and $\g^{\otimes k}:= \g \otimes \overset{k}{\cdots} \otimes \g$ for $k \geq 1$. The bilateral ideal $\mathcal{R} \subset T(\g)$ spanned by the elements $X \otimes Y - Y \otimes X$ gives rise to the {\it symmetric algebra} $S(\g)$ of $\g$ through the quotient
\[ S(\g) := T(\g) / \mathcal{R}. \]

The dual $\g^{*}$ of $\g$ is canonically a Poisson manifold by means of the so-called \textit{Lie--Poisson} structure
\begin{equation}
\set{f, g} (\theta) = \langle [\dd f_{\theta}, \dd g_{\theta}], \theta \rangle, \qquad f, g \in C^{\infty}(\g^{*}), \qquad \theta \in \g^{*}
\label{eq:app_liepoisson}
\end{equation}
where $\langle \cdot, \cdot \rangle$ denotes the canonical pairing between $\g$ and $\g^{*}$ and the differential mappings $\dd f_{\theta}$ and $\dd g_{\theta}$ are considered as elements of $\g$ in a natural way (see \cite{Vaisman1994} for more details). Now, the generators $X_{1}, \ldots, X_{r}$ can be taken as global coordinates on $\g^{*}$, giving rise to the brackets 
\[ \set{X_{\alpha}, X_{\beta}} = C_{\alpha \beta}^{\gamma} X_{\gamma}. \]
As an $\R$-algebra, $S(\g)$ is  isomorphic to the algebra $\R[X_{1}, \ldots, X_{r}]$ of polynomials on the elements of $\g$. Thus, from the isomorphism $\g \simeq (\g^{*})^{*}$ we see that the elements of $S(\g)$ can be naturally interpreted as polynomial functions on the dual vector space $\g^{*}$ of $\g$. Consequently, the Lie--Poisson bracket \eqref{eq:app_liepoisson} induces a Poisson algebra structure on $S(\g)$.  

Finally, we recall that $S(\g)$ can be endowed with a Poisson coalgebra structure through the latter Poisson bracket together with the coproduct $\Delta: S(\g) \to S(\g) \otimes S(\g)$ specified on the generators of $\g$ by 
\[ \Delta(X_{\alpha}) =  X_{\alpha} \otimes 1 + 1 \otimes X_{\alpha}, \qquad 1 \leq \alpha \leq r. \]


\end{document}